\documentclass[twocolumn]{aastex631}

\usepackage{amsmath}

\shorttitle{Protocluster selection with ODIN }
\shortauthors{Ramakrishnan et al.}
\graphicspath{{./}{}}

\begin{document}

\newcommand{\ie}{$i.e.$,}
\newcommand{\mpc}{Mpc}
\newcommand{\Lya}{Ly$\alpha~$}
\newcommand{\msun}{M$_{\odot}$}
\newcommand{\cmtwo}{cm$^{-2}$}
\newcommand{\kms}{\,km~s$^{-1}$}      
\newcommand{\minpoint}{\mbox{$'\mskip-4.7mu.\mskip0.8mu$}}
\newcommand{\secpoint}{\mbox{$''\mskip-7.6mu.\,$}}
\newcommand{\sqdeg}{\mbox{${\rm deg}^2$}}
\newcommand{\squig}{\sim\!\!}
\newcommand{\subsun}{\mbox{$_{\twelvesy\odot}$}}
\newcommand{\et}{{\it et al.}~}
\newcommand{\er}[2]{$_{-#1}^{+#2}$}
\newcommand{\cgsflux}{ergs cm$^{-2}$ s$^{-1}$ Hz$^{-1}$}
\newcommand{\sdrel}{($1+\delta_{LAE}$)$~$}
\def\h50{\, h_{50}^{-1}}
\def\hbl{km~s$^{-1}$~Mpc$^{-1}$}
\def\ltsima{$\; \buildrel < \over \sim \;$}
\def\simlt{\lower.5ex\hbox{\ltsima}}
\def\gtsima{$\; \buildrel > \over \sim \;$}
\def\simgt{\lower.5ex\hbox{\gtsima}} 
\def\arcsec{$''$}
\def\arcmin{$'$}
\newcommand{\ebv}{$E$($B-V$)}

\newcommand{\cpurple}    {red!75!green!50!blue}
\newcommand{\unitcgssb}  {erg\,s$^{-1}$\,cm$^{-2}$\,arcsec$^{-2}$}
\newcommand{\unitcgslum} {erg\,s$^{-1}$}
\newcommand{\sqarcsec}   {arcsec$^2$}


\title{ODIN: Identifying Protoclusters and Cosmic Filaments Traced by Ly$\alpha$-emitting Galaxies}

\author[0000-0002-9176-7252]{Vandana Ramakrishnan}
\affiliation{Department of Physics and Astronomy, Purdue University, 525 Northwestern Avenue, West Lafayette, IN 47907, USA}

\author[0000-0003-3004-9596]{Kyoung-Soo Lee}
\affiliation{Department of Physics and Astronomy, Purdue University, 525 Northwestern Avenue, West Lafayette, IN 47907, USA}

\author[0000-0003-0570-785X]{Maria Celeste Artale}
\affiliation{Departamento de Ciencias Fisicas, Universidad Andres Bello, Fernandez Concha 700, Las Condes, Santiago, Chile}

\author[0000-0003-1530-8713]{Eric Gawiser}
\affiliation{Physics and Astronomy Department, Rutgers, The State University, Piscataway, NJ 08854}

\author[0000-0003-3078-2763]{Yujin Yang}
\affiliation{Korea Astronomy and Space Science Institute, 776 Daedeokdae-ro, Yuseong-gu, Daejeon 34055, Republic of Korea}

\author[0000-0001-9521-6397]{Changbom Park}
\affiliation{Korea Institute for Advanced Study, 85 Hoegi-ro, Dongdaemun-gu, Seoul 02455, Republic of Korea}

\author[0000-0001-6320-261X]{Yi-Kuan Chiang}
\affiliation{Institute of Astronomy and Astrophysics, Academia Sinica (ASIAA), Taipei 10617, Taiwan}

\author[0000-0002-1328-0211]{Robin Ciardullo} 
\affiliation{Department of Astronomy \& Astrophysics, The Pennsylvania State University, University Park, PA 16802, USA} \affiliation{Institute for Gravitation and the Cosmos, The Pennsylvania State University, University Park, PA 16802, USA}

\author[0000-0002-4928-4003]{Arjun Dey}
\affiliation{NSF’s National Optical-Infrared Astronomy Research Laboratory, 950 N. Cherry Ave., Tucson, AZ 85719, USA}

\author[0000-0001-6842-2371]{Caryl Gronwall}
\affiliation{Department of Astronomy \& Astrophysics, The Pennsylvania
State University, University Park, PA 16802, USA}
\affiliation{Institute for Gravitation and the Cosmos, The Pennsylvania
State University, University Park, PA 16802, USA}

\author[0000-0002-4902-0075]{Lucia Guaita}
\affiliation{Departamento de Ciencias Fisicas, Universidad Andres Bello, Fernandez Concha 700, Las Condes, Santiago, Chile}

\author[0000-0003-3428-7612]{Ho Seong Hwang}
\affiliation{Department of Physics and Astronomy, Seoul National University, 1 Gwanak-ro, Gwanak-gu, Seoul 08826, Republic of Korea}
\affiliation{SNU Astronomy Research Center, Seoul National University, 1 Gwanak-ro, Gwanak-gu, Seoul 08826, Republic of Korea}

\author[0009-0003-9748-4194]{Sang Hyeok Im}
\affiliation{Department of Physics and Astronomy, Seoul National University, 1 Gwanak-ro, Gwanak-gu, Seoul 08826, Republic of Korea}

\author[0000-0002-2770-808X]{Woong-Seob Jeong}
\affiliation{Korea Astronomy and Space Science Institute, 776 Daedeokdae-ro, Yuseong-gu, Daejeon 34055, Republic of Korea}

\author[0009-0002-3931-6697]{Seongjae Kim}
\affiliation{Korea Astronomy and Space Science Institute, 776 Daedeokdae-ro, Yuseong-gu, Daejeon 34055, Republic of Korea}

\author[0000-0001-6270-3527]{Ankit Kumar}
\affiliation{Departamento de Ciencias Fisicas, Universidad Andres Bello, Fernandez Concha 700, Las Condes, Santiago, Chile}

\author[0000-0002-6810-1778]{Jaehyun Lee}
\affiliation{Korea Astronomy and Space Science Institute, 776 Daedeokdae-ro, Yuseong-gu, Daejeon 34055, Republic of Korea}\affiliation{Korea Institute for Advanced Study, 85 Hoegi-ro, Dongdaemun-gu, Seoul 02455, Republic of Korea}

\author[:0000-0001-5342-8906]{Seong-Kook Lee}
\affiliation{SNU Astronomy Research Center, Seoul National University, 1 Gwanak-ro, Gwanak-gu, Seoul 08826, Republic of Korea}\affiliation{Department of Physics and Astronomy, Seoul National University, 1 Gwanak-ro, Gwanak-gu, Seoul 08826, Republic of Korea}

\author[0009-0008-4022-3870]{Byeongha Moon}
\affiliation{Korea Astronomy and Space Science Institute, 776 Daedeokdae-ro, Yuseong-gu, Daejeon 34055, Republic of Korea}

\author[0000-0001-9850-9419]{Nelson Padilla}
\affiliation{Instituto de Astronomía Teórica y Experimental (IATE), CONICET-UNC, Laprida 854, X500BGR, Córdoba, Argentina}

\author[0000-0001-8592-2706]{Alexandra Pope}
\affiliation{Department of Astronomy, University of Massachusetts, Amherst, MA 01003, USA}

\author[0000-0001-8245-7669]{Roxana Popescu}
\affiliation{Department of Astronomy, University of Massachusetts, Amherst, MA 01003, USA}

\author[0009-0000-7651-3900]{Akriti Singh}
\affiliation{European Southern Observatory, Alonso de Córdova 3107, Vitacura, Santiago de Chile, Chile}\affiliation{Departamento de Ciencias Fisicas, Universidad Andres Bello, Fernandez Concha 700, Las Condes, Santiago, Chile}

\author[0000-0002-4362-4070]{Hyunmi Song}
\affiliation{Department of Astronomy and Space Science, Chungnam National University, 99 Daehak-ro, Yuseong-gu, Daejeon, 34134, Republic of Korea}

\author[0000-0001-6162-3023]{Paulina Troncoso}
\affiliation{Escuela de Ingeniería, Universidad Central de Chile, Avenida Francisco de Aguirre 0405, 171-0614 La Serena, Coquimbo, Chile}

\author[0000-0001-5567-1301]{Francisco Valdes}
\affiliation{NSF's National Optical-Infrared Astronomy Research Laboratory, 950 N. Cherry Avenue, Tucson, AZ 85719, USA}

\author[0000-0001-6047-8469]{Ann Zabludoff}
\affiliation{Steward Observatory, University of Arizona, 933 North Cherry Avenue, Tucson, AZ 85721, USA}

\begin{abstract}

To understand the formation and evolution of massive cosmic structures, studying them at high redshift, in the epoch when they formed the majority of their mass is essential. The One-hundred-deg$^2$ DECam Imaging in Narrowbands (ODIN) survey is undertaking the widest-area narrowband program to date, to use Ly$\alpha$-emitting galaxies (LAEs) to trace the large-scale structure (LSS) of the Universe {on the scale of 10 - 100 cMpc} at three cosmic epochs. In this work, we present results at $z$ = 3.1 based on early ODIN data in the COSMOS field. 
We identify and characterize protoclusters and cosmic filaments using multiple methods and discuss their strengths and weaknesses. We then compare our observations against the IllustrisTNG suite of cosmological hydrodynamical simulations. The two are in excellent agreement, 
with a similar number and angular size of structures identified above a specified density threshold. We are able to recover the simulated protoclusters with $\log$(M$_{z=0}$/$M_\odot$) $\gtrsim$ 14.4 in $\sim$ 60\% of the cases. With these objects we show that the descendant masses of the protoclusters in our sample can be estimated purely based on our 2D measurements, finding a median $z$ = 0 mass of $\sim10^{14.5}$M$_\odot$. The lack of information on the radial extent of each protocluster introduces a $\sim$0.4~dex uncertainty in its descendant mass. Finally, we show 
that the recovery of the cosmic web in the vicinity of protoclusters is both efficient and accurate. The similarity of our observations and the simulations imply that our structure selection is likewise robust and efficient, demonstrating that LAEs are reliable tracers of the LSS. 

\end{abstract}


\section{Introduction} \label{sec:intro}

According to the hierarchical theory of structure formation, matter is organized into a {\it cosmic web}, comprised of linear filaments intersecting at nodes of high density surrounded by vast voids \citep{bond96, Springel2005, Bolyan_Kolchin2009}. This large-scale structure (LSS) determines how much cold gas is available to a galaxy and the likelihood of a merger or interaction with another galaxy, thereby acting as one of the fundamental drivers of galaxy evolution. 


Out to $z \approx 1.5$, redshift surveys and other observational techniques have enabled the selection of samples of galaxies inhabiting clusters, groups, and filaments \citep[e.g.,][]{Eisenhardt2008, Koyama2014, Rykoff2014, Tempel2014, Bleem2015, Rykoff2016, Malavasi2017, Hayashi2018, Gonzalez2019}. These studies show that galaxies in cluster or group environments tend to be older and more massive, and are more likely to have ceased star formation than those in the field \citep[e.g.,][]{Peng2010, Quadri2012, van_der_Burg2013}. Filaments may have a weaker but similar effect and may be responsible for pre-processing galaxies that are falling into cluster- or group environments \citep[e.g.,][]{Sarron2019,Salerno2019}.


At Cosmic Noon ($z \gtrsim 2$), when the global star formation rate reached its peak \citep{Madau2014}, these environmental effects are predicted to be even more dramatic. In the high-density regions within the LSS, the accretion rates of infalling gas and the incidence of galaxy interactions are expected to be greatest {\citep[see discussion in][and references therein]{Lemaux2022}}, fostering both enhanced in-situ star formation and black hole activity. A popular hypothesis is that highly dissipative gas-rich mergers help the efficient feeding of gas into the central black hole and trigger active galactic nuclei (AGNs), which may ultimately quench the star formation activity \citep[e.g.,][]{hopkins06}. These expectations are indeed in line with the heightened SF and AGN activities found in a handful of {\it protocluster} systems \citep[e.g.,][]{Casey2015,wang16, Umehata2015,Oteo2018,Harikane2019,Lemaux2022} as well as the emergence of quenched galaxies in such environments \citep[e.g.,][]{Shi2021,Ito2023}. 

Yet, the role that the LSS environment plays in galaxy formation at Cosmic Noon remains under-explored. Our limited knowledge is due to a combination of factors, including the observational limitations of measuring precise redshifts of faint, high-redshift galaxies, the inherent scarcity of massive cosmic structures, and our incomplete grasp of the indicators for the locations of these structures. The lack of readily identifiable signatures—such as a hot intracluster medium and/or a concentration of quiescent galaxies—in young, yet-to-be-virialized structures of mostly star-forming galaxies leads to a strong reliance on spectroscopy for finding protoclusters. While the lack of large, statistical samples prevent us from disentangling the effects of cosmic variance from general properties of protocluster galaxies, studies from heterogeneously selected samples can lead to seemingly conflicting conclusions. Although cosmic filaments connected to these protoclusters likely play a vital role in replenishing fresh gas for sustained star formation, such medium-density features are even more difficult to discern than dense protocluster cores.

The One-hundred-deg$^2$ DECam Imaging in Narrowbands \citep[ODIN,][]{Lee2024} survey is designed to obtain large and \emph{uniformly selected} samples of protoclusters and filaments at three cosmic epochs ($z = 2.4$, 3.1 and 4.5) using Ly$\alpha$-emitting galaxies (LAEs) as tracers of underlying matter distribution. As the most common electron transition in the universe, Ly$\alpha$ emission traces ionized and/or excited gas from star formation, black hole activity, and the gravitational collapse of dark matter halos. A large fraction of low-luminosity star-forming galaxies \citep[which dominate the cosmic star formation rate density:][]{reddy09} show Ly$\alpha$ emission \citep{stark10}. These LAEs tend to have lower stellar masses, younger population ages, less internal extinction than systems selected via their broadband colors, and low galaxy bias \citep{Gawiser2007,Kusakabe2018,weiss21,White2024}. These traits make LAEs a most efficient tracer of the underlying dark matter distribution, one which can be used to constrain cosmology \citep[e.g.,][]{gebhardt21,White2024} and the most massive cosmic structures \citep{Dey2016,huang22}. 
Upon completion, $\approx$600 protoclusters are expected to be discovered by ODIN, which will facilitate robust statistical investigations of the cosmic evolution of protoclusters and their galaxy inhabitants.

In this paper, we use the early ODIN data in the COSMOS field to present our selection of protoclusters and cosmic filaments. Building on the results presented in \citet{Ramakrishnan2023}, we calibrate and fine-tune our LSS detection methods by carrying out careful comparisons with cosmological hydrodynamical simulations. The outline of this paper is as follows. In Section~\ref{sec:data}, we give details of the observational and simulation data. In Sections~\ref{sec:sd_maps} and \ref{sec:protoclusters}, we describe how we construct LAE surface density maps and how we use them to detect protoclusters and filaments. We validate our procedures and interpret the results using the mock data created from the simulations in Section~\ref{sec:comparison_with_TNG}. Finally, the properties of our observationally selected structures are discussed in Section \ref{sec:discussion} followed by a summary of our findings in Section~\ref{sec:summary}. Throughout this paper, we assume a Planck cosmology \citep{Planck2016_cosmology}: $\Omega_{\rm \Lambda} = 0.6911$, $\Omega_{\rm b} = 0.0486$, $\Omega_{\rm m} = 0.3089$, $H_{0} = 100\,h\,{\rm km}\,{\rm s}^{-1}\,{\rm Mpc}^{-1}$ and $h=0.6774$. Distances are given in units of comoving Mpc (cMpc) unless noted otherwise. 


\section{Observational and simulation data} \label{sec:data}

\subsection{The ODIN survey} \label{subsec:odin}

The ODIN survey is conducting the widest-area deep narrowband imaging program to date, using three custom narrowband filters ($N419$, $N501$, and $N673$) to identify redshifted \Lya emission. In this work, we make use of the Year 1 ODIN data taken with the $N501$ filter ($\lambda_C$/$\Delta\lambda$ $=$ 5014/75~\AA; $\bar{z}$/$\Delta z$ $=$ 3.12/0.06) in the extended COSMOS field. {The data has a resolution of 0$\farcs$27~pix$^{-1}$ and a 5$\sigma$ magnitude limit of 25.4 AB over a total area of $\sim$ 9 deg$^2$, allowing us to detect Ly$\alpha$ emission down to a line flux of $\sim$ 2.1 $\times$ 10$^{-17}$ ergs s$^{-1}$ cm$^{-2}$. 
As shown in Table \ref{tab:imaging}, our narrowband data is similar in depth to the deepest surveys undertaken at similar redshifts while being $\sim$ 4 times or larger in area.} For more details about the survey fields and observing strategy, we refer interested readers to \citet{Lee2024}.

\begin{figure*}
    \centering
    \includegraphics[width=\linewidth]{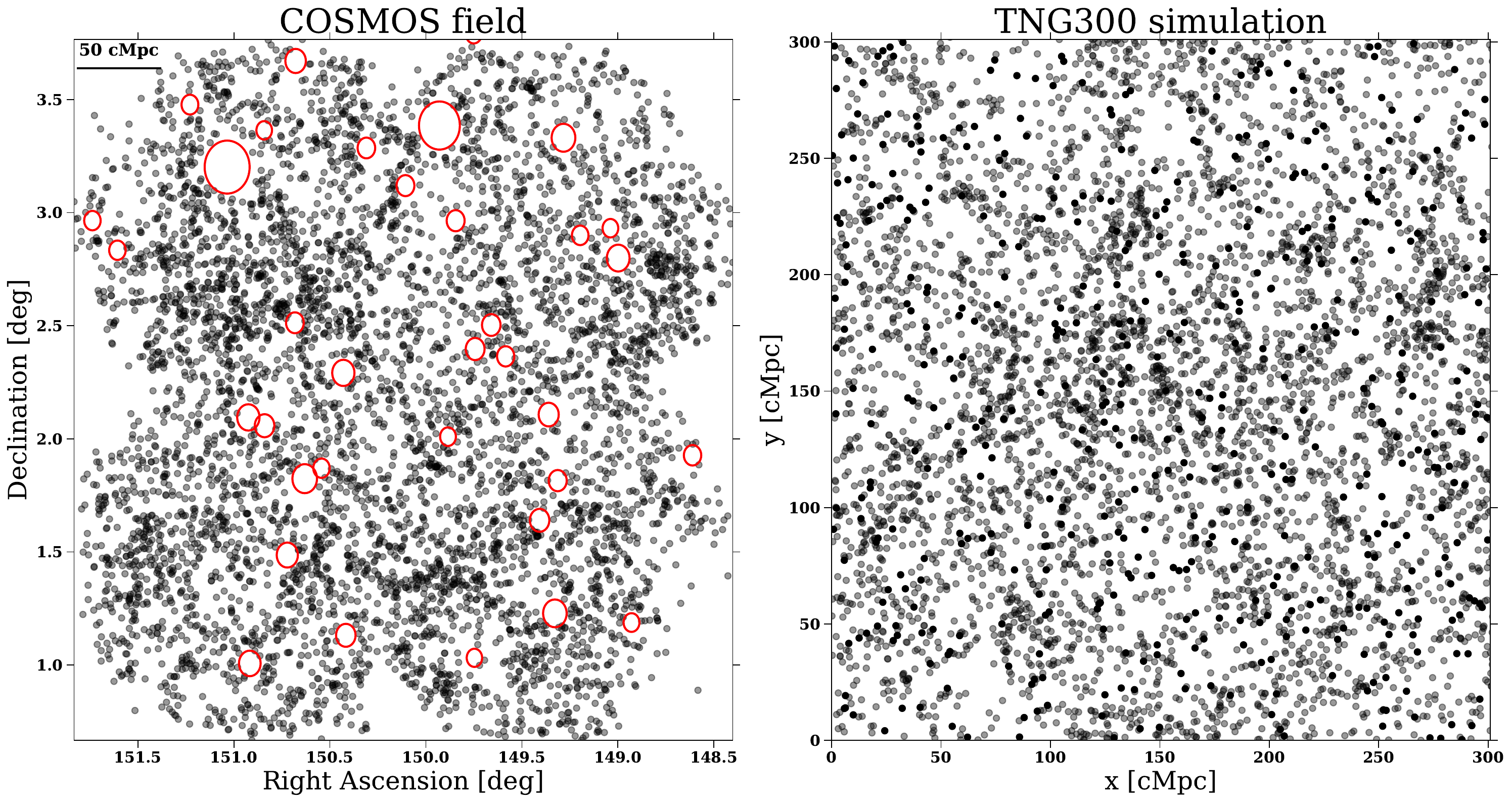}
    \caption{Positions of the observed ODIN LAEs (left panel, see text in Section \ref{subsec:odin}) and the mock LAEs from a randomly chosen slice of TNG300 (right panel, see text in Section \ref{subsec:tng}). In the left panel, the holes left by bright stars are shown by red circles.}
    \label{fig:lae_positions}
\end{figure*}

{The positions of the ODIN LAEs are shown in the left panel of Figure~\ref{fig:lae_positions}.} The method for selecting these LAEs is detailed in \citet{Firestone2023}. Briefly, we identify LAEs as $N501$-detected sources exhibiting a narrowband excess over the continuum, corresponding to a rest-frame equivalent width of 20~\AA. The continuum magnitude is calculated as a weighted combination of the magnitude in two broadband filters ($g$ and $r$ for $N501$ LAEs). {We reject sources flagged for saturated pixels or other image defects and those close to bright stars, with the total excluded area being $\sim$ 1.5 deg$^2$. The resultant gaps left in the LAE source distribution are highlighted in Figure~\ref{fig:lae_positions}. The largest of these is $\sim$ 144~arcmin$^2$ in area but there are only two such holes in the field. The remainder are much smaller ($\lesssim$ 36~arcmin$^2$).} The continuum is estimated using broadband data from the Hyper Suprime-Cam Subaru Strategic Program \citep[HSC SSP:][]{Aihara2018a, Aihara2018b} second data release \citep{Aihara2019}. 

{The final LAE sample comprises 5,691 sources over $\approx$7.5~deg$^2$, corresponding to a surface density of 0.21 arcmin$^{-2}$. This is comparable to other narrowband surveys with similar filter widths and narrowband depths (see Table \ref{tab:imaging}).}    
{Spectroscopic follow-up efforts for ODIN LAEs are ongoing, but sufficient spectra have been taken to obtain a preliminary estimate of the purity of our LAE sample. Of the sources that yielded a redshift, $\approx$97\% are confirmed as LAEs (V. Ramakrishnan et al.\ in prep.).} 

\def\etal{{et~al.\null}}

\begin{deluxetable*}{lcclcccc} \label{tab:imaging}
\tablecaption{{Existing Samples of $z=3.1$ LAEs}}
\tablehead{\colhead{Sample} & \colhead{\# of} &\colhead{Total Area} &\colhead{Depth} &
\colhead{EW$_0$} & \colhead{$\Delta z$}     &\colhead{\# of} &\colhead{Surface density}\\
&\colhead{Fields}   &\colhead{(deg$^2$)} &\colhead{(ergs~s$^{-1}$)}
&\colhead{(\AA)} & &\colhead{LAEs} &\colhead{(arcmin$^{-2}$)}}
\startdata
\textbf{This work} & \textbf{1} & \textbf{9.0} & \textbf{1.9} $\mathbf{\times~10^{42}}$ & \textbf{20} & \textbf{0.059} & \textbf{5,691} & \textbf{0.21}\\
\citet{Ciardullo2002} &1 &0.13   &4.5 $\times~10^{42}$  &20 &0.045 &9 & 0.02 \\
\citet{Hayashino2004} &1 &0.21   &4.0 $\times~10^{42}$  &38 &0.063 &283 & 0.37\\
\citet{Gronwall2007}  &1 &0.28   &1.3 $\times~10^{42}$  &20 &0.024 &259 & 0.26\\
\citet{Ouchi2008}     &1 &0.98   &1   $\times~10^{42}$  &64 &0.061 &356 & 0.10 \\
\citet{Matsuda2009}   &1 &0.20   &1.3 $\times~10^{42}$  &20 &0.061 &127 & 0.18 \\
\citet{Ciardullo2012} &1 &0.28   &2.4 $\times~10^{42}$  &20 &0.047 &199 & 0.20 \\
\citet{Yamada2012}   &4 &2.42   &1.5 $\times~10^{42}$  &46 &0.063 & 2,161 & 0.25 \\
\citet{Bielby2016} &5 &1.25   &1.0   $\times~10^{42}$  &65 &0.063 &643 & 0.14\\
\citet{Matthee2017} &1 &0.85   &1.0   $\times~10^{43}$  &12 &0.082 &65 & 0.02\\
\enddata
\end{deluxetable*}


\subsection{Building Mock ODIN Observations with TNG} \label{subsec:tng}

To build a concrete framework in which we can interpret our observations, we use the IllustrisTNG300-1 simulation \citep[hereafter TNG300:][]{Nelson2019,Pillepich2018a,Pillepich2018b} and define our mock LAE samples. {TNG300 provides the largest volume (302.6~cMpc on a side) of all available simulations of the IllustrisTNG suite.} 
Given the rarity of massive galaxy (proto)clusters, this is especially crucial for our work.
All TNG simulations assume the Planck cosmology \citep{Planck2016_cosmology}.
The surface area viewed along the X, Y, or Z direction is $\approx$90,000 cMpc$^2$, well matched to the angular extent of the ODIN data in the COSMOS field, $\approx$120,000 cMpc$^2$, and the simulation box is several times larger than the ODIN radial extent of $\sim$60~cMpc at $z=3.1$.  
TNG300 has a baryon mass resolution of $1.1 \times 10^7 M_\odot$ and a dark matter mass resolution of 5.9 $\times$ 10$^7$ $M_\odot$. In our analysis we employ galaxies with stellar mass greater than 10$^7$ M$_\odot$, corresponding to a halo mass of $\gtrsim$ 10$^9$ M$_\odot$; as we do not use any stellar or gas physics in our analysis, the resolution is sufficient for our purposes. 

In constructing mock LAE samples, our primary goal is to mimic the spatial distribution of the ODIN LAEs as closely as possible so that we can evaluate their utility as tracers of the LSS\null. Understanding the complex behavior of Ly$\alpha$ radiative transfer requires vastly higher resolution simulations and thus is outside the scope of this work. 
Our selection of mock LAEs from among the TNG300 galaxies is hence based only on their stellar mass; we do not attempt a selection based on the star formation rate or modeled \Lya flux or equivalent width as in, e.g., \citet{Dijkstra2010,Ravi2024}. We describe our procedure in detail below.

First, we match the 75~\AA\ (60 cMpc at $z$ = 3.1) full-width-at-half-maximum of $N501$ by creating cosmic `slices' from the TNG300 $z=3$ snapshot with $\approx$80~cMpc in line-of-sight thickness. The filter and thus the window function are not a perfect top hat in shape. To emulate this effect, we assign the LAE selection probability to match the shape of the filter transmission function. In all things being equal, the probability of being selected as a mock LAE is 1, 0.5, and 0 at the distance of 0, 30, and 40~cMpc from the center of the slice, respectively. In practice, the probability of an LAE being detected at a given position along the redshift direction also depends on its line luminosity, as bright LAEs are more likely to be detected when they fall on the wings of the filter than fainter ones, and hence are detected over a larger volume. However, since we are interested only in the average number density of \emph{all} LAEs, irrespective of line luminosity, this does not significantly affect our analysis. 




Second, we aim to reproduce the small- and large-scale clustering of LAEs. The galaxy bias for $z\sim 3$ LAEs is relatively low at $b\lesssim 2$ \citep[][]{Gawiser2007,Ouchi2008,White2024}, suggesting that LAEs have low stellar mass content and are hosted by low-mass halos \citep[e.g.,][]{Guaita2010,Kusakabe2018}. To match the small-scale clustering, we must simultaneously match the LAE overdensity distribution across the field; the details of this measurement are presented in Section~\ref{sec:sd_maps}. Motivated by the findings of \citet{Hagen2014}, {we select the mock LAEs such that their stellar masses obey} 
a lognormal distribution: i.e., $\log M_\ast/M_\odot$ is a Gaussian function, with mean and standard deviation ($\mu$, $\sigma$).  Lowering $\mu$ values would shift the host halos to lower masses (thus lower bias). A larger scatter $\sigma$ (while fixing $\mu$) would permit both higher- and lower-mass halos to host LAEs, thereby changing how LAEs trace these halos. We find that $\mu = 8.75$ and $\sigma=0.75$ yield the best fit to our surface density measurements. The left panel of Figure~\ref{fig:real_vs_tng_hist} shows the resultant distribution of the LAE overdensity relative to the observations. Our best-fit parameters are fully consistent with the stellar mass distribution of LAEs based on SED fitting reported by \citet{Hagen2014}, log$(M_*/M_\odot$) = 8.97$^{+0.60}_{-0.71}$. They are also similar to the result of \citet{Vargas2014}, who find log$(M_*/M_\odot)$ = 8.45$^{+0.72}_{-0.67}$ for LAEs at $z$ = 2.1. 

Finally, we match the measured sky density of LAEs. The $N501$ LAE surface density is 0.21~arcmin$^{-2}$ {(see Section \ref{subsec:odin})}. In comparison, the number of TNG300 galaxies selected based on the above criteria is typically ten times greater. Additionally, we assume 10\% of our LAEs may be contaminants. This number is based on the fraction of our spectroscopic targets in the two faintest bins ($N501$=24.5--25.0 and 25.0--25.5~AB) that yielded a redshift, which is 94.3\% and 85.9\%, respectively (V.\ Ramakrishnan et al.\ in prep.). By doing so, we are assuming that all sources that did not result in a redshift are interlopers. Our spectroscopic success rate (i.e., the fraction of sources that yield a redshift for which it is within the expected range) is 97\%, so our estimate is conservative. Changing it to 5\% does not change our results. From the selected TNG300 galaxies, we randomly choose a subset whose number is equal to 90\% of the LAE density; assuming that LAE sample contaminants are unclustered, the remaining 10\% is drawn as a purely random distribution within the same region.

To compare directly with our observations, we utilize the progenitors of 30 of the most massive galaxy clusters in the final simulation box of TNG300, ranked based on their total stellar mass. The stellar mass, halo mass, and group ID of these clusters at $z$ = 0 are found in \citet{Andrews2024}. We identify the protoclusters by tracing the main progenitor branch of the merger trees of the $z$ = 0 clusters to their $z$ = 3 predecessors. The descendant masses of the protoclusters range from $1.5 \times 10^{15}M_\odot$ for the most massive system to $2 \times 10^{14}M_\odot$ for the least massive. We generate cosmic slices, aligning the midpoint of each with that of a protocluster using the periodic boundary conditions of the simulation. Slicing the TNG300 volume along the X, Y and Z axes produces 90 slices for comparison. In a given slice, typically a few other protoclusters from the top 30 most massive sample are included although, depending on their positions in the redshift direction, they may be only partially represented. {The positions of the mock LAEs in a randomly chosen slice are shown in the right panel of Figure \ref{fig:lae_positions}.}

\section{Tracing large-scale structure} \label{sec:sd_maps}

This section aims to delineate the large-scale structure in our observations by constructing the LAE surface density maps. We employ two methods; first, smoothing over the LAE positions with a fixed Gaussian kernel; and second, constructing the Voronoi diagram of the LAEs. These two methods are summarized below but are discussed in more detail in \citet{Ramakrishnan2023}. 

\subsection{Gaussian smoothing} \label{subsec:gauss}

The simplest approach to measure the surface density across a field involves smoothing over the positions of the galaxies within it using a fixed-size kernel. We employ a two-dimensional Gaussian kernel with FWHM 10~cMpc. This FWHM is decided following the methodology of \citet{Badescu2017}, also utilized in \citet{Ramakrishnan2023}. The kernel size is chosen such that the resultant surface density map maximizes the total probability at the positions of the LAEs in the real data. This is achieved through a leave-one-out cross-validation, where the likelihood of finding a point at the location $\vec{r_j}$ of the $j$th data point is estimated as:
\begin{equation}
    p(\vec{r_j}) = \sum_{i\neq j} \frac{1}{\sqrt{2\pi}\sigma}\exp{\frac{-(\vec{r_i}-\vec{r_j})^2}{2\sigma^2}}
\end{equation}
The optimum $\sigma$ value is the one which maximizes $\prod_{j}p(\vec{r_j})$. 


\begin{figure*}
    \centering
    \includegraphics[width=\linewidth]{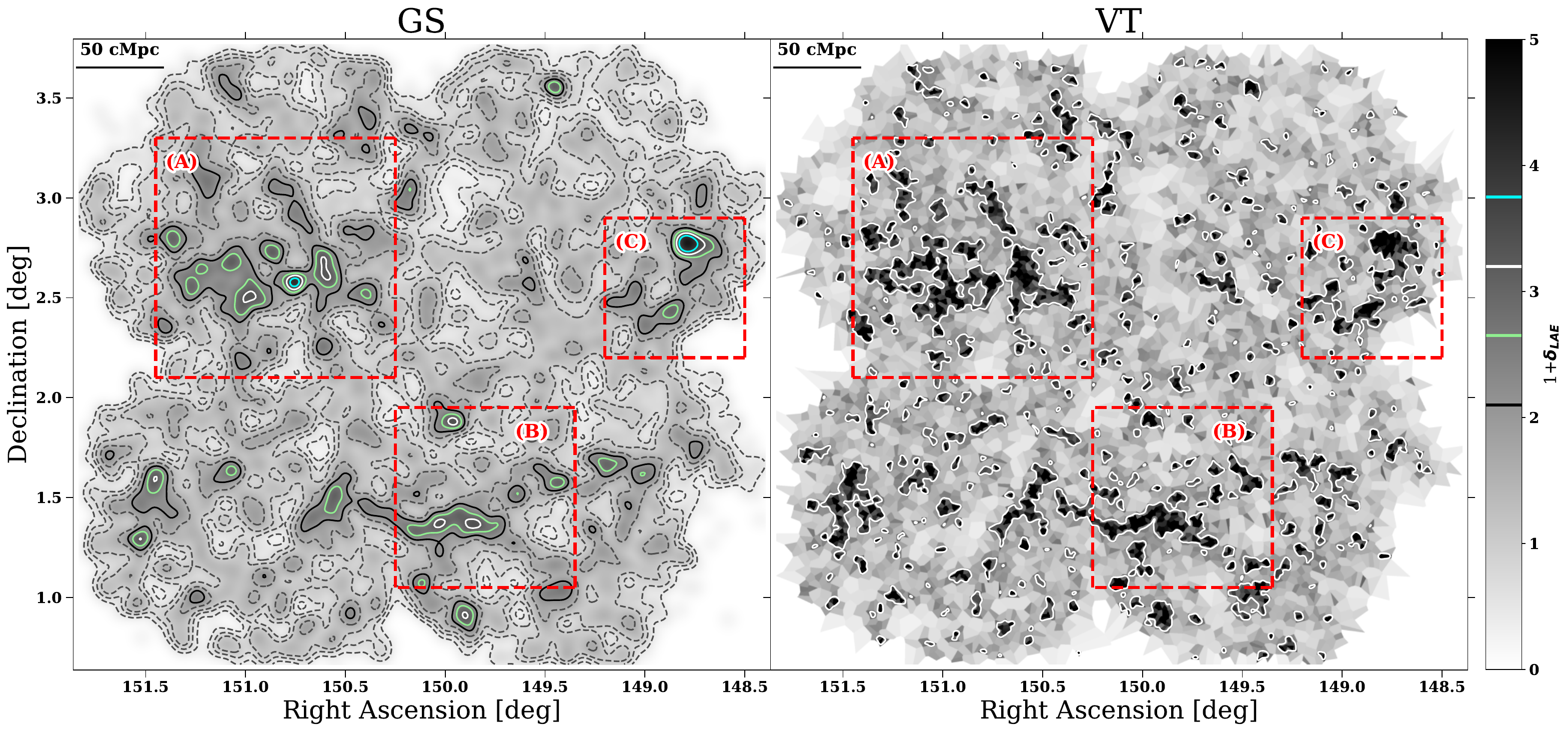}
    \caption{\emph{Left:} Gaussian smoothed (GS) LAE surface density map. Black, green, white, and cyan contours indicate LAE overdensities of 2, 3, 4, and 5$\sigma$ significance respectively; the corresponding values of \sdrel are marked on the color bar. The three largest overdensity complexes are highlighted in red. \emph{Right:} VT-based density map. White contours indicate overdensities of 3$\sigma$ significance. The LSS revealed by the surface density maps is highly clumpy and irregular. Non-spherical/irregular overdensities are better delineated at fixed significance in the VT map than in the GS map. The clover shape of the surface density maps arises from the HSC-SSP data, which consists of four Deep and one central UltraDeep pointing in the COSMOS field.}
    \label{fig:sd_maps}
\end{figure*}

Before creating the LAE surface density map, we fill in the holes left by removing the sources near bright stars and image defects with uniformly distributed random points with surface density matched to that of the LAEs. {We do not find significant differences between the surface density maps produced before and after carrying out this procedure}. After convolving with the kernel, the overdensity map is computed by dividing the $\Sigma_{LAE}$ map by the mean surface density, $\bar{\Sigma}_{\rm LAE}$:
\begin{equation}
    (1 + \delta_{LAE}) = \frac{\Sigma_{LAE}}{\bar{\Sigma}_{LAE}}
    \label{eq:lae_od}
\end{equation}
The mean and standard deviation are determined by fitting the LAE surface density distribution 
with a Gaussian function, $\exp[-(\Sigma_{LAE} - \overline{\Sigma}_{LAE})^2/2\sigma^2]$. The fit is restricted to within $\pm 1.5\sigma$ (after iterative sigma-clipping) to ensure that our estimate is not biased by the presence of multiple high LAE overdensities, which show up at the high \sdrel end of the distribution. 


The left panel of Figure~\ref{fig:sd_maps} shows the resultant \sdrel map, which we will refer to as the GS map, hereafter. 
It shows overdense regions that are both strongly clustered and highly irregular in shape, consistent with expectations from the hierarchical theory of structure formation, in which smaller structures continuously merge to form larger ones, a phenomenon supported by hydrodynamical simulations \citep[e.g.,][]{Bolyan_Kolchin2009}. The three most prominent overdensity complexes, highlighted by red boxes and labeled as complexes~A, B, and C, were discussed in detail in \citet{Ramakrishnan2023}. 

Smoothing with a fixed kernel is most effective at identifying structures with a size and shape similar to that of the kernel itself. Thus, the Gaussian smoothing method may not adequately capture non-isotropic features, potentially resulting in an underestimation of the significance of many observed structures. To address this, we explore tessellation-based methods in Section~\ref{subsec:voronoi}.

\subsection{Voronoi tessellation} \label{subsec:voronoi}

Tessellation-based density estimates offer the advantage of being scale-independent and do not assume any specific shape or size for the underlying structures. The two algorithms most commonly used in the literature are the Voronoi tessellation \citep[VT; e.g.,][]{Dey2016,Cucciati2018,Lemaux2018,Hung2020} and the Delaunay tessellation \citep[e.g.,][]{Malavasi2021, Sarron2021}. \citet{Darvish2015} find, through analysis of a simulated dataset, that the Delaunay tessellation fares more poorly at estimating the `true' surface density value at a given point as compared to the VT. Additionally, the method tends to overestimate the surface density in overdense regions. Considering these results, we opt for the Voronoi tessellation algorithm as outlined below.

In a two-dimensional case, the VT divides a plane into distinct cells based on the positions of a set of generating points e.g., the sky locations of LAEs. The Voronoi cell of each generating point includes all regions in the plane that are closer to it than to any other generating point. Consequently, the area of each cell is a measure of surface density: cells associated with LAEs in overdense regions will be small due to the proximity of numerous other LAEs, whereas those in underdense regions will be larger. Since each cell contains a single LAE, its LAE surface density $\Sigma_i$ is given by, 
\begin{equation}
    \Sigma_i = \frac{1}{A_i}
\end{equation}
where $A_i$ is the area of the $i$th cell. As before, the masked regions are filled in before constructing the Voronoi diagram.

In Figure~\ref{fig:sd_maps}, we show the VT \sdrel map together with the GS map. While both maps detect the most significant structures, the VT method detects them at higher significance. As expected, the method also identifies a greater number of overdensities by capturing irregular or anisotropic features. 




\section{Features of the LSS in observations: protoclusters and filaments} \label{sec:protoclusters}

Galaxy protoclusters represent some of the most striking features of large-scale structure at high redshift and are expected to be observed as significant galaxy overdensities spanning $\approx$10~cMpc in scale \citep{Chiang2013}. According to the hierarchical theory of structure formation, massive halos hosting protoclusters are connected to filaments of the cosmic web \citep[e.g.,][]{Kuchner2022} along which pristine gas is being accreted to feed star formation. Indeed, these predictions align qualitatively with the features observed in our \sdrel map. Multiple overdensities with angular scales of 5--10~cMpc are evident in Figure~\ref{fig:sd_maps} (right), frequently clustered together to form {\it complexes} comprising 2 to 5 adjacent overdensities. Additionally, regions of high overdensity, where \sdrel $>$ 3, are interconnected by more moderate `bridges' with \sdrel $=$ 2--3.  Several features show a distinctly linear morphology, reminiscent of cosmic filaments \citep[for example, the northwestern corner of ``Complex~A'' and the southeastern corner of ``Complex~C'', see][for more detail]{Ramakrishnan2023}. 

Motivated by these observations, our next objective is to pinpoint the locations of protoclusters and filaments of the cosmic web using LAEs as tracers. By directly comparing our observations with {predictions based on the TNG300 mock LAEs}, 
we will also test how robust our protocluster candidates are and measure their key properties.


\subsection{Selecting protoclusters from density maps} \label{subsec:SEP}

Protoclusters are generally defined as regions that will evolve into a virialized structure with masses $\gtrsim$ 10$^{14} M_\odot$ by $z = 0$ \citep[e.g.,][]{Overzier2016}. 
In distant look-back times, protoclusters are loosely bound regions comprised of multiple dark matter halos. This poses a challenge in measuring their physical extent. 
%
In this study, we define a protocluster as a region exhibiting significant size and overdensity, unlikely to occur by chance alone, and containing sufficient mass that could plausibly collapse to form a single halo with mass greater than or equal to 10$^{14} M_\odot$ by $z = 0$.


According to this operational definition, selecting protoclusters entails identifying contiguous regions above a pre-set overdensity threshold and minimum area. This procedure mirrors source detection in pixelated astrophysical images. Therefore, adopting approaches similar to those outlined by \citet{Lemaux2018,Hung2020,Sarron2021}, we utilize source detection software on our density maps for protocluster selection.

We begin by pixelating the surface density maps by interpolating them over a two-dimensional grid of positions with pixel size 3\farcs6. Our grid size, corresponding to 115~ckpc at $z = 3.1$, is small enough to clearly delineate the boundaries of the structures, as it is two orders of magnitude smaller than the typical size of a protocluster. For source detection, we use \textsc{SEP} \citep{Barbary2016}, a Python implementation of the \textsc{SExtractor} \citep{Bertin1996} software. The number of detected structures depends strongly on the detection threshold ({\tt DETECT\_THRESH}) and minimum area ({\tt DETECT\_MINAREA}). Our goal is to optimize these parameters to maximize the identification of robust candidates while minimizing the inclusion of spurious objects resulting from chance alignments of LAEs. 

\begin{figure*}
    \centering
    \includegraphics[width=0.8\linewidth]{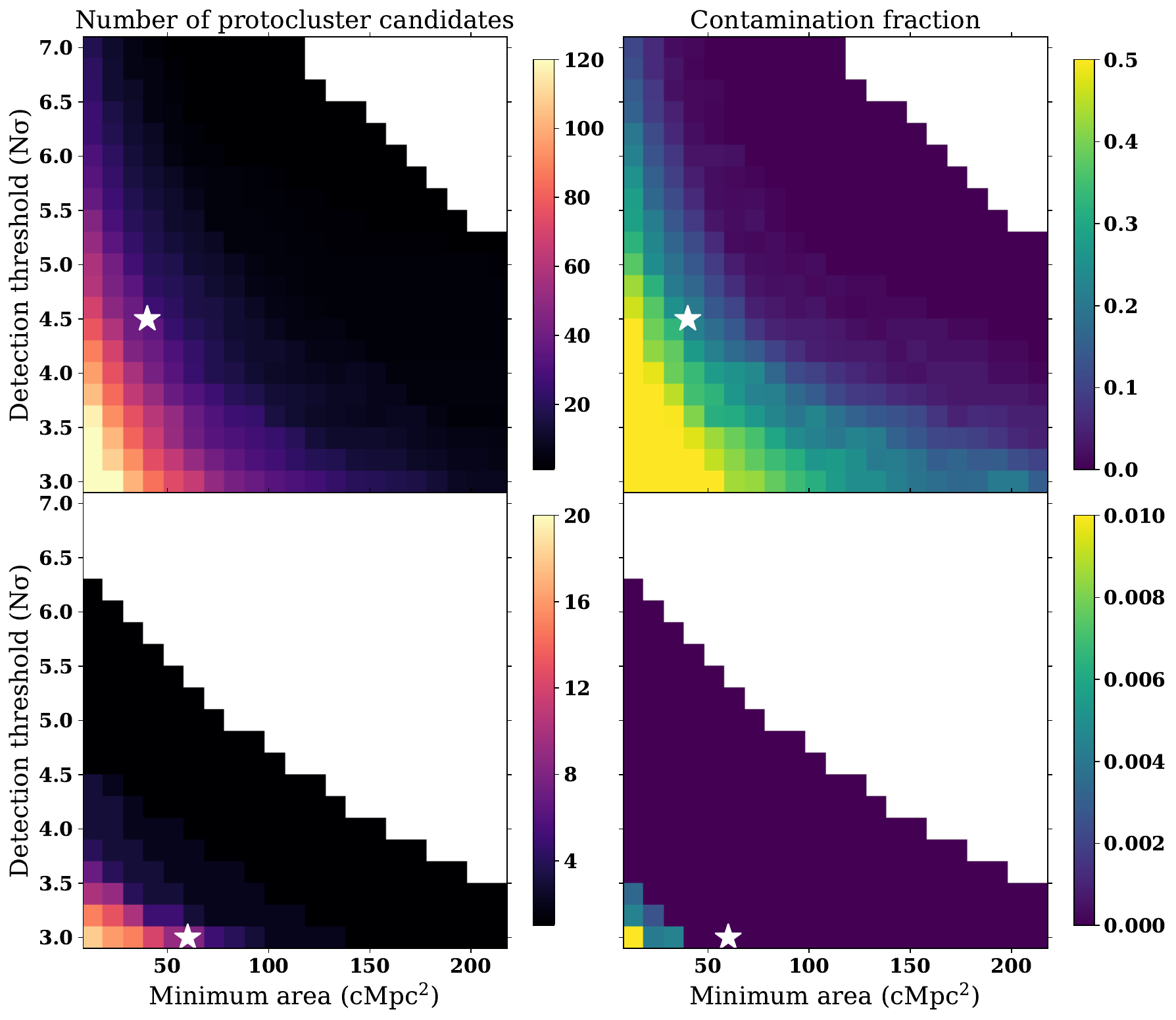}
    \caption{Number of candidates and contamination fraction for protoclusters selected from the VT map (top) and GS map (bottom), for different density thresholds and minimum areas. The density threshold is expressed in units of N$\sigma$ above the field surface density. White stars indicate the detection settings adopted for this work. While the VT map suffers from higher contamination than the GS map, it also detects far more candidates.}
    \label{fig:false_frac_all}
\end{figure*}

\begin{figure*}
    \centering
    \includegraphics[width=0.85\linewidth]{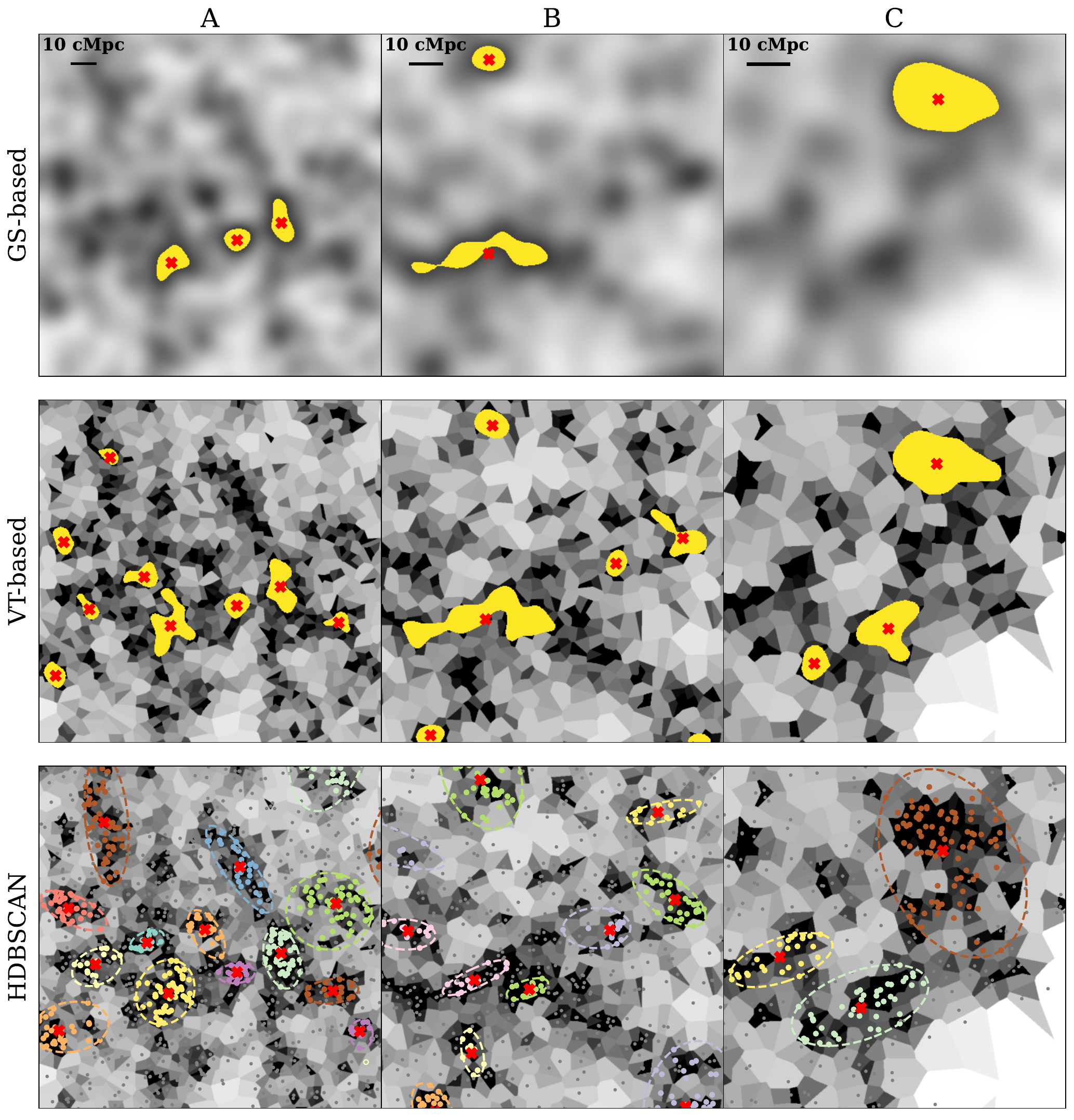}
    \caption{Protocluster candidates selected from the GS map (top row), the VT map (middle), and with HDBSCAN (bottom). Red crosses indicate the geometric centers of the selected overdensities. Yellow swathes indicate the extent of the protoclusters in the upper two rows. In the bottom row, colored points indicate LAEs classified as belonging to a cluster by HDBSCAN, enclosed by an ellipse for clarity, while gray points are classified as noise.}
    \label{fig:submap_comparison}
\end{figure*}

To assess contamination, we randomly select a subset of \emph{all} continuum and line-emitting sources detected within the $N501$ image, matching the number {and narrowband magnitude distribution} of LAEs in the field. We then generate GS and VT maps for these random points following the same procedure as for the LAE surface density maps. Since these points are distributed across a wide range of redshift, any overdensity observed in these `random maps' is unlikely to be genuine but rather the result of chance alignments. Hence, the contamination fraction for a specific set of detection parameters can be approximated by dividing the number of sources detected in the random map by the number detected in the LAE surface density map. This evaluation is depicted in the top and bottom right panels of Figure~\ref{fig:false_frac_all} for the VT and GS map, respectively, with the contamination fraction averaged over 30 iterations.

For the GS-selected protoclusters, the contamination fraction is low for all sets of detection parameters used. Our fiducial setup, {\tt DETECT\_THRESH} of $3\sigma$ and {\tt DETECT\_MINAREA} of 4600~pixels ($\simeq$ 60 cMpc$^2$), yields a contamination fraction of $<$ 0.1\% where $\sigma$ is the (sigma-clipped) standard deviation of $\delta_{LAE}$. Compared to the GS map, the pixel-to-pixel density fluctuations are much greater in the VT map, introducing much higher noise in structure detection. {This likely indicates that the scale at which the statistics are being sampled is too small. We mitigate this effect by smoothing the map with a 5~cMpc FWHM Gaussian kernel. On average, this kernel interpolates over $\sim$ 4 LAEs. It is thus large enough to reduce the noise but small enough to avoid adjacent structures blending into one considering that a typical diameter of a protocluster at $z\sim3$ is 10~cMpc.} We choose {\tt DETECT\_THRESH}  of $4.5\sigma$ and {\tt DETECT\_MINAREA} of 3000 pixels ($\simeq$ 40 cMpc$^2$), yielding the contamination rate of 20\%.

Figure~\ref{fig:submap_comparison} (top and middle rows) shows the protoclusters identified from the GS and VT-based maps within the highlighted Complexes A, B, and C\null. The GS map predominantly captures the largest structures, with smaller and more irregular formations remaining undetected; nonetheless, the identified structures exhibit high robustness. In contrast, the VT map reveals numerous structures overlooked by the GS map but is also subject to higher noise levels and is more susceptible to chance alignments of LAEs. The detailed comparison of different detection methods is presented in Section~\ref{sec:discussion}.

\subsection{Selecting protoclusters with HDBSCAN} \label{subsec:hdbscan}

\begin{figure*}
    \centering
    \includegraphics[width=\linewidth]{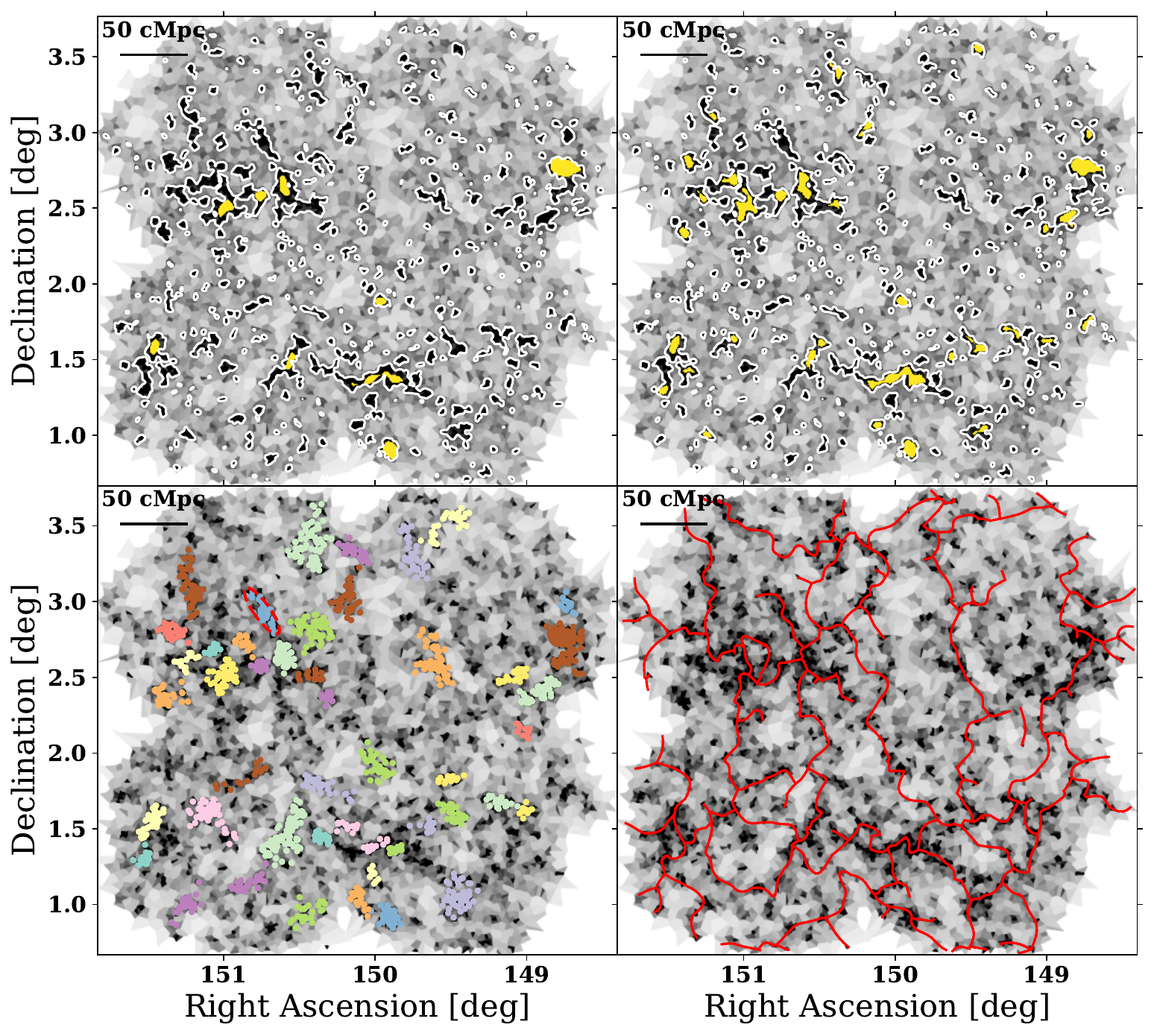}
    \caption
    {Protoclusters selected from the GS map (top left), VT map (top right), with HDBSCAN (bottom left) and filaments {identified with DisPerSE with a persistence threshold of 2$\sigma$} (bottom right) overlaid on the VT map. The red dashed region in the bottom left panel highlights a structure that is missed by the VT and GS maps but is detected with HDBSCAN. The filaments correspond to regions of moderate ($\sim$ 2-3$\sigma$) overdensity, and connect protoclusters together. }
    \label{fig:filaments}
\end{figure*}

Density-based clustering techniques, as a subset of unsupervised machine learning algorithms \citep{Kriegel2011}, discern clusters by identifying regions of high point density in data space, set apart by areas of low density.  These methods can be applied to pinpoint protoclusters from the clustering of the LAEs. Here, we employ the Hierarchical Density-Based Spatial Clustering of Applications with Noise (HDBSCAN) algorithm \citep{Campello2013}.  HDBSCAN offers the advantage of detecting clusters with varying densities, regardless of their shape, provided they exceed a specified size threshold.  Briefly, this algorithm measures density at each location, finding clusters as peaks in the density distribution separated by troughs. Density estimation is based on the distance to the $K$th nearest neighbor.

In HDBSCAN, each point is classified into {\it clusters} or {\it noise} based on two parameters: the $K$th nearest neighbor distance and the minimum cluster size. The latter dictates the minimum number of data points constituting a cluster. In the Python implementation of the HDBSCAN algorithm \citep[{\tt hdbscan};][]{McInnes2017}, these parameters are denoted as {\tt min\_samples} and {\tt min\_cluster\_size}, respectively. In our setup, we opt for values of 15 and 10, respectively (i.e., a minimum cluster size of 15 LAEs and the 10th nearest neighbor distance; see Appendix~\ref{appendix:hdbscan}). Additionally, we require that each cluster exhibit a median surface density above a specified threshold. This surface density is calculated as the number of LAEs divided by the enclosed area, computed as the sum of the areas of all Voronoi cells encompassing the cluster members. We set the threshold at 0.23~arcmin$^{-2}$ (i.e., $\gtrsim 1\sigma$ above the field mean {calculated from the VT map}), which yields a contamination rate of $\approx$20\% similar to that for the VT sample. The HDBSCAN-detected protoclusters are illustrated in the bottom row of Figure~\ref{fig:submap_comparison}. Symbols of the same color denote membership within the same structure, while gray circles represent LAEs outside of protoclusters. They are also shown in Figure~\ref{fig:filaments}.

\subsection{Cosmic filaments with DisPerSE} \label{subsec:filaments}

The cosmic web at low redshift ($z$ $\lesssim$ 1) has been investigated in detail through multiple surveys \citep[e.g.][]{Alpaslan2014,Tempel2014,Malavasi2017}. These studies have yielded valuable insight into the complex interplay of filaments and the clusters at the nodes of the cosmic web. Notably, \citet{Sarron2019} and \citet{Salerno2019} found evidence that galaxies infalling into clusters along filaments have a higher passive fraction than those being accreted from other directions. 
Moreover, several studies observed that on scales of 5--10~ cMpc, massive and quiescent galaxies are preferentially located closer to filaments than low-mass and star-forming galaxies \citep{Malavasi2017,Kraljic2017,Laigle2018}. This indicates that filaments may be pre-processing the galaxies being accreted along them.
At higher redshift, {however,} the role of the cosmic web on galaxy formation remains largely {observationally} unconstrained.  Motivated by
the discernible presence of filamentary structures in the LAE surface density map, we aim to detect cosmic web filaments in our observations in this section. We explore the reliability of our filament detection in Section~\ref{subsec:tng_fils}.

To identify cosmic filaments, we use the Discrete Persistent Structure Extractor \citep[DisPerSE;][]{Sousbie2011a,Sousbie2011b}, a widely utilized tool in both observational \citep[e.g.,][]{Sarron2019, Malavasi2017, Malavasi2020} and simulation studies \citep[e.g.,][]{Byrohl2023,Im2024}. DisPerSE utilizes the Delaunay Tessellation Field Estimator \citep{Schaap2000, van_de_Weygaert2009, Cautun2011} to create a discrete representation of the density field based on a given set of points. It then locates critical points within this density field approximation and defines filaments as arcs connecting saddle points to maxima. The filaments are constructed by creating short segments tangent to the gradient of the density field at each point. To account for data noise, DisPerSE employs the concept of persistence, which measures the density contrast between critical points defining a filament. Persistence represents the range of density thresholds over which a filament connecting two critical points remains significant relative to the noise. This significance is expressed in units of $\sigma$, where an $N\sigma$ filament corresponds to a probability under a 1D Gaussian distribution. We stress that this method of measuring the persistence level does not imply the use of a Gaussian distribution within DisPerSE but rather is purely an expression of the likelihood of the filaments arising from noise. For example, a persistence level of 3$\sigma$ (2$\sigma$) means that the extracted filament has a 99.7 (95.5)\% probability of being a true feature. 

After filling in the voids left by star masks, we run DisPerSE {on the positions of the LAEs} with a persistence of 2$\sigma$. {We note that there is no significant difference between the filaments recovered with and without filling in the voids, as shown in \citet{Ramakrishnan2023}.} The resultant filament network is illustrated in the bottom right panel of Figure~\ref{fig:filaments} as red curves superimposed on the VT density map in greyscale. The figure reveals a complex web extending over the entire field. Several previously noted linear overdensities (e.g., within complexes A and C) are part of this network. Moreover, the filaments appear to converge at the locations of the extended complexes, interconnecting the individual protoclusters within. Indeed, every protocluster is connected to at least one filament, with several positioned at the convergence of multiple filaments. Our findings are fully consistent with the hierarchical picture wherein protoclusters occupy the nodes of the cosmic web.

Given that the line-of-sight window function set by the $N501$ filter is substantially larger than the dimensions of individual protoclusters and cosmic filaments, our 2D-based detection algorithm is expected to include false positives arising from random noise fluctuations. In Section~\ref{subsec:tng_fils}, we show that while false detection - in particular, of filaments - does indeed occur, all filaments around overdense protoclusters are robustly identified. Protocluster detection is even more secure. 



\section{Cosmic structures in simulations: building expectations with  TNG} \label{sec:comparison_with_TNG}

Cosmological hydrodynamical simulations, such as TNG300, offer invaluable insights not readily accessible through direct observations. By facilitating connections between observable traits of protoclusters and cosmic filaments with more fundamental attributes and allowing us to track their evolution across cosmic time, simulations provide a crucial context in which we understand our data. In this section, 
we compare the ODIN protocluster and filament samples with those derived from the TNG300 simulation to gain an understanding of the physical properties of the structures detected in Section~\ref{sec:protoclusters}. 

\begin{figure*}
    \centering
    \includegraphics[width=0.9\linewidth]{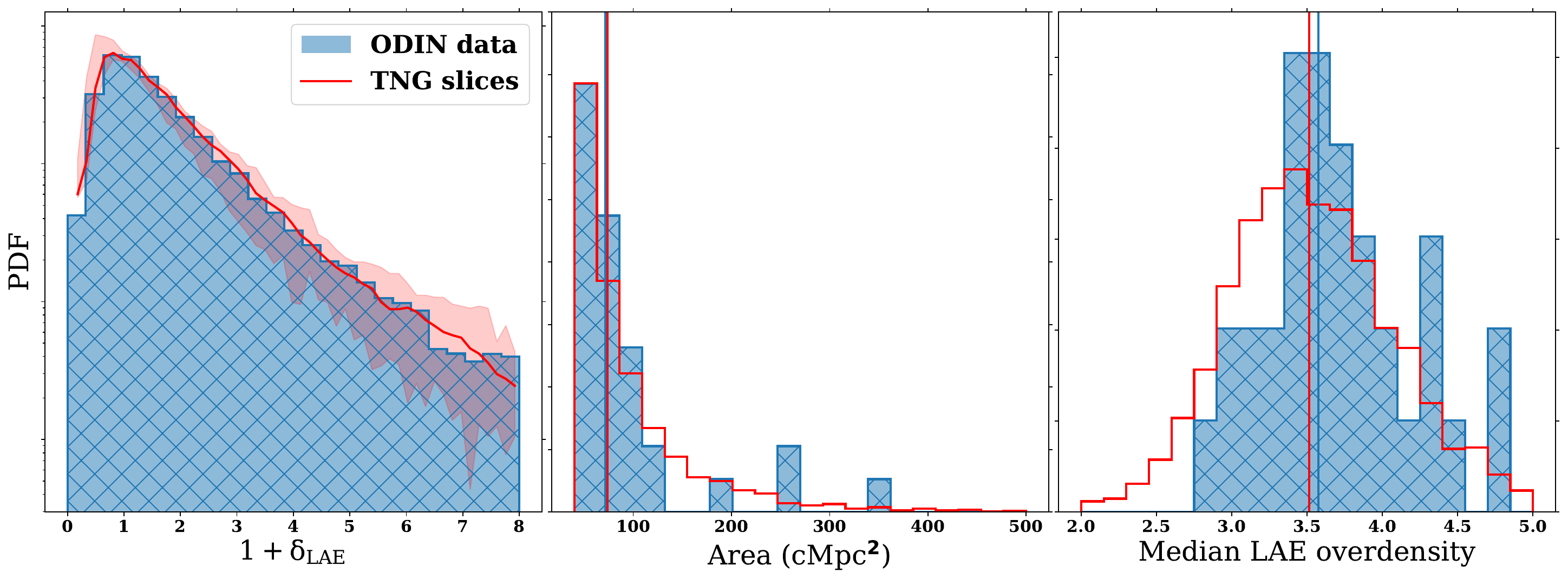}
    \caption{\emph{Left:} LAE surface density distribution of the TNG300 slices, in comparison to that of the observed LAEs. The red line shows the median distribution across all 90 slices while the shaded region shows the spread. The observed and simulated LAE distributions are very similar, particularly at the high end which represents massive structures.
    \emph{Middle and right:} Area and median LAE overdensity ($\delta_{LAE}$) for the protocluster candidates detected in the 90 TNG300 slices compared to those of the observationally detected protocluster candidates. Blue/red vertical lines indicate the median value for the ODIN data/TNG300. The properties of the structures are similar in the observations and simulation.}
    \label{fig:real_vs_tng_hist}
\end{figure*}

\begin{figure*}[t]
    \centering
    \gridline{\fig{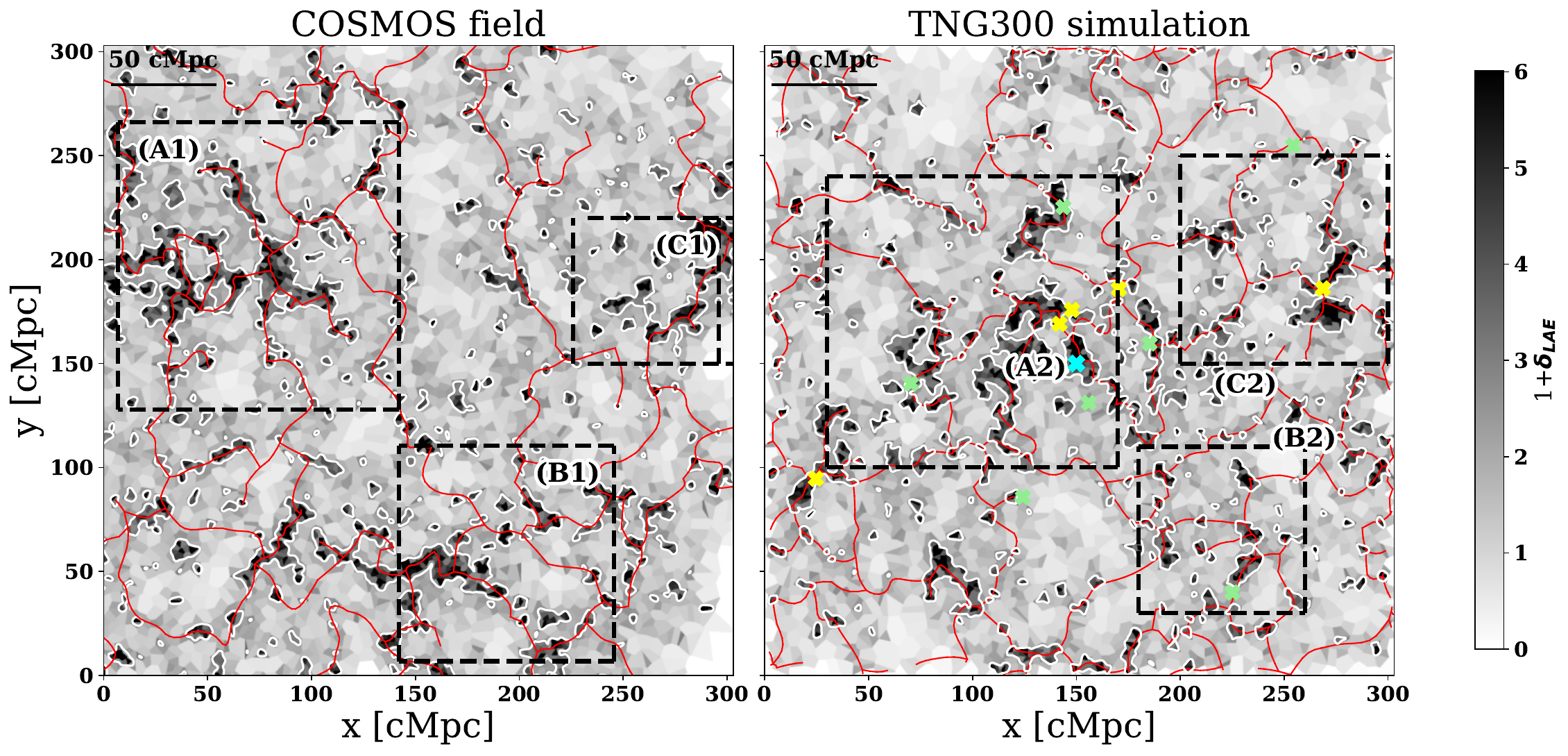}{\linewidth}{}}
    \vspace{-1em}
    \gridline{\fig{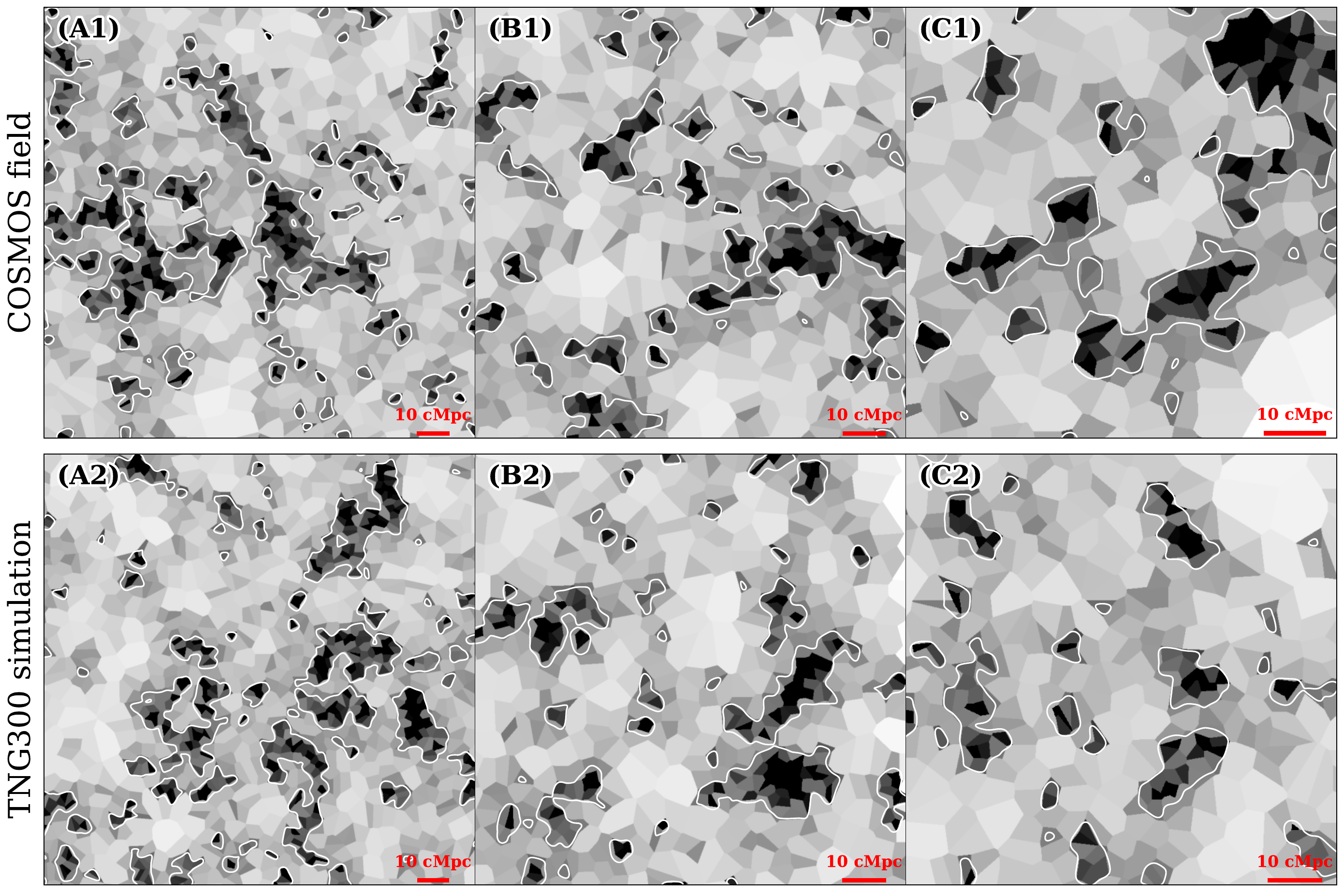}{0.8\linewidth}{}}    
    \caption{VT-based surface density map of our observed LAEs versus that of a 60 cMpc thick slice of the TNG300 simulation box at $z$ $=$ 3. {The positions of individual LAEs/mock LAEs are shown in Figure \ref{fig:lae_positions}.} Red lines in the top panels represent filaments selected with DisPerSE {with a persistence threshold of 2$\sigma$}. The cyan cross in the top right panel indicates the position of the massive protocluster on which the slice is centered (group ID = 13 at $z$ = 0), while yellow crosses show the other massive protoclusters in the same volume (group IDs = 1, 2, 7, 11, 23 at $z$ = 0). Green crosses show the progenitors of less massive clusters within the slice. Some overdensity complexes (A1, B1, C1; A2, B2, C2) are highlighted and zoomed in on the bottom panels. The two maps show visible similarities.}
    \label{fig:real_vs_tng_maps}
\end{figure*}

\subsection{Comparison of observations and simulations} \label{subsec:slices}

In the left panel of Figure~\ref{fig:real_vs_tng_hist}, we compare the LAE surface density distribution of the real data with the TNG300 slices. As detailed in Section~\ref{subsec:tng}, the latter are constructed from TNG300 and are designed to match the ODIN filter transmission as well as the LAE surface density and clustering bias. Multiple realizations are averaged over to result in the surface density distribution shown for the simulation.  
The high-end tail of the LAE surface density distribution, which represents the field's highest-density regions corresponding to protoclusters and filaments, is nearly perfectly reproduced. 

In Figure~\ref{fig:real_vs_tng_maps}, we show the VT map of both data and one $z=3$ TNG300 slice side by side. The displayed slice is chosen at random and is centered on a massive protocluster with descendant mass 5.5 $\times$ 10$^{14}$ M$_\odot$ (group ID=13 at $z$ = 0: cyan cross). It also contains five additional structures (yellow crosses) that will evolve into a cluster with mass greater than 2 $\times$ 10$^{14}$ by $z=0$, as well as several objects that will evolve into less massive clusters (green crosses). Indeed the majority of the largest overdensities seen within the slice are associated with protoclusters. The two maps are remarkably similar in that both display structures of similar sizes and irregular morphologies, arranged into extended complexes. We highlight three such similar regions (Complexes A1, B1, and C1 in the real data, and A2, B2, and C2 in the simulations, shown by dashed-line boxes). These mock structures are also connected by cosmic web filaments. 

The distributions of the total transverse area and median LAE surface density of protoclusters detected in real data and simulation span a similar range, demonstrating good agreement given the constraints of modeling.
This is illustrated in Figure~\ref{fig:real_vs_tng_hist}. The median number of protoclusters (averaged over 90 TNG slices) is 37 compared to 33 detected in our data. They correspond to the protocluster surface density of $(4.1 \pm 0.6) \times 10^{-4}$ and $(3.3 \pm 0.6) \times 10^{-4}$~cMpc$^{-2}$, respectively. The uncertainties represent the shot noise only and thus should be considered as a lower limit.  

One caveat is that the possible effects of radiative transfer on the observability of LAEs is ignored. In practice, the presence of dense gas and dust, particularly in the cores of protoclusters, may hide the LAEs in these regions \citep[e.g.,][]{Shimakawa2017,Momose2021,huang22}. This might lead us to either underestimate the LAE surface density in the cores of our observational protocluster candidates or incorrectly pinpoint their centers. Since the median densities of the observed and simulated LAEs are similar, we expect that any underestimation of LAE surface density is relatively small. However, we cannot rule out the possibility of mislocating the centers, and there is some evidence suggesting discrepancies between the locations of overdense regions identified by LAEs and those identified by Lyman Break galaxies \citep[e.g.,][]{Shi2019}. Nevertheless, the excellent agreement between our data and simulations suggests that the massive cosmic structures identified with current and future ODIN data are robust.

\subsection{Calibration of descendant mass estimate} \label{subsec:mass_calib}



Estimations of the descendant mass allow us to establish direct connections between high-redshift protoclusters and present-day galaxy clusters, thereby tracing the evolution of massive cosmic structures and their galaxy populations across cosmic time. Following the methodology outlined by \citet{Steidel1998}, we compute the descendant mass of ODIN protoclusters by assuming that the mass contained within the overdensity region will collapse into a cluster-sized halo by $z = 0$, with a total mass $M_{z=0}$ given by
\begin{align}\label{eq:mass} 
    M_{z=0} = \rho_{m}V_{PC} = \rho_{0,z}(1+\delta_{m})V_{PC} 
\end{align}
where $\rho_{m}$ and $\delta_{m}$ are density and overdensity of matter, respectively. $V_{PC}$ is the protocluster volume, and $\rho_{0,z}$ is the mean density of the Universe at redshift $z$. If the galaxy bias, $b_{g}$, is known, Equation~\ref{eq:mass} becomes:
\begin{equation}
    M_{z=0} = \left(1+\frac{\delta_{g}}{b_{g}}\right) \rho_{0,z}V_{PC}
\end{equation}
where $\delta_{g}$ is galaxy overdensity. 
We assume $b_{g} = 1.8$ \citep{Gawiser2007,White2024} and fix $\delta_g$ to the median LAE overdensity within the protocluster region. Since we do not have any redshift information for the majority of our LAEs, we assume that the size of a protocluster in the line-of-sight direction is similar to that in the transverse direction and approximate the volume of a protocluster from its area ($A_{PC}$) as $V_{PC}$ $=$ $A_{PC}^{1.5}$: i.e., the shape of a protocluster is approximated as a cube. $M_{z=0}$ is then estimated as,
\begin{equation}
    M_{z=0} = \left(1+\frac{\delta_g}{b_g}\right)\rho_{0,z}~A_{PC}^{1.5}
    \label{eqn:today_mass}
\end{equation}



The relation $V_{PC} = A_{PC}^{1.5}$ is a good estimate for an isotropic structure. As we have already seen, many of our protoclusters are strikingly non-spherical in shape, and thus the above relation may be a poor approximation. If a protocluster is extended in the transverse dimension, it would be observed as a more modest overdensity with a larger angular extent. Conversely, a structure stretched along the line-of-sight direction is expected to be more compact with a larger overdensity. Using TNG-detected structures, we quantify the detection rate and the uncertainty in mass estimates due to the lack of information in the third dimension. 


Figure~\ref{fig:real_vs_tng_hist} shows that there is a good agreement between TNG-selected and observational protoclusters for both angular sizes (middle) and median LAE overdensity (right). 
We also assess the likelihood of the most 30 massive cluster progenitors to be selected as a protocluster candidate when viewed along the X-, Y-, and Z directions. To this end, we only consider the structures for which: i) the measured (angular) position center lies within 10~cMpc of the true center; and ii) the line-of-sight position center is within 25~cMpc of the center of a given TNG slice. For reference, the FWHM of all slices is 60~cMpc and a typical size of a protocluster is $\approx$10~cMpc.

\begin{figure*}
    \centering
    \includegraphics[width=\linewidth]{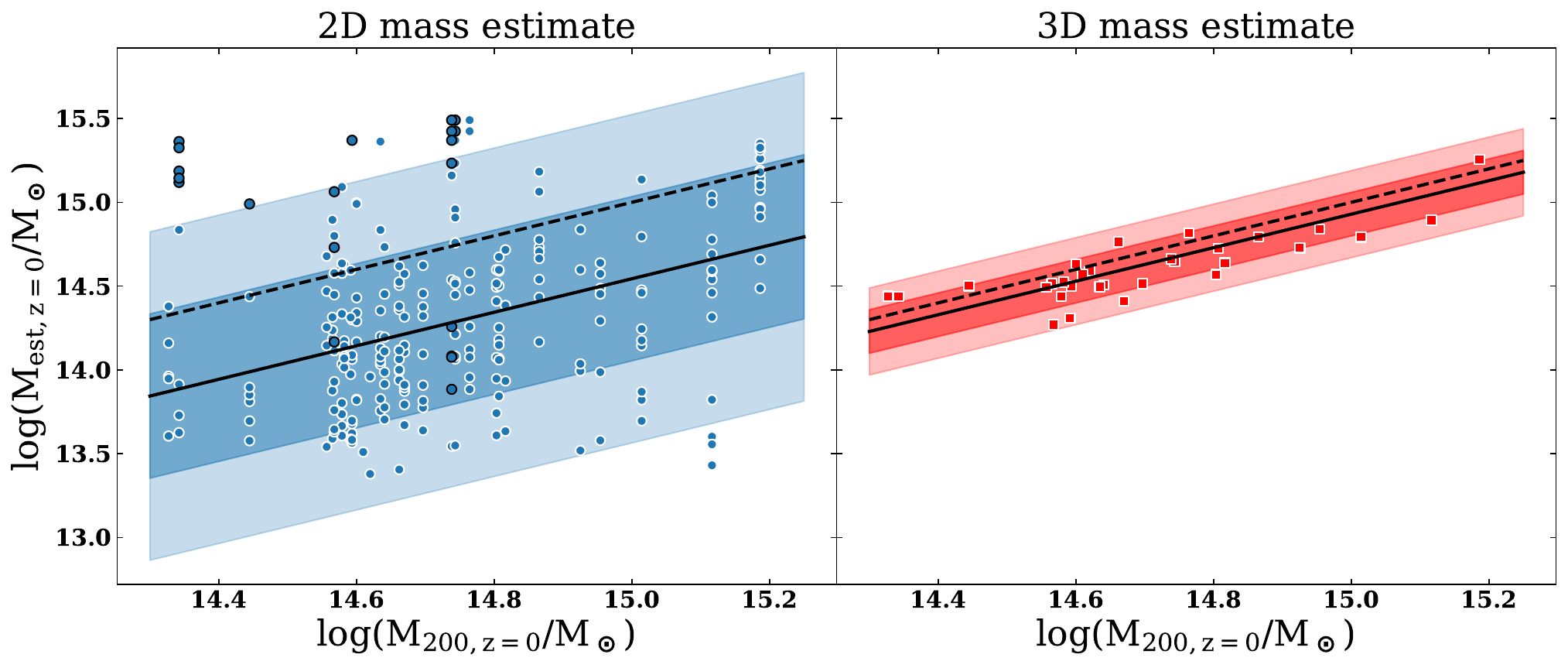}
    \caption{$M_{z=0}$ values estimated from Equation \ref{eqn:today_mass} compared to the true values for the progenitors of the 30 most massive clusters in the TNG300 box. The left panel (blue points) indicates the masses estimated from the 2D projection of the LAE overdensity corresponding to the protocluster, with the LAEs matched in number density to observations. The dashed line indicates a one-to-one relation between estimated and true descendant masses, while the solid line indicates a relation of $M_{est,z=0}$ = 0.35$M_{200,z=0}$. Highlighted points show cases where the protocluster was merged with another, more massive object by the detection. The right panel (red points) indicates the masses estimated from the 3D positions of the cluster members (see Section \ref{subsec:mass_calib} for more details). Shaded regions indicate the range of the 1 and 2$\sigma$ scatter of the estimated masses. While the 2D mass estimate is subject to considerable scatter, primarily arising from the loss of information in the $z$ direction, it correlates well with the true descendant mass.}
    \label{fig:mass_estimate}
\end{figure*}

Based on this definition, the median recovery rate of the protoclusters is 60\%. The progenitor of the most massive cluster is always detected while the progenitors of the next 4 most massive clusters are detected $>$ 90\% of the time. As expected, the likelihood of finding smaller structures depends on the sightlines being favorable or adverse to robust detection. If we lower the threshold for our VT-based protocluster selection from 4.5$\sigma$ to 3.5$\sigma$, the recovery rate of the protoclusters increases to 80\%. The higher success rate comes at the price of much greater contamination of $\sim$45\%, compared to $\approx$20\% for our fiducial setup.

In the left panel of Figure~\ref{fig:mass_estimate}, we compare the {\it estimated} and true descendant mass, $M_{est,z=0}$ and $M_{200,z=0}$ for the recovered clusters in the 90 slices, {where $M_{200}$ refers to the mass contained within a spherical region whose mean density is 200 times the critical density of the Universe.} 
The figure illustrates that $M_{est,z=0}$  tends to underestimate the true mass about 80\% of the time. The median $M_{est,z=0}/M_{true,z=0}$ ratio is {0.35$^{+0.67}_{-0.23}$}, as indicated by the solid black line. 
This implies that the cosmic volume that ends up in a galaxy cluster by $z=0$ is much greater than what we identify in the data as significant galaxy overdensities likely associated with a protocluster. 

Our findings are consistent with the results from \citet{Chiang2013}, who, based on the semianalytical models implemented in the Millennium simulations, found that about 40\% of the total mass at $z=0$, $M_{\rm today}$, is enclosed within a 6 (8)~cMpc-radius sphere for Virgo- and Coma-sized protoclusters with $M_{\rm today}=(3-10) \times 10^{14} M_\odot$ and $> 10^{15} M_\odot$, respectively. In comparison, the median effective radius of our simulated protoclusters (computed as ($A_{PC}/\pi$)$^{0.5}$) is $\sim$ 5~cMpc. To account for this effect, we correct the estimated masses by a factor of 0.35. 




Figure~\ref{fig:mass_estimate} shows that the 2D-based mass estimates yield a considerable scatter of 0.49~dex. The most extreme outliers arise from cases where two or more protoclusters are merged together during detection, particularly for the protoclusters with $M_{200,z=0} \leq$ 10$^{14.5}$ $M_\odot$; these points are highlighted with a black outline in Figure~\ref{fig:mass_estimate}. Even excluding these cases, the scatter remains high at $\sim$ 0.44~dex.
We ascertain that the scatter is a result of the loss of information in the $z$ direction as follows.
We perform the 3D Voronoi tessellation of the 3D volume centered on the 30 massive clusters, this time, using \emph{all} the galaxies with $M_{\rm star} > 10^7 M_\odot$. Using the subhalo merger trees of the $z=0$ cluster members, we also identify the `member galaxies' in the $z=3$ snapshot. 
The LAE overdensity and the protocluster region are computed similarly to the 2D case but Voronoi cells are now 3D polyhedra instead of 2D polygons, while galaxy overdensity is measured as the median LAE overdensity of the Voronoi cells containing member galaxies.  $V_{PC}$ is obtained by summing over these Voronoi cells.

The resultant mass estimate is shown in the right panel of Figure~\ref{fig:mass_estimate}. As expected, the scatter of the 3D $z=0$ mass estimates is considerably smaller, at $\sim$ 0.15 dex, than in the 2D case, once we have eliminated all uncertainties regarding the cluster membership and the information on the third dimension. The estimated descendant mass is also significantly less underestimated than in the 2D case, being $\sim$ 85\% of the true value. It is noteworthy that the {\it intrinsic} scatter of 0.15~dex is comparable to 0.2~dex estimated by  \citet{Chiang2013} in converting galaxy overdensity measured within a (15 cMpc)$^3$ cubic window into descendant mass. 

When we compare the `true' volume found with the 3D Voronoi tessellation to that derived from the projected protocluster area, we find that the scatter of the latter ($\approx$0.52~dex) is sufficient to account for the entirety of the uncertainty in the mass estimate. This suggests that the uncertainty in $M_{z=0}$ is dominated by the volume estimation and that the isotropic assumption is less than ideal for protoclusters at high redshift. This result is not surprising, but it does quantitatively demonstrate the need for large-scale spectroscopic follow-up of protocluster candidates. However, we also emphasize that the 2D mass estimation is correct in an \emph{average} sense, and statistical analyses for a large sample of protocluster candidates that make use of this mass estimate will be robust.


\begin{figure*}
    \centering
    \includegraphics[width=\linewidth]{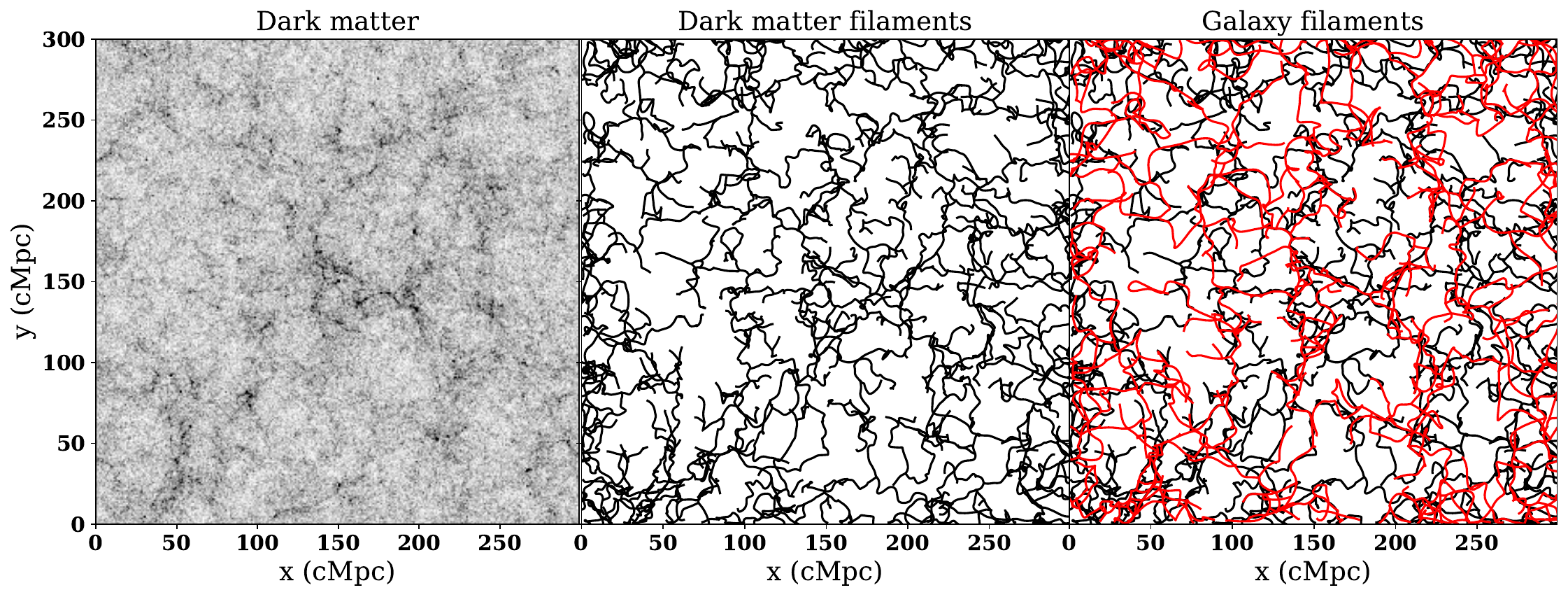}
    \caption{\emph{Left:} Distribution of dark matter in a randomly chosen TNG300 slice. \emph{Middle:} The filament network identified with DisPerSE, using 0.1\% of the dark matter particles and with a persistence threshold of 6$\sigma$. \emph{Right:} The filament network identified with DisPerSE using galaxies (red), again with a persistence threshold of 6$\sigma$, overlaid on that traced by dark matter. Although the former is less detailed than the latter, there is excellent correspondence between the two.}
    \label{fig:dm_vs_gal_fils}
\end{figure*}

\begin{figure*}[h]
    \centering
    \includegraphics[width=0.9\linewidth]{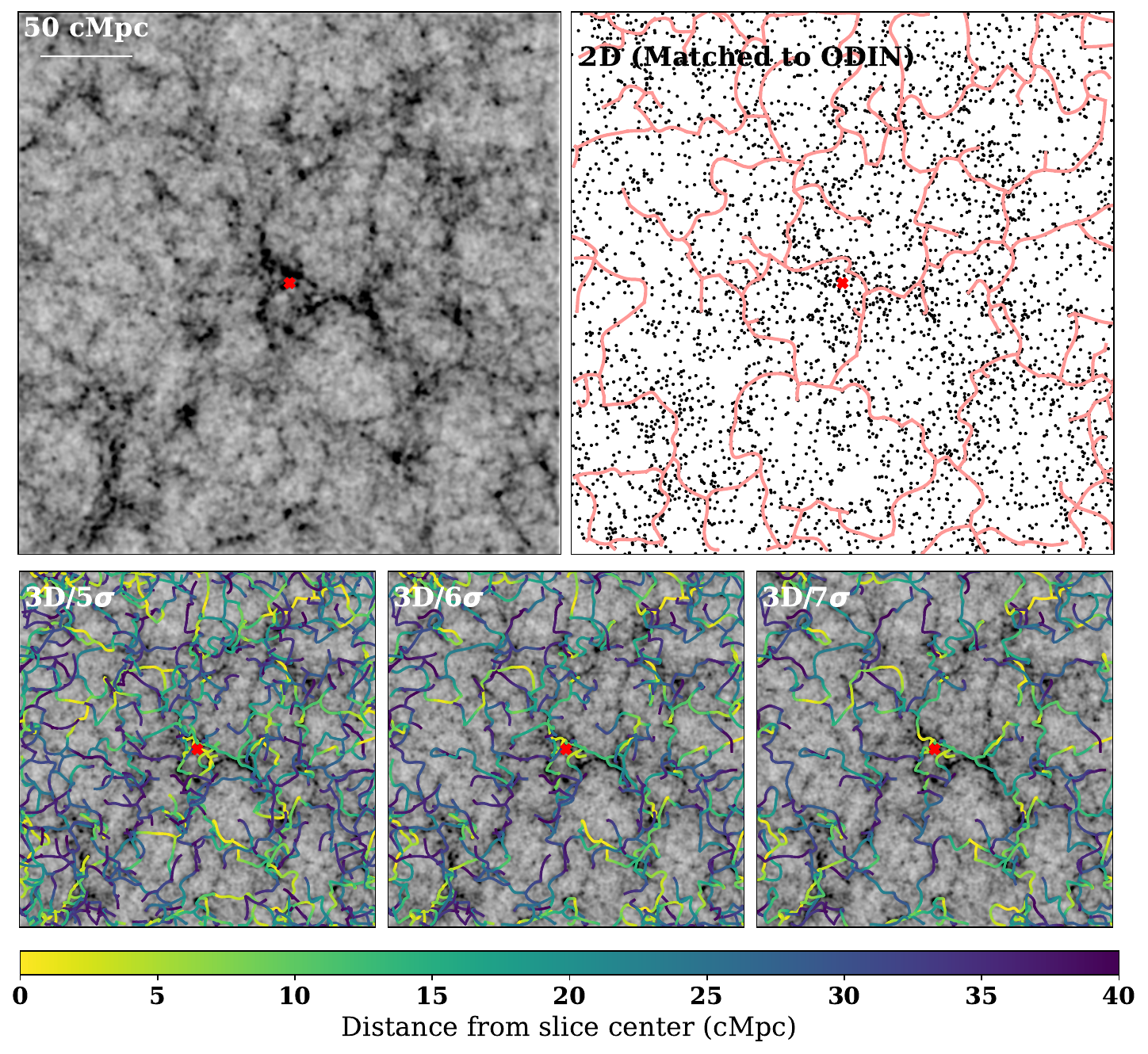}
    \caption{\emph{Top left}: The dark matter distribution for a randomly selected slice, centered on a protocluster with descendant mass log$(M_{200}/M_\odot$) = 15.18 (red cross; group ID = 0 within the $z$ = 0 TNG300 simulation box). \emph{Top right:} The filaments (pink lines) extracted by DisPerSE for this slice are based on the mock LAE positions, which are shown by black dots. This network is comparable to the cosmic web selected with ODIN data. \emph{Bottom:} The filaments extracted by DisPerSE for the same slice using the 3D positions of the slice galaxies with 10$^7$ $M_\odot$ $<$ M$_*$ $<$ 10$^{12}$ $M_\odot$, with varying persistence thresholds. The filaments are color-coded by their separation from the center of the slice along the line-of-sight direction. At large distances from the slice center, the filaments are truncated by the slice boundaries. The number of such truncated filaments reduces at higher persistence thresholds.}
    \label{fig:tng_fils}
\end{figure*}

\subsection{The fidelity of filament recovery} \label{subsec:tng_fils}

At high redshift, active star formation and galaxy growth taking place in protoclusters are likely supported by cool gas transported along filaments of the cosmic web \citep[e.g.,][]{Dekel2009,Daddi2021}. The wide area coverage of ODIN presents a unique opportunity to study protoclusters in concert with the surrounding cosmic web, provided that we can accurately recover the filaments. As they are even more extended structures than protoclusters, the effect of projection along the line-of-sight 
on filament recovery must be considered separately for filaments. In this section, we use the TNG300 slices to examine how well the observed 2D filament network corresponds to the true underlying 3D network. We focus specifically on filaments in the vicinity of protoclusters, to understand how likely it is for a filament that appears near a protocluster in projection to be physically connected to the protocluster.


We make use of the 90 slices of the TNG300 full box at $z$ $=$ 3 described in Section \ref{subsec:tng}. For each slice, we use DisPerSE to extract filaments using the positions of the mock LAEs matched to ODIN, projected along the slice axis. This is henceforth referred to as the `2D skeleton' or `2D filaments'. To create a comparison set of filaments that represents the `true' cosmic web, we also run DisPerSE on the 3D positions of the slice galaxies with 10$^7$ $M_\odot$ $<$ M$_*$ $<$ 10$^{12}$ $M_\odot$, henceforth referred to as the `3D skeleton' or `3D filaments'. We use a persistence threshold of 2$\sigma$ to extract the 2D skeleton, identical to that used on the observed LAEs. For the 3D skeleton, we use multiple persistence thresholds of 5, 6, and 7$\sigma$ to compare the 2D skeleton against 3D features of varying significance. 

{How well does the 3D skeleton traced by galaxies correspond to the underlying distribution of dark matter? \citet{Im2024} quantitatively analyze this question for the Horizon Run 5 simulation. They find that while the filaments outlined by all the simulated galaxies are less detailed than the filaments traced by dark matter, the former nevertheless align closely with the latter. Though we do not attempt to replicate the full analysis of \citet{Im2024}, we find similar behavior in TNG300, as demonstrated in Figure~\ref{fig:dm_vs_gal_fils} where we show the filaments traced by dark matter and the 3D skeleton traced by galaxies in a randomly chosen slice. 
}

\newcommand{\dskela}{$d_{2d \rightarrow 3d}$}
\newcommand{\dskelb}{$d_{3d \rightarrow 2d}$}

\begin{figure*}[th!]
    \centering
    \includegraphics[width=5.5in]{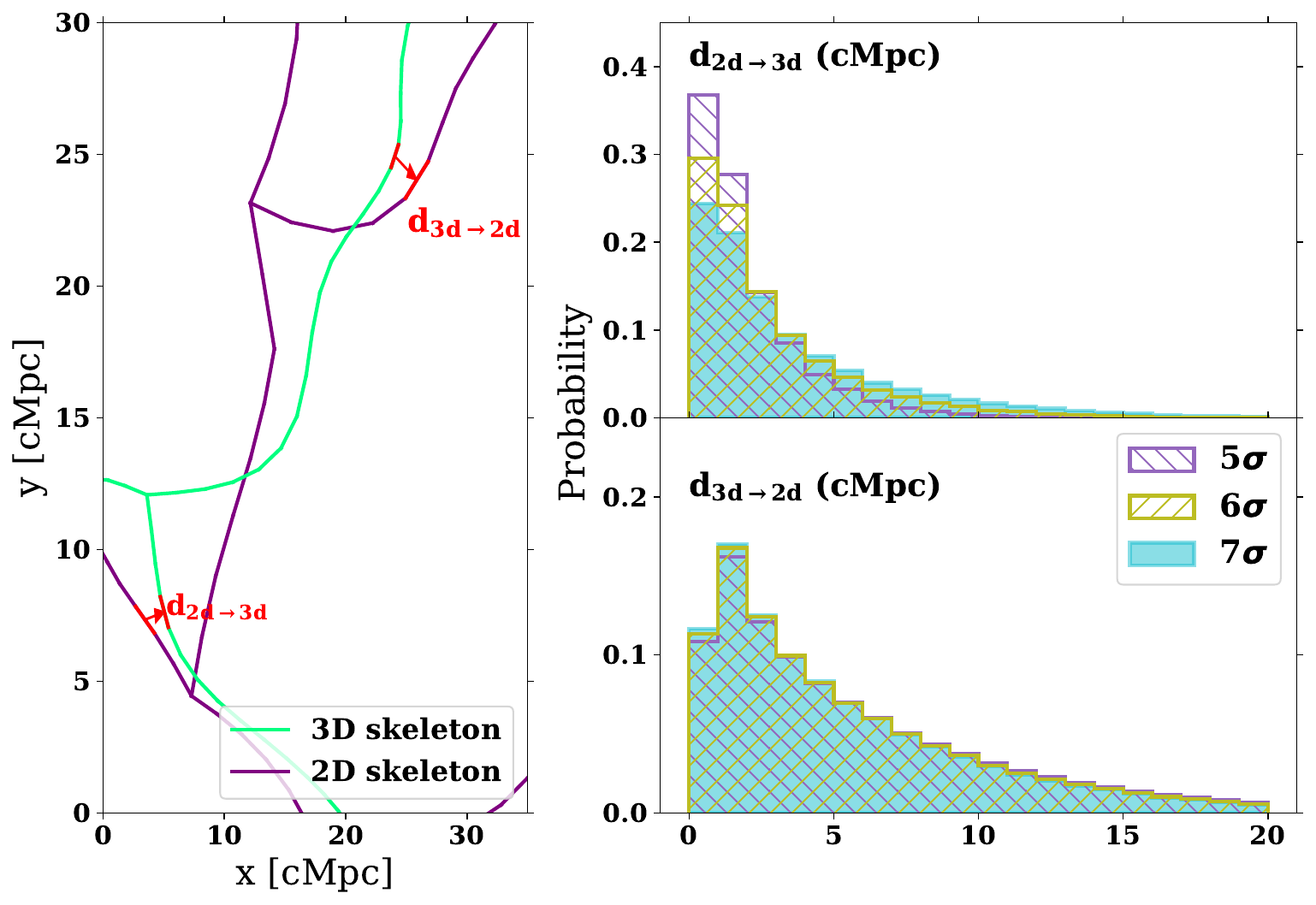}
    \caption{\emph{Left:} An example comparison of the 2D and 3D skeletons, illustrating how \dskela\ and \dskelb\ are measured. \emph{Right:} Comparison of the 2D and 3D skeleton using the separation of the segments, \dskela\ (top) and \dskelb\ (bottom). The separations are calculated across all 90 slices of 80~cMpc thickness from the $z=3$ TNG300 simulation. In the majority of cases, both \dskela\ and \dskelb\ are less than 5~cMpc, regardless of the specific persistence threshold.}
    \label{fig:dskel_seps}
\end{figure*}

In Figure~\ref{fig:tng_fils}, we show the dark matter distribution in one of the TNG300 slices overlaid with the 2D and 3D skeletons. The similarity between the DisPerSE filaments (both 2D and 3D) and the underlying dark matter distribution is clear, as is the fact that the 2D and 3D skeletons mirror each other closely. However, the 2D skeleton does not perfectly reproduce the true 3D cosmic web. To quantify the extent to which the cosmic web can be recovered by our observational data, we compare the 2D filaments to the 3D skeleton. We do this following a similar procedure to 
\citet{Sarron2019} \citep[see also][]{Laigle2018,Kuchner2022}, by matching the individual segments out of which DisPerSE constructs the 2D and 3D skeletons (see Section \ref{subsec:filaments}) to each other. We measure the (projected) separation between the midpoints of each segment of the 2D skeleton and those of the 3D skeleton. We denote the minimum distance \emph{from} a segment of the 2D filament network \emph{to} one of the 3D network as \dskela, and the inverse as \dskelb. This mapping is illustrated in the left panel of Figure~\ref{fig:dskel_seps}. The former measurement enables us to determine which of the 2D filaments represent true features of the 3D cosmic web, while the latter allows us to determine which features of the underlying filament network are successfully recovered by observations.

In the top right panel of Figure~\ref{fig:dskel_seps}, we show histograms of \dskela\ for the three persistence thresholds. Hydrodynamical simulations have found that the typical radius of a filament at $z \sim 3$ is 2--3 pMpc (8 -- 12 cMpc: \citealt{Zhu2021}, see also \citealt{Im2024}). The median values of \dskela\ are well below this range, being 2.0, 2.7, and 4~cMpc respectively for the 3D skeletons found with persistence 5, 6, and 7$\sigma$. The values are less than 5~cMpc (1.25 pMpc) for the majority of the 2D segments. We refer to the segments with \dskela~$< 5$~cMpc as `matched 2D segments', meaning that they have a `match' in the 3D skeleton. The fraction of matched segments decreases with increasing persistence threshold, from $\sim$ 80\% for a persistence threshold of 5$\sigma$ to $\sim$ 60\% when the persistence threshold is 7$\sigma$. This is reasonable; as the persistence threshold is raised, the extracted network is increasingly restricted to only the most significant filaments, and finer features are lost, leading to a greater number of 2D segments going unmatched.      
Irrespective of the adopted persistence value, 55\% of the 3D segments have 2D matches within \dskelb~$<$ 5~cMpc.
We conclude that 2D skeleton serves as a good representation of the underlying cosmic web even though it faithfully recovers the individual features  $\approx$55\% of the time.  
%
%

\begin{figure*}
    \centering
    \includegraphics[width=\linewidth]{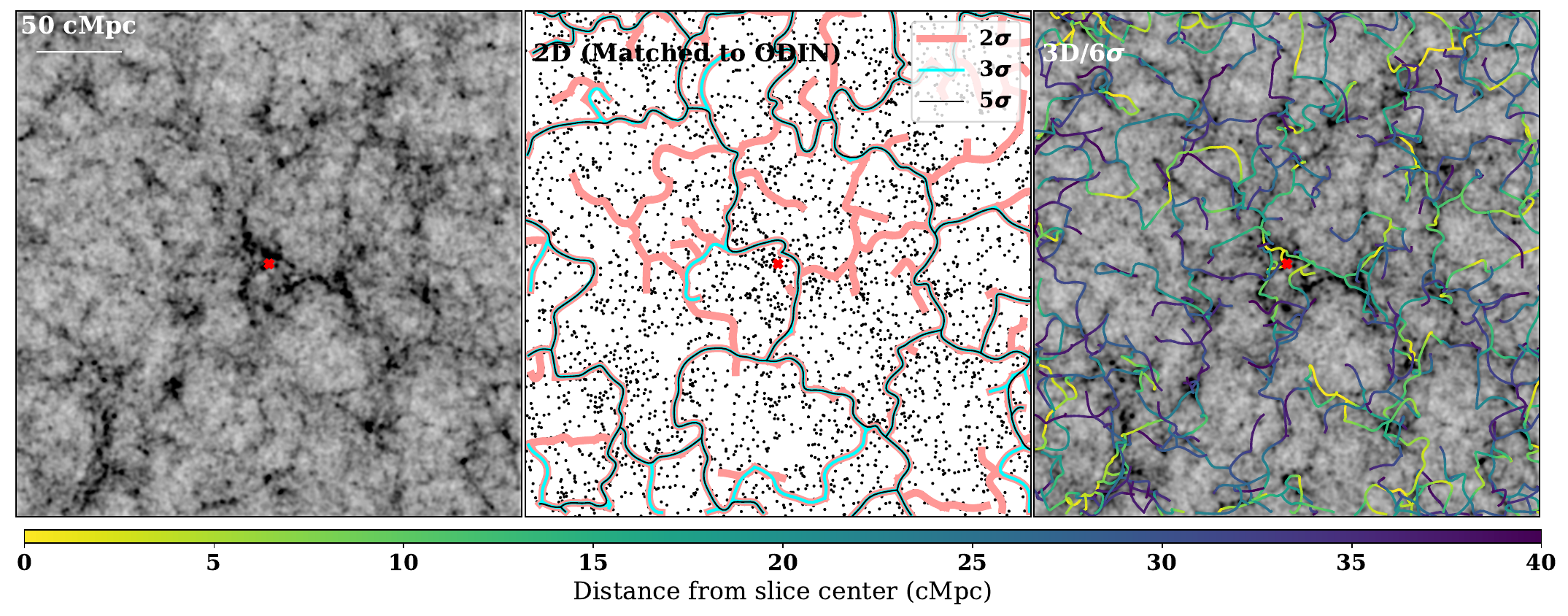}
    \caption{Similar to Figure \ref{fig:tng_fils}, but showing the 2D skeletons extracted with multiple persistence thresholds.}
    \label{fig:tng_fils_2}
\end{figure*}

Next, we examine whether the fraction of matched 2D segments depends on the persistence threshold used to extract the 2D skeleton. If the matched fraction increases substantially with increasing persistence threshold, it would indicate that our chosen persistence threshold is too generous. Reassuringly, we find that the matched fraction only rises marginally with the persistence threshold, increasing by $\lesssim$ 5\% from a persistence threshold of 2$\sigma$ to 5$\sigma$. This is visualized in Figure~\ref{fig:tng_fils_2}, where we show the 2D skeletons extracted with persistence thresholds of 2, 3, and 5$\sigma$ in comparison to the dark matter distribution and the 3D skeletons. The majority of the 2D filaments extracted with a lower persistence threshold, which do not appear with a higher one, can be visually matched to the true features of the 3D skeleton.

\begin{figure}
    \centering
    \includegraphics[width=0.9\linewidth]{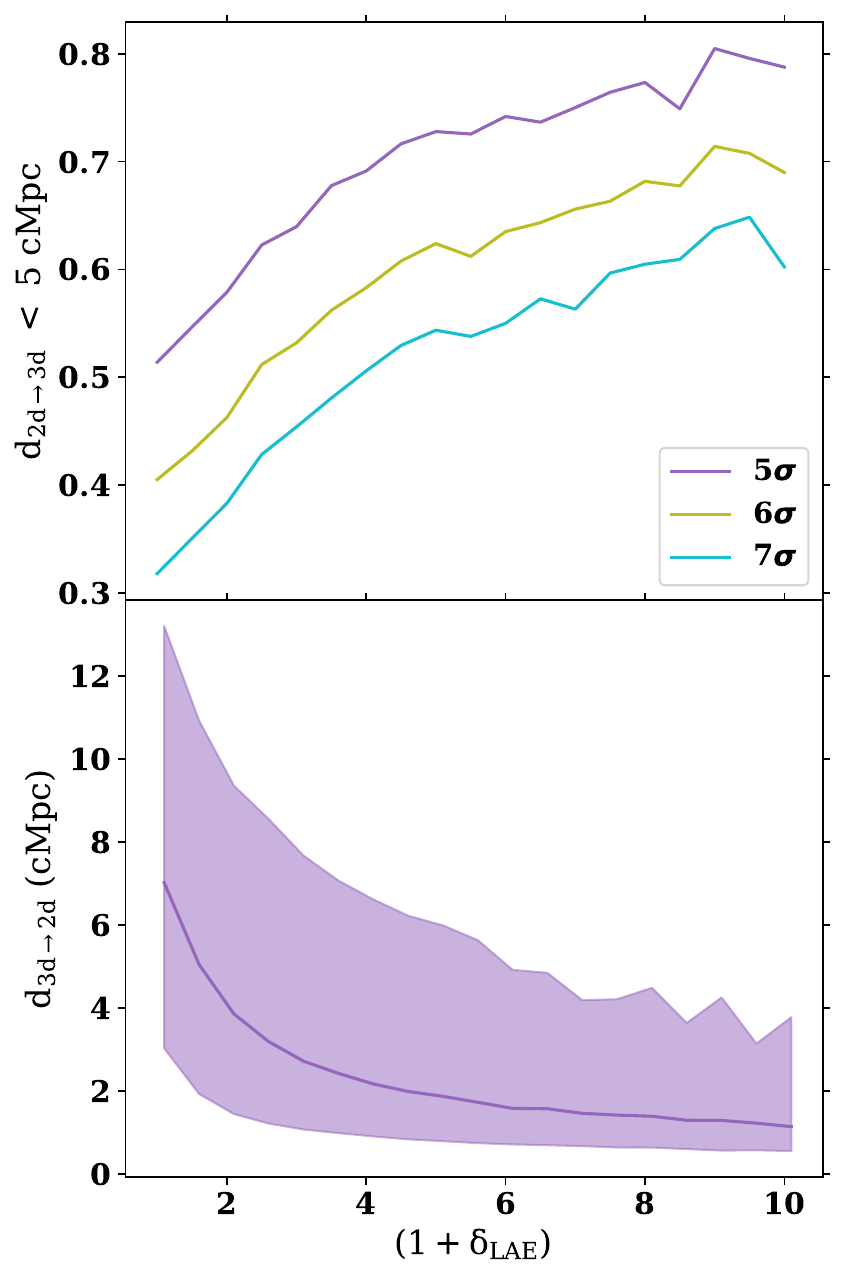}
    \caption{\emph{Top:} The fraction of 2D filament segments with a match with the 3D skeleton (detected with various persistence thresholds), as a function of the LAE surface density at the position of the filament segment. Regardless of the persistence threshold used to select the 3D skeleton, the fraction of matched 2D segments increases with increasing surface density. \emph{Bottom:} \dskelb\ as a function of the (2D) LAE surface density for the 5$\sigma$ threshold 3D skeleton; the other two skeletons give virtually identical results. The solid line indicates the median value while the shaded region indicates the spread as measured from the 16-84 percentile scatter.}
    \label{fig:matched_vs_sd}
\end{figure}

Intriguingly, the fraction of matched 2D segments increases with increasing surface density for all cases, as shown in the top panel of Figure~\ref{fig:matched_vs_sd}, suggesting that the higher the observed LAE surface density is, the more likely the detected filament is to be an accurate representation of the 3D cosmic web. 
We also examine the converse, i.e. the probability that a filament within an overdense region is recovered by observations. On considering the 3D skeleton within 10 cMpc of a massive protocluster, we find that the fraction of matched 3D segments is considerably higher than in average regions, increasing from a median of $\sim$ 55\% to a median of $\sim$ 86\% percent in all three persistence cases. The matched fraction is even higher if we restrict the region under consideration to within 5 cMpc of the protoclusters, being 100\% in the majority of slices. 

\begin{figure*}
    \centering
    \includegraphics[width=0.9\linewidth]{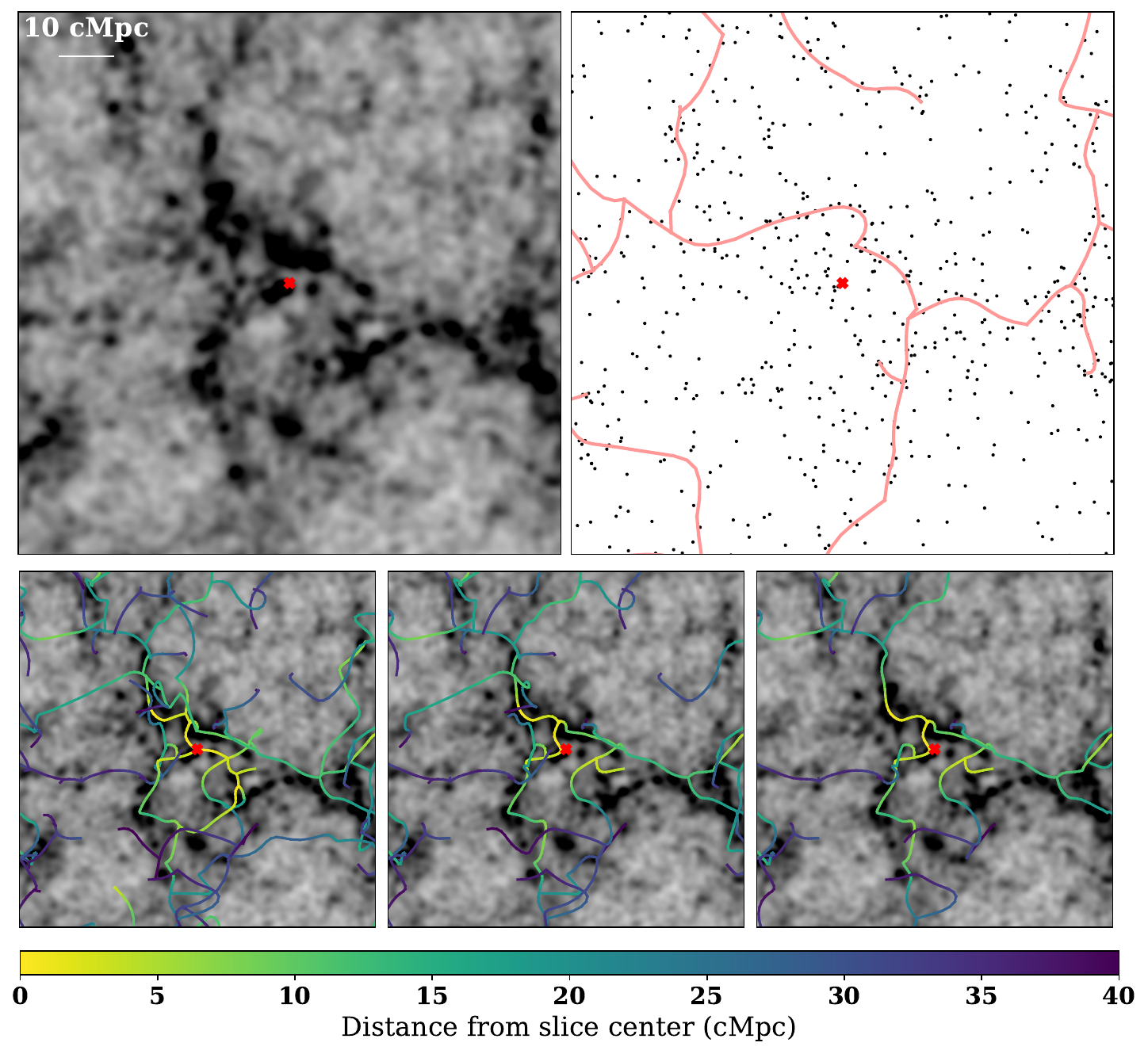}
    \caption{As Figure \ref{fig:tng_fils}, but zoomed in on the region surrounding a massive protocluster}
    \label{fig:fil_zoom}
\end{figure*}

In the bottom panel of Figure~\ref{fig:matched_vs_sd}, we show \dskelb\ as a function of the (2D) LAE surface density in the slices. To avoid confusion, we only show the result for the 5$\sigma$ threshold skeleton, as the remaining two persistence thresholds give nearly identical results. With increasing LAE surface density, \dskelb\ decreases rapidly; i.e., in dense protocluster regions, not only is the recovery rate of the 3D skeleton considerably higher but also the 2D skeleton resembles the 3D skeleton more closely. This is illustrated in Figure~\ref{fig:fil_zoom} where we show the 2D and 3D skeletons for the same slice as in Figure~\ref{fig:tng_fils} but zoom in on the massive protocluster. These results show that the ODIN data will be capable of studying protoclusters within the context of the surrounding cosmic web.

\section{Discussion} \label{sec:discussion}

\subsection{Comparison of  structure detection methods} \label{subsec:method_comparison}

In Section~\ref{sec:protoclusters}, we based our protocluster selection on the GS and VT maps and the  HDBSCAN algorithm. These methods resulted in 9, 33, and 47 structures, respectively. 
We now wish to evaluate the similarities and differences between these methods. 

The false positive rate in the GS method is very low for all sets of detection parameters and is $\sim$ 0 for our fiducial criteria. The 9 GS-based protoclusters may thus be considered the most secure candidates. All GS structures are present in the VT sample with similar morphologies. In all cases, the separation between the centers determined by the GS and VT method is within 2.5~cMpc.
For all but one that is present in both GS and VT samples, the measured area is slightly greater (by a factor of $\sim$ 1.1 - 1.8) in the latter. 
This is because irregular overdensities are generally detected at higher significance in the VT map. 


All structures in the GS sample are also present in the HDBSCAN sample. 
One is split into two structures by HDBSCAN as shown in the middle column of Figure~\ref{fig:submap_comparison}. For the remaining 8, the structures have similar morphologies in both the GS and HDBSCAN analyses; however, the measured area in HDBSCAN is larger than that estimated from the GS method in by a factor of $\sim$ 2 - 5. 

%


More interesting is the comparison between the structures detected from the VT map and using HDBSCAN.
Figure~\ref{fig:filaments} shows the locations of the VT (upper right) and HDBSCAN samples (lower left). 
Only two of the 33 protocluster candidates in the VT sample are not detected by HDBSCAN\null. As for the remaining 31, one is recovered as two separate objects and two are merged into a single structure. For all the protoclusters detected by both the VT and HDBSCAN methods, the latter yields larger measured areas than the former, typically by more than a factor of 2.


Why are the VT-based structures smaller and fewer in number compared to the HDBSCAN-selected ones? Visual inspection of Figure~\ref{fig:filaments} suggests that many HDBSCAN-selected structures with no VT counterpart are more elongated and filamentary. In particular, the filamentary arm previously highlighted in Complex~A is identified by HDBSCAN (highlighted in red in the figure) but not from the VT map. If we relax the overdensity threshold for the VT detection from $4.5\sigma$ to $3.5\sigma$, the number of HDBSCAN-selected protoclusters that overlap one or more VT-based ones increases from 30 to all 47. The areas of the two sets of protoclusters become more {similar}. However, the downside of lowering the detection threshold for the VT map is the increased contamination rate. We conclude that HDBSCAN fares better in recovering lower surface density features with a more elongated morphology, but with the downside that it is more difficult to ascribe a physical meaning to the detection parameters for this method ($K$th nearest neighbor and minimum cluster size) than for those for the GS and VT methods (minimum area and density threshold).

\subsection{Descendant masses of the protocluster candidates}

{The number densities of the VT and HDBSCAN protocluster samples (which we expect to be more complete than the GS  sample) are 5.5 $\times$ 10$^{-6}$~cMpc$^{-3}$ and 7.9 $\times$ 10$^{-6}$~cMpc$^{-3}$, respectively. The number density of clusters at $z$ = 0 in  TNG300 is 1.2 $\times$ 10$^{-5}$~cMpc$^{3}$, implying that our protocluster sample has a completeness of $\sim$ 50\%. Purely based on the comoving number density, i.e. by assuming that each of our protocluster candidates evolves into an independent galaxy cluster by $z$ $=$ 0, the descendant masses are estimated to be $\gtrsim$ 2 $\times$ 10$^{14}$~M$_\odot$ and 1.6 $\times$ 10$^{14}$~M$_\odot$, respectively. In practice, some of our structures will likely merge into a single more massive cluster. Thus, the masses quoted above represent a lower limit.}

\begin{figure*}
    \centering
    \includegraphics[width=\linewidth]{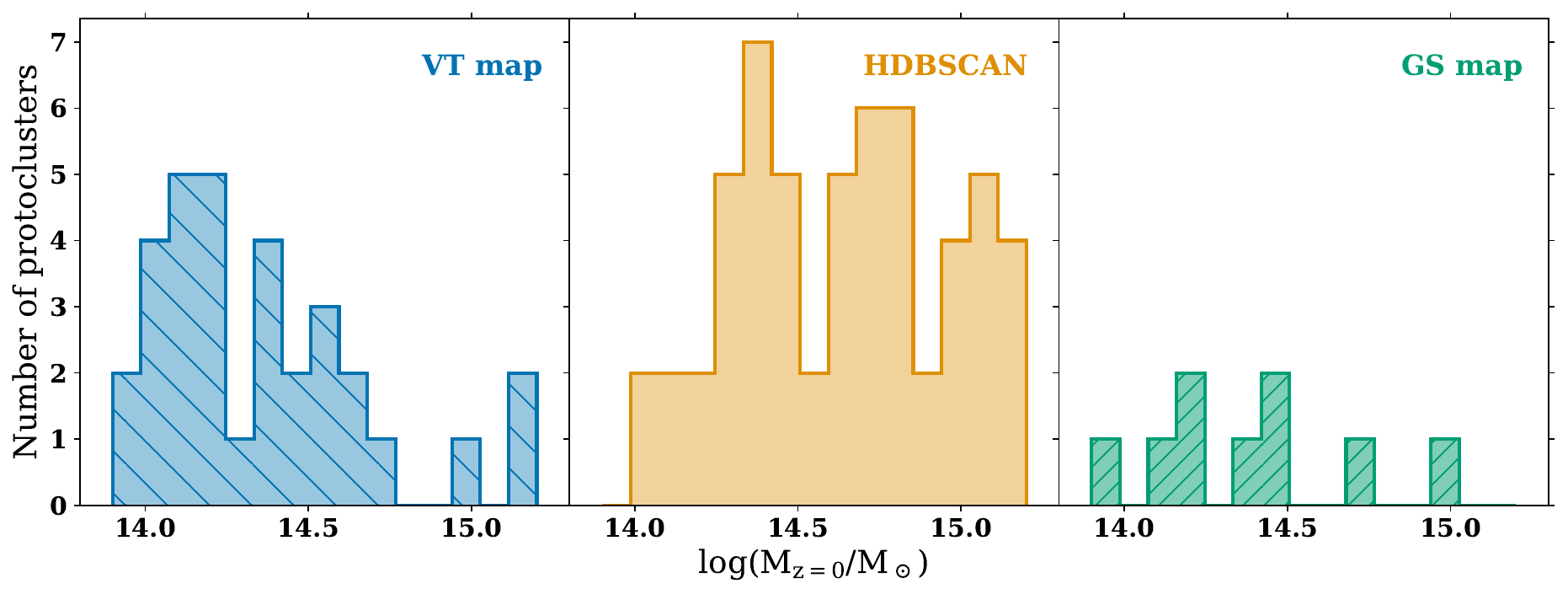}
    \caption{Descendant masses of the individual protoclusters for all three samples, after applying appropriate corrections to the mass calculated with Equation \ref{eqn:today_mass}.}
    \label{fig:halo_masses}
\end{figure*}

We now estimate the descendant masses of {individual} protocluster candidates. We showed in Section~\ref{subsec:mass_calib} that for our choice of protocluster detection parameters, Equation~\ref{eqn:today_mass} underestimates the descendant masses of the structures selected from the VT map by a factor of $\sim$ 3. A similar analysis for the GS and HDBSCAN methods (see Appendix~\ref{appendix:b}) shows that the masses of the former are underestimated by a factor of $\sim$ 2, and the latter by a factor of $\sim$ 1.15. In Figure~\ref{fig:halo_masses}, we show histograms of the descendant masses after making the appropriate corrections. The median masses are $\log(M_{z=0}/M_\odot) = 14.35$, 14.75, and 14.52 for the VT, HDBSCAN, and GS protoclusters, respectively, suggesting that our protocluster candidates will evolve into moderately massive `Virgo-type' clusters \citep{Chiang2013}. {The median values and the overall range spanned by the descendant masses are in reasonable agreement with the lower limits inferred from the number density. This again highlights the completeness of our protocluster sample.}

\subsection{Comparison with other works}

Several studies have made use of LAEs to trace the LSS at $z$ $\sim$ 2 - 4 \citep[e.g.,][]{Hayashino2004, Yamada2012, Lee2014, Badescu2017, Kikuta2019,huang22}. ODIN reaches a comparable \Lya line flux to these past surveys ($\sim$ 1 - 9 $\times$ 10$^{-17}$ ergs s$^{-1}$ cm$^{-2}$), but over a considerably larger contiguous area - \citet{Yamada2012} and \citet{Kikuta2019} have the largest area coverage at 1.38 \sqdeg\ and 1.1 \sqdeg, respectively, while the remainder survey areas of $\sim$ 0.2 - 0.5 \sqdeg. ODIN is also unusual in that our protocluster candidates are discovered blindly, with no prior information available; \citet{Badescu2017} and \citet{Kikuta2019} target the vicinity of a pair of \Lya nebulae and a hyperluminous QSO, respectively, while \citet{Hayashino2004, Yamada2012,Lee2014,huang22} used narrowband imaging to follow up on previously identified galaxy overdensities.

{It is worth noting that the structure studied by \citet{Yamada2012}, located at $z$ = 3.09 in the SSA22 field, is strikingly similar in its size and clumpy morphology to our Complex A. The entirety of the 1.38~\sqdeg\ region they survey is overdense in LAEs, of which a region of $\sim$ 60~cMpc$^2$ was previously studied by \citet{Steidel2000} and \citet{Hayashino2004} and comprises the core of the protocluster. 
The core region has been shown to host multiple extended and luminous \Lya nebulae \citep{Matsuda2004}, a trait shared by Complex A \citep[see][]{Ramakrishnan2023}, suggesting that the two systems are similar.}

{The protocluster studies using different techniques include \citet{Toshikawa2016}, who used $u$-dropout galaxies to identify five protoclusters  at $z$ $\sim$ 3 over 4 deg$^2$ of the fields covered by the Canada-France-Hawaii Telescope Legacy Survey \citep{Sawicki2019}; \citet{Toshikawa2018} use $g$-dropout galaxies to find 216 protoclusters at $z$ $\sim$ 3.8 over the 200 deg$^2$ of the HSC-SSP Wide fields. These samples have number densities of $\sim$ 1 $\times$ 10$^{-7}$ and 1.5 $\times$ 10$^{-6}$, respectively. \citet{Chiang2014} use galaxies with photometric redshifts from the COSMOS/UltraVISTA catalog to find 26 protoclusters at 1.6 $<$ z $<$ 3.1 over an area of $\sim$ 1.2 deg$^2$, corresponding to a number density of 1.2 $\times$ 10$^{-6}$ cMpc$^{-3}$. These works identify overdensities of similar sizes (5--15~cMpc) and, in the case of \citet{Toshikawa2018}, similar descendant mass ($\sim$ 1 - 2 $\times$ 10$^{14}$ M$_\odot$). However, the redshift uncertainty for the tracers used by these works is much larger than that of our LAEs. This may lead to density peaks being diluted by interlopers, resulting in lower number densities.}



\section{Conclusions} \label{sec:summary}

The ODIN survey is the largest-area deep field narrowband survey undertaken to date. By enabling the selection of LAEs, which are low-mass, star-forming galaxies and are well-localized in redshift space over a wide contiguous area, ODIN makes it possible to comprehensively trace the large-scale structure at high redshift. In this paper, we have used the early ODIN data taken in the COSMOS field with the $N501$ filter to compile a sample of protoclusters and cosmic filaments.

We create LAE surface density maps using two methods - by smoothing over the LAE positions with a fixed-size Gaussian kernel (GS map, Section \ref{subsec:gauss}) and by constructing the Voronoi diagram of the LAEs (VT map, Section \ref{subsec:voronoi}). We select protocluster candidates from the surface density maps (Section \ref{subsec:SEP}) and apply the density-based clustering algorithm HDBSCAN to the LAE positions (Section \ref{subsec:hdbscan}). We also select filaments of the cosmic web (Section \ref{subsec:filaments}). We assess the reliability of our structure detection by comparing our observations against the results obtained with a carefully created mock sample of LAEs from the IllustrisTNG300-1 hydrodynamical simulation (Section \ref{sec:comparison_with_TNG}). Our main results and conclusions are as follows:\\

\noindent 1. The large-scale structure revealed by our surface density maps is distinctly clumpy and irregular, with overdense regions clustered together in extended complexes. The VT map identifies filaments as regions of moderate LAE overdensity (\sdrel = 2--3) connecting the regions of highest density (\sdrel $>$ 3). These features are in line with the expectations of hierarchical structure formation.\\

\noindent 2. The three methods of identifying protoclusters that we explore all have their strengths and weaknesses. The VT map identifies a greater number of objects but also suffers from a higher contamination rate than the GS map. The GS map recovers only the most significant overdensities but is almost free of interlopers. The HDBSCAN map recovers the maximum number of structures, however, the input parameters are less straightforward to interpret or assign physical meaning to than for the surface density maps.\\

\noindent 3. The surface density maps of the $z$ = 3 TNG300 simulation box display remarkably similar features to those seen in the observations. The number and size of the overdensities selected with our fiducial detection parameters are likewise similar in the two. As shown in Figure \ref{fig:real_vs_tng_maps}, many of the prominent overdensities in the TNG300 surface density maps correspond to the progenitors of $z$ = 0 clusters.\\

\noindent 4. The simulation shows that we can successfully recover $\sim$ 60\% of the protoclusters with descendant mass $\gtrsim 2 \times 10^{14} M_\odot$. We find that the descendant mass of the simulated protoclusters can be estimated purely based on their area and LAE overdensity measured in 2D. The lack of information in the redshift direction introduces a scatter of $\sim$ 0.4 dex on the measurement.\\

\noindent 5. By comparing the filaments identified in 2D and 3D in TNG300, we show that despite projection effects, the cosmic web recovered in 2D is a close representation of the true LSS in regions with high LAE surface density. In the vicinity of protoclusters, the 3D filament network is almost perfectly recovered.\\

\noindent 6. On estimating the descendant masses of our observed protocluster samples, we find that they span the range log$(M_{z=0}/M_\odot)$ $\sim$ 14.0 - 15.0, with the median of the VT, GS and HDBSCAN samples being 14.35, 14.52 and 14.75 respectively. The majority of our protoclusters are thus likely to evolve into intermediate-mass `Virgo-type' clusters.\\

Our results establish the robustness of our protocluster and filament samples and demonstrate that LAEs are reliable and efficient tracers of large-scale structures at high redshift. Upon completion, the ODIN survey will allow us to select secure samples of massive structures that are nearly ten times larger than those presented here and span three cosmic epochs. ODIN will thus be well placed to grant insight into the formation and evolution of cosmic structures near Cosmic Noon.

\facility{Blanco}
\software{SExtractor \citep{Bertin1996}, SEP \citep{Barbary2016}, hdbscan \citep{McInnes2017}, Astropy \citep{astropy:2013,astropy:2018,astropy:2022}}

\begin{acknowledgments}
The authors acknowledge financial support from the National Science Foundation under Grant Nos. AST-2206705, AST-2408359, and AST-2206222, and from the Ross-Lynn Purdue Research Foundations. 
This material is based upon work supported by the National Science Foundation Graduate Research Fellowship Program under Grant No. DGE-2233066 to NF.
MCA and AK acknowledge financial support from 
ANID/Fondo 2022 ALMA/31220021. MCA acknowledges support from ANID BASAL project FB210003.
The Institute for Gravitation and the Cosmos is supported by the Eberly College of Science and the Office of the Senior Vice President for Research at the Pennsylvania State University.
HSH acknowledges the support from the National Research Foundation of Korea grant funded by the Korean government (MSIT) (No. 2021R1A2C1094577).
J.L. is supported by the National Research Foundation of Korea (NRF-2021R1C1C2011626).
SL is supported by the National Research Foundation of Korea grant funded by the Korean government (MSIT; Nos. 2020R1I1A1A01060310, 2020R1A2C3011091, 2021M3F7A1084525).
Based on observations at Cerro Tololo Inter-American Observatory, NSF’s NOIRLab (Prop. ID 2020B-0201; PI: K.-S. Lee), which is managed by the Association of Universities for Research in Astronomy under a cooperative agreement with the National Science Foundation.
\end{acknowledgments}

\appendix

\section{Parameter selection for HDBSCAN} \label{appendix:hdbscan}

As stated in Section \ref{subsec:hdbscan}, HDBSCAN is a density-based clustering algorithm that separates points into `clusters' and `noise'. HDBSCAN performs this classification by estimating the density at each data point from its $K$th nearest neighbor distance, referred to as the `core distance'. Points in dense(sparse) regions will have relatively small(large) core distances, and the core distance of a data point is thus inversely proportional to its density.

HDBSCAN is an extension of the DBSCAN algorithm. DBSCAN selects clusters by imposing a user-defined density threshold. This is implemented by placing a ceiling on the core distance/$K$th nearest neighbor distance, that is, by requiring that there be $K$ points within a distance $\epsilon$ or less of a given point. DBSCAN begins with a point that satisfies the requirement of having a core distance less than or equal to $\epsilon$ and classifies it and its neighbors within $\epsilon$ as a single cluster. If any of these neighbors also pass the density criterion, their neighbors within $\epsilon$ are assigned to the same cluster as well. This continues until there are no more points within the cluster that have a core distance less than $\epsilon$, at which point the cluster is closed. DBSCAN then selects an unassigned point with $K$ or more neighbors within $\epsilon$ and repeats the process. This continues until no points that pass the density threshold remain, at which juncture all remaining data points are classified as noise. 

The approach of DBSCAN, of applying a constant density/core distance threshold, is restrictive. The optimal value of $\epsilon$ is difficult to identify and can vary from region to region within the data. HDBSCAN circumvents the difficulty of selecting a single density threshold by making use of \emph{hierarchical clustering}; that is, it creates a hierarchy of clusters by applying successively higher density thresholds and then selects the optimal set of clusters from this hierarchy. This makes it ideal for selecting clusters of varying density from a single dataset, as in our case.

\begin{figure}
    \centering
    \includegraphics[width=\linewidth]{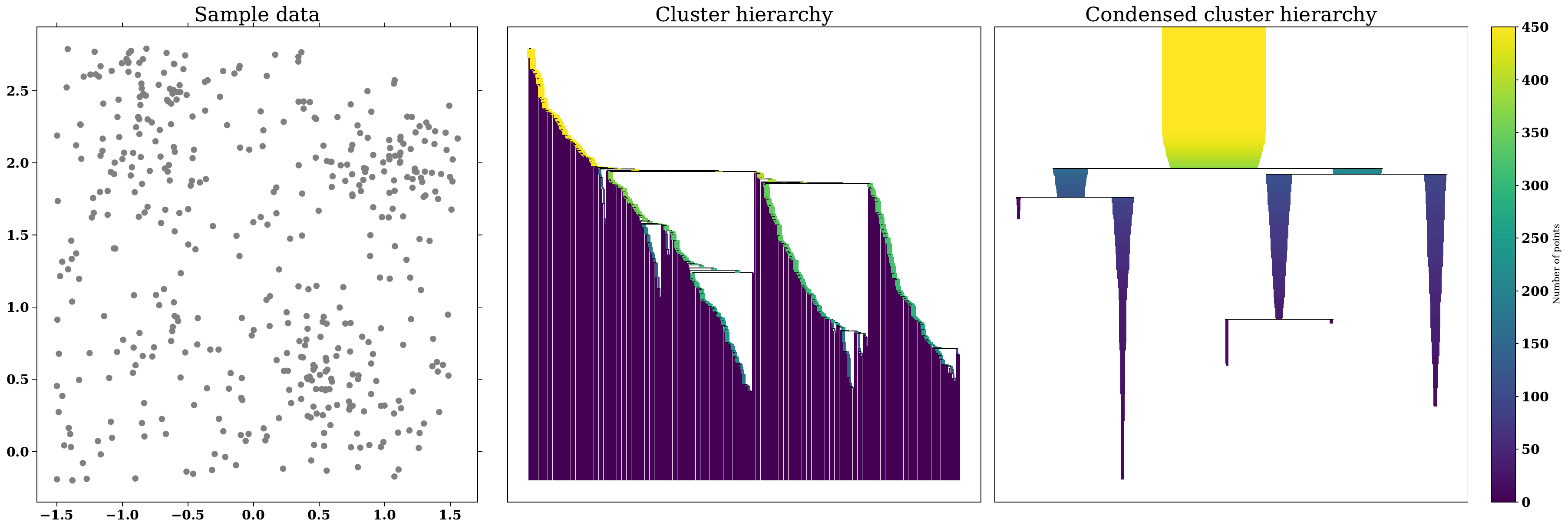}
    \caption{The cluster hierarchy (middle panel) and condensed cluster hierarchy (right panel) constructed by HDBSCAN for a sample dataset (left panel). The sample dataset contains three regions of densely clustered points along with random noise. The cluster hierarchy is constructed by applying successive density cuts. Each branch of the tree represents a cluster of points selected above some density cut. The condensed cluster hierarchy shows the cluster hierarchy after clusters smaller than the minimum cluster size are discarded.}
    \label{fig:appendix_1}
\end{figure}

As it constructs a cluster hierarchy, HDBSCAN does not require the user to supply the parameter $\epsilon$. Instead, the parameters which govern the cluster selection are $K$, as before (which determines how density is measured) and an additional parameter, the minimum cluster size. HDBSCAN uses the minimum cluster size to condense the cluster hierarchy - any cluster in the hierarchy that has fewer points than this size is discarded. This is shown in Figure \ref{fig:appendix_1} for a sample dataset. From this condensed hierarchy HDBSCAN selects the final set of clusters.

\begin{figure}
    \centering
    \includegraphics[width=0.7\linewidth]{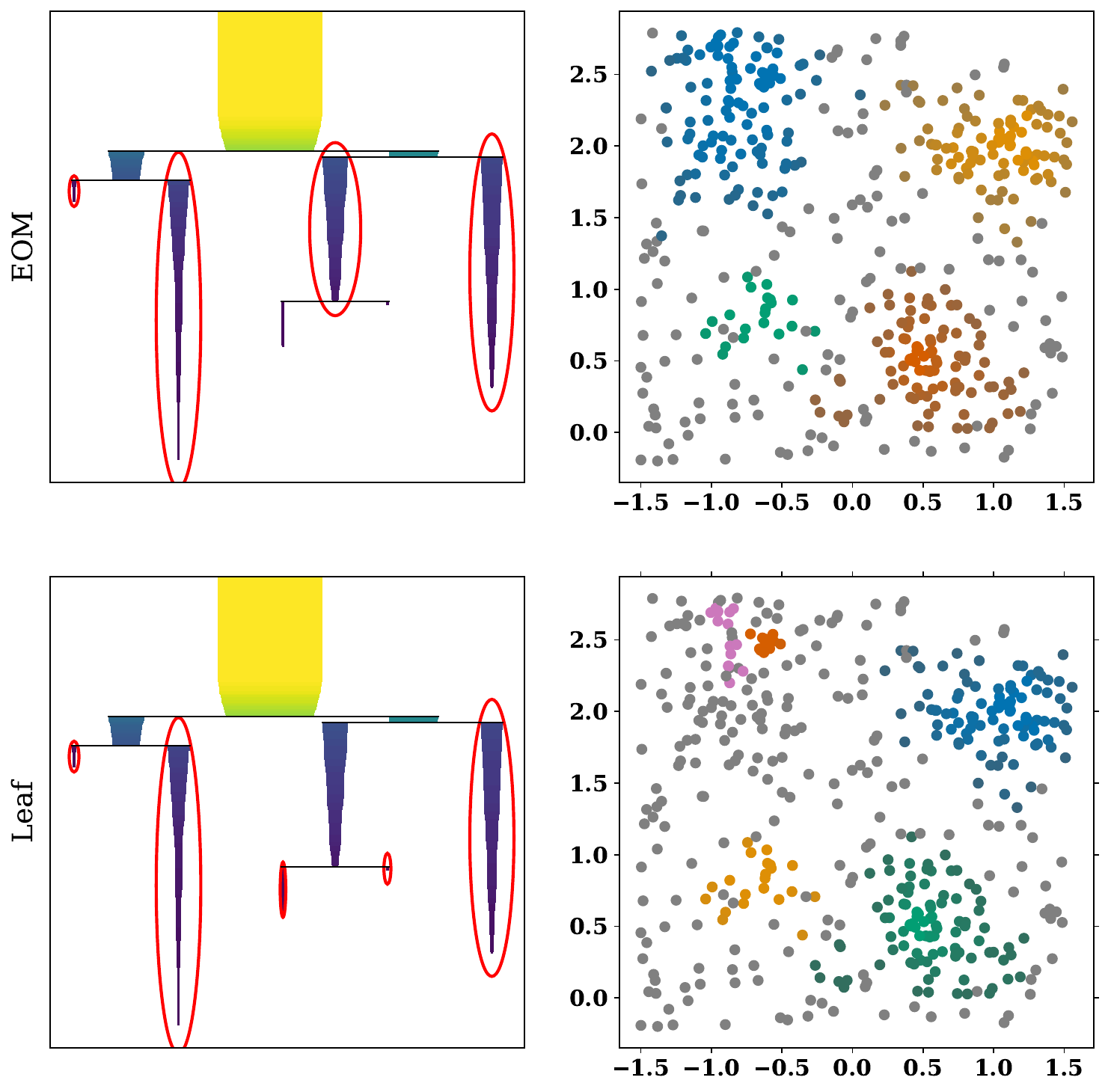}
    \caption{The selection of clusters from the condensed cluster hierarchy with the excess-of-mass (EOM, top row) and leaf (bottom row) methods. Selected branches of the cluster hierarchy are circled in red in the left panels and the corresponding set of cluster points are shown in color in the right panel, while noise points are gray. The EOM method selects the clusters that persist for the greatest range of density cuts, while the leaf method selects the `leaves' of the cluster hierarchy, i.e. the smallest surviving clusters.}
    \label{fig:appendix_2}
\end{figure}

The \textsc{hdbscan} library provides two ways of selecting the final set of clusters. The first method, the `excess-of-mass' method, selects the most stable set of clusters, defined as those that persist for the greatest range of density thresholds. The second method, the `leaf' method, selects the smallest surviving clusters in the hierarchy. These two methods are illustrated in Figure \ref{fig:appendix_2} for the same sample dataset.

Based on the above description, there are two main parameters that govern the clustering based on HDBSCAN, namely $K$ and the minimum cluster size. By determining which nearest-neighbour distance will be used for estimating the density, $K$ controls how finely the density distribution is sampled, in essence acting as a kind of smoothing scale.
Changing the minimum cluster size affects the condensed cluster hierarchy by changing the number of points required to be considered a cluster, that is, more points are discarded as noise and the cluster selection is more conservative. Setting either of these parameters to too low a value will result in many noise peaks being erroneously selected as clusters, whereas setting them to be too large may result in structures being blended together. Our key consideration in choosing the parameters is ensuring that nearby structures are separated to the extent possible, which we enforce as described below. 

\begin{figure}
    \centering
    \includegraphics[width=0.5\linewidth]{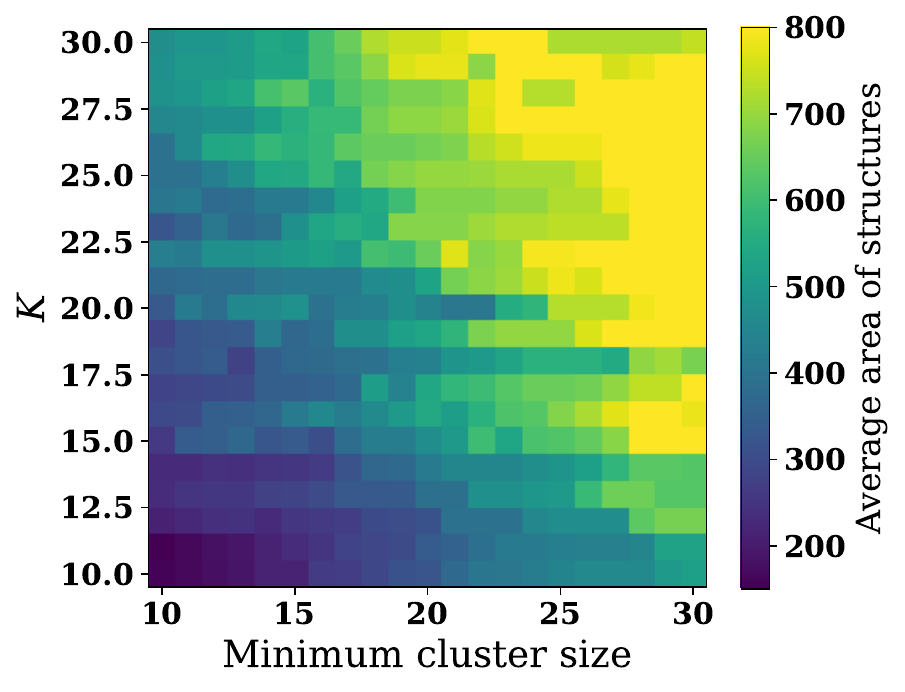}
    \caption{Average area of the structures detected by HDBSCAN, for various combinations of the $K$th nearest neighbour used to measure density ({\tt min\_samples}) and minimum number of points to constitute a cluster ({\tt min\_cluster\_size}). The expected area of a protocluster is $\sim$ 100 - 200 cMpc$^2$; areas much larger than this suggest that multiple structures are being blended together by the algorithm.}
    \label{fig:hdbscan_areas}
\end{figure}

We run HDBSCAN on the positions of the LAEs with {\tt min\_samples} and {\tt min\_cluster\_size} both in the range 10--30. We observe that the excess-of-mass method tends to blend structures irrespective of the specific parameter values used, thus we choose to use the leaf method in our cluster selection. As discussed in the main text, we additionally require that the clusters be above a surface density threshold determined such that the contamination fraction is $\lesssim$ 0.2. We estimate the mean area of the structures selected for each combination of parameters, as shown in Figure~\ref{fig:hdbscan_areas}. With increasing values of {\tt min\_samples} and {\tt min\_cluster\_size}, the mean area of the detected structures increases, indicating that smaller structures are indeed being blended. We select our final parameters to be {\tt min\_cluster\_size} = 10 and {\tt min\_samples} = 15, which as shown in Figure~\ref{fig:hdbscan_areas} yields a protocluster sample with a mean area of $\sim$ 200~cMpc$^2$. This choice is based on the physical motivation that the scale of a protocluster is expected to be $\sim$ 10--15~cMpc \citep{Chiang2013}, corresponding to a projected area (for an isotropic structure) of $\sim$ 100--225~cMpc$^2$. We make use of a value of {\tt min\_samples} on the larger end of the allowed range because while over-smoothing the density distribution is undesirable, it is equally undesirable for the distribution to be too finely sampled as it becomes very noisy. 

\section{Calibration of the mass estimation for GS map and HDBSCAN}\label{appendix:b}

In Section~\ref{subsec:mass_calib}, we evaluated the reliability of the estimated descendant massses for individual protocluster candidates selected from the VT map. We found that the descendant masses were on average underestimated by a factor of $\sim$ 3, which we attributed to the underestimation of the protocluster volume by our detection algorithm. Here we similarly examine the accuracy of the descendant mass estimates for the protocluster candidates selected from the GS map and with HDBSCAN. 

\begin{figure}
    \includegraphics[width=5.in]{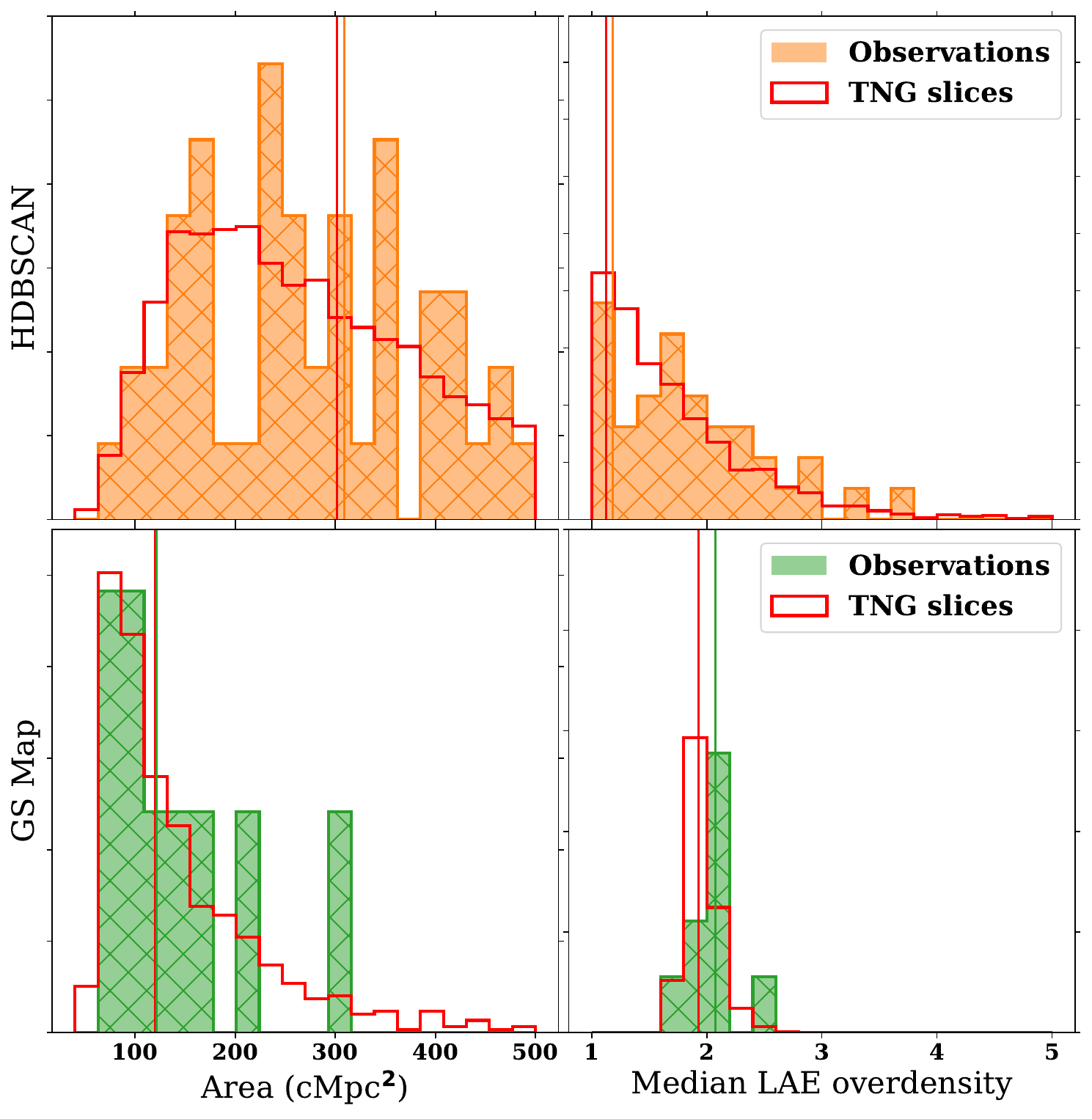}
    \caption{Area and median LAE overdensity ($\delta_{LAE}$) for the structures selected with HDBSCAN (top) and from the GS map (bottom) in the TNG300 slices compared to those in observations. Solid lines indicate the median values of each property. As with the structures selected from the VT map, the properties of the objects in the simulation and observations are similar.}
    \label{fig:real_vs_tng_4}
\end{figure}

We once again use the 90 slices constructed in Section~\ref{subsec:tng}. Analogously to the middle and right panels of Figure~\ref{fig:real_vs_tng_hist}, we compare the properties of the structures selected with HDBSCAN (selected from the GS map) from these slices to those of the observed protoclusters in the top (bottom) panel of Figure~\ref{fig:real_vs_tng_4}. The structures remain similar in projected area and median LAE overdensity between theory and simulations, reaffirming the fact that our observational protocluster candidates are reasonable. 

As in Section~\ref{subsec:mass_calib}, for each slice, we consider those of the 30 massive cluster progenitors lying within the slice which are successfully recovered. The recovery rate of the HDBSCAN is $\sim$ 70\%, slightly higher than that of the VT map. This is in accordance with the fact that the HDBSCAN candidates are the most numerous. By contrast, the recovery rate of the GS map is very low, $\sim$ 20\%, in line with our observation that only the most prominent structures are recovered.

\begin{figure}
    \centering
    \includegraphics[width=\linewidth]{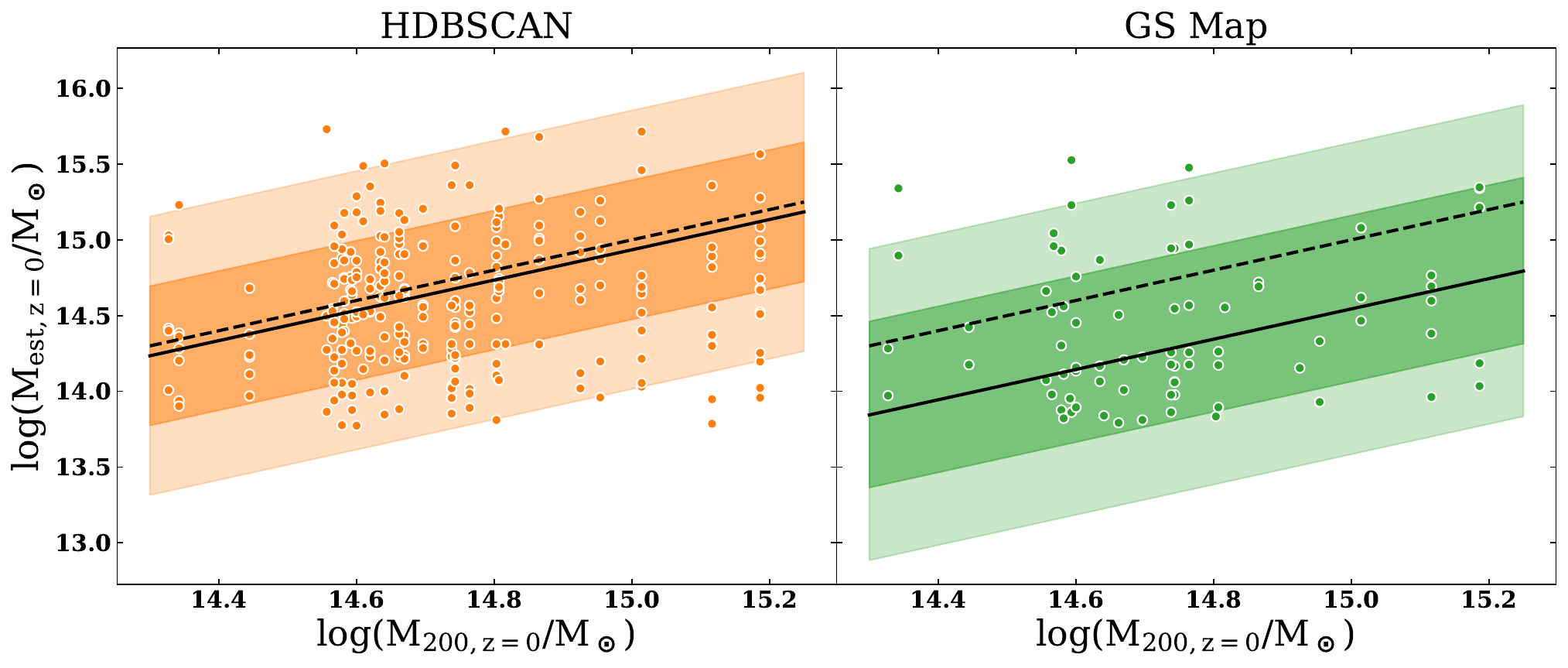}
    \caption{Descendant mass estimated according to Equation \ref{eqn:today_mass} compared to the true descendant mass for structures selected with HDBSCAN (left) and from the GS map (right). The dashed lines represent a 1-to-1 relation of estimated and true mass whereas the solid lines represent the median ratio of the two.}\label{fig:m_est_to_m_true_2}
\end{figure}

We compare the estimated and true descendant masses for recovered clusters for both methods in Figure~\ref{fig:m_est_to_m_true_2}. We find that the descendant masses of the protoclusters are underestimated by a factor of $\sim$ 2 with the GS method, and by a factor of $\sim$ 1.15 with HDBSCAN. In both cases, the scatter is similar to that observed with the VT map, $\sim$ 0.5~dex.

\bibliography{refs}

\begin{thebibliography}{}
\expandafter\ifx\csname natexlab\endcsname\relax\def\natexlab#1{#1}\fi
\providecommand{\url}[1]{\href{#1}{#1}}
\providecommand{\dodoi}[1]{doi:~\href{http://doi.org/#1}{\nolinkurl{#1}}}
\providecommand{\doeprint}[1]{\href{http://ascl.net/#1}{\nolinkurl{http://ascl.net/#1}}}
\providecommand{\doarXiv}[1]{\href{https://arxiv.org/abs/#1}{\nolinkurl{https://arxiv.org/abs/#1}}}

\bibitem[{{Aihara} {et~al.}(2018{\natexlab{a}}){Aihara}, {Arimoto},
  {Armstrong}, {Arnouts}, {Bahcall}, {Bickerton}, {Bosch}, {Bundy}, {Capak},
  {Chan}, {Chiba}, {Coupon}, {Egami}, {Enoki}, {Finet}, {Fujimori}, {Fujimoto},
  {Furusawa}, {Furusawa}, {Goto}, {Goulding}, {Greco}, {Greene}, {Gunn},
  {Hamana}, {Harikane}, {Hashimoto}, {Hattori}, {Hayashi}, {Hayashi},
  {He{\l}miniak}, {Higuchi}, {Hikage}, {Ho}, {Hsieh}, {Huang}, {Huang},
  {Ikeda}, {Imanishi}, {Inoue}, {Iwasawa}, {Iwata}, {Jaelani}, {Jian},
  {Kamata}, {Karoji}, {Kashikawa}, {Katayama}, {Kawanomoto}, {Kayo}, {Koda},
  {Koike}, {Kojima}, {Komiyama}, {Konno}, {Koshida}, {Koyama}, {Kusakabe},
  {Leauthaud}, {Lee}, {Lin}, {Lin}, {Lupton}, {Mandelbaum}, {Matsuoka},
  {Medezinski}, {Mineo}, {Miyama}, {Miyatake}, {Miyazaki}, {Momose}, {More},
  {More}, {Moritani}, {Moriya}, {Morokuma}, {Mukae}, {Murata}, {Murayama},
  {Nagao}, {Nakata}, {Niida}, {Niikura}, {Nishizawa}, {Obuchi}, {Oguri},
  {Oishi}, {Okabe}, {Okamoto}, {Okura}, {Ono}, {Onodera}, {Onoue}, {Osato},
  {Ouchi}, {Price}, {Pyo}, {Sako}, {Sawicki}, {Shibuya}, {Shimasaku},
  {Shimono}, {Shirasaki}, {Silverman}, {Simet}, {Speagle}, {Spergel},
  {Strauss}, {Sugahara}, {Sugiyama}, {Suto}, {Suyu}, {Suzuki}, {Tait},
  {Takada}, {Takata}, {Tamura}, {Tanaka}, {Tanaka}, {Tanaka}, {Tanaka},
  {Terai}, {Terashima}, {Toba}, {Tominaga}, {Toshikawa}, {Turner}, {Uchida},
  {Uchiyama}, {Umetsu}, {Uraguchi}, {Urata}, {Usuda}, {Utsumi}, {Wang}, {Wang},
  {Wong}, {Yabe}, {Yamada}, {Yamanoi}, {Yasuda}, {Yeh}, {Yonehara}, \&
  {Yuma}}]{Aihara2018a}
{Aihara}, H., {Arimoto}, N., {Armstrong}, R., {et~al.} 2018{\natexlab{a}},
  \pasj, 70, S4, \dodoi{10.1093/pasj/psx066}

\bibitem[{{Aihara} {et~al.}(2018{\natexlab{b}}){Aihara}, {Armstrong},
  {Bickerton}, {Bosch}, {Coupon}, {Furusawa}, {Hayashi}, {Ikeda}, {Kamata},
  {Karoji}, {Kawanomoto}, {Koike}, {Komiyama}, {Lang}, {Lupton}, {Mineo},
  {Miyatake}, {Miyazaki}, {Morokuma}, {Obuchi}, {Oishi}, {Okura}, {Price},
  {Takata}, {Tanaka}, {Tanaka}, {Tanaka}, {Uchida}, {Uraguchi}, {Utsumi},
  {Wang}, {Yamada}, {Yamanoi}, {Yasuda}, {Arimoto}, {Chiba}, {Finet},
  {Fujimori}, {Fujimoto}, {Furusawa}, {Goto}, {Goulding}, {Gunn}, {Harikane},
  {Hattori}, {Hayashi}, {He{\l}miniak}, {Higuchi}, {Hikage}, {Ho}, {Hsieh},
  {Huang}, {Huang}, {Imanishi}, {Iwata}, {Jaelani}, {Jian}, {Kashikawa},
  {Katayama}, {Kojima}, {Konno}, {Koshida}, {Kusakabe}, {Leauthaud}, {Lee},
  {Lin}, {Lin}, {Mandelbaum}, {Matsuoka}, {Medezinski}, {Miyama}, {Momose},
  {More}, {More}, {Mukae}, {Murata}, {Murayama}, {Nagao}, {Nakata}, {Niida},
  {Niikura}, {Nishizawa}, {Oguri}, {Okabe}, {Ono}, {Onodera}, {Onoue}, {Ouchi},
  {Pyo}, {Shibuya}, {Shimasaku}, {Simet}, {Speagle}, {Spergel}, {Strauss},
  {Sugahara}, {Sugiyama}, {Suto}, {Suzuki}, {Tait}, {Takada}, {Terai}, {Toba},
  {Turner}, {Uchiyama}, {Umetsu}, {Urata}, {Usuda}, {Yeh}, \&
  {Yuma}}]{Aihara2018b}
{Aihara}, H., {Armstrong}, R., {Bickerton}, S., {et~al.} 2018{\natexlab{b}},
  \pasj, 70, S8, \dodoi{10.1093/pasj/psx081}

\bibitem[{{Aihara} {et~al.}(2019){Aihara}, {AlSayyad}, {Ando}, {Armstrong},
  {Bosch}, {Egami}, {Furusawa}, {Furusawa}, {Goulding}, {Harikane}, {Hikage},
  {Ho}, {Hsieh}, {Huang}, {Ikeda}, {Imanishi}, {Ito}, {Iwata}, {Jaelani},
  {Kakuma}, {Kawana}, {Kikuta}, {Kobayashi}, {Koike}, {Komiyama}, {Li},
  {Liang}, {Lin}, {Luo}, {Lupton}, {Lust}, {MacArthur}, {Matsuoka}, {Mineo},
  {Miyatake}, {Miyazaki}, {More}, {Murata}, {Namiki}, {Nishizawa}, {Oguri},
  {Okabe}, {Okamoto}, {Okura}, {Ono}, {Onodera}, {Onoue}, {Osato}, {Ouchi},
  {Shibuya}, {Strauss}, {Sugiyama}, {Suto}, {Takada}, {Takagi}, {Takata},
  {Takita}, {Tanaka}, {Terai}, {Toba}, {Uchiyama}, {Utsumi}, {Wang}, {Wang}, \&
  {Yamada}}]{Aihara2019}
{Aihara}, H., {AlSayyad}, Y., {Ando}, M., {et~al.} 2019, \pasj, 71, 114,
  \dodoi{10.1093/pasj/psz103}

\bibitem[{{Alpaslan} {et~al.}(2014){Alpaslan}, {Robotham}, {Driver}, {Norberg},
  {Baldry}, {Bauer}, {Bland-Hawthorn}, {Brown}, {Cluver}, {Colless}, {Foster},
  {Hopkins}, {Van Kampen}, {Kelvin}, {Lara-Lopez}, {Liske}, {Lopez-Sanchez},
  {Loveday}, {McNaught-Roberts}, {Merson}, \& {Pimbblet}}]{Alpaslan2014}
{Alpaslan}, M., {Robotham}, A. S.~G., {Driver}, S., {et~al.} 2014, \mnras, 438,
  177, \dodoi{10.1093/mnras/stt2136}

\bibitem[{{Andrews} {et~al.}(2024){Andrews}, {Artale}, {Kumar}, {Lee},
  {Florek}, {Anand}, {Cerdosino}, {Ciardullo}, {Firestone}, {Gawiser},
  {Gronwall}, {Guaita}, {Hong}, {Hwang}, {Lee}, {Lee}, {Padilla}, {Park},
  {Popescu}, {Ramakrishnan}, {Song}, {Vivanco C{\'a}diz}, \&
  {Vogelsberger}}]{Andrews2024}
{Andrews}, M., {Artale}, M.~C., {Kumar}, A., {et~al.} 2024, arXiv e-prints,
  arXiv:2410.08412, \dodoi{10.48550/arXiv.2410.08412}

\bibitem[{{Astropy Collaboration} {et~al.}(2013){Astropy Collaboration},
  {Robitaille}, {Tollerud}, {Greenfield}, {Droettboom}, {Bray}, {Aldcroft},
  {Davis}, {Ginsburg}, {Price-Whelan}, {Kerzendorf}, {Conley}, {Crighton},
  {Barbary}, {Muna}, {Ferguson}, {Grollier}, {Parikh}, {Nair}, {Unther},
  {Deil}, {Woillez}, {Conseil}, {Kramer}, {Turner}, {Singer}, {Fox}, {Weaver},
  {Zabalza}, {Edwards}, {Azalee Bostroem}, {Burke}, {Casey}, {Crawford},
  {Dencheva}, {Ely}, {Jenness}, {Labrie}, {Lim}, {Pierfederici}, {Pontzen},
  {Ptak}, {Refsdal}, {Servillat}, \& {Streicher}}]{astropy:2013}
{Astropy Collaboration}, {Robitaille}, T.~P., {Tollerud}, E.~J., {et~al.} 2013,
  \aap, 558, A33, \dodoi{10.1051/0004-6361/201322068}

\bibitem[{{Astropy Collaboration} {et~al.}(2018){Astropy Collaboration},
  {Price-Whelan}, {Sip{\H{o}}cz}, {G{\"u}nther}, {Lim}, {Crawford}, {Conseil},
  {Shupe}, {Craig}, {Dencheva}, {Ginsburg}, {Vand erPlas}, {Bradley},
  {P{\'e}rez-Su{\'a}rez}, {de Val-Borro}, {Aldcroft}, {Cruz}, {Robitaille},
  {Tollerud}, {Ardelean}, {Babej}, {Bach}, {Bachetti}, {Bakanov}, {Bamford},
  {Barentsen}, {Barmby}, {Baumbach}, {Berry}, {Biscani}, {Boquien}, {Bostroem},
  {Bouma}, {Brammer}, {Bray}, {Breytenbach}, {Buddelmeijer}, {Burke},
  {Calderone}, {Cano Rodr{\'\i}guez}, {Cara}, {Cardoso}, {Cheedella}, {Copin},
  {Corrales}, {Crichton}, {D'Avella}, {Deil}, {Depagne}, {Dietrich}, {Donath},
  {Droettboom}, {Earl}, {Erben}, {Fabbro}, {Ferreira}, {Finethy}, {Fox},
  {Garrison}, {Gibbons}, {Goldstein}, {Gommers}, {Greco}, {Greenfield},
  {Groener}, {Grollier}, {Hagen}, {Hirst}, {Homeier}, {Horton}, {Hosseinzadeh},
  {Hu}, {Hunkeler}, {Ivezi{\'c}}, {Jain}, {Jenness}, {Kanarek}, {Kendrew},
  {Kern}, {Kerzendorf}, {Khvalko}, {King}, {Kirkby}, {Kulkarni}, {Kumar},
  {Lee}, {Lenz}, {Littlefair}, {Ma}, {Macleod}, {Mastropietro}, {McCully},
  {Montagnac}, {Morris}, {Mueller}, {Mumford}, {Muna}, {Murphy}, {Nelson},
  {Nguyen}, {Ninan}, {N{\"o}the}, {Ogaz}, {Oh}, {Parejko}, {Parley}, {Pascual},
  {Patil}, {Patil}, {Plunkett}, {Prochaska}, {Rastogi}, {Reddy Janga},
  {Sabater}, {Sakurikar}, {Seifert}, {Sherbert}, {Sherwood-Taylor}, {Shih},
  {Sick}, {Silbiger}, {Singanamalla}, {Singer}, {Sladen}, {Sooley},
  {Sornarajah}, {Streicher}, {Teuben}, {Thomas}, {Tremblay}, {Turner},
  {Terr{\'o}n}, {van Kerkwijk}, {de la Vega}, {Watkins}, {Weaver}, {Whitmore},
  {Woillez}, {Zabalza}, \& {Astropy Contributors}}]{astropy:2018}
{Astropy Collaboration}, {Price-Whelan}, A.~M., {Sip{\H{o}}cz}, B.~M., {et~al.}
  2018, \aj, 156, 123, \dodoi{10.3847/1538-3881/aabc4f}

\bibitem[{{Astropy Collaboration} {et~al.}(2022){Astropy Collaboration},
  {Price-Whelan}, {Lim}, {Earl}, {Starkman}, {Bradley}, {Shupe}, {Patil},
  {Corrales}, {Brasseur}, {N{"o}the}, {Donath}, {Tollerud}, {Morris},
  {Ginsburg}, {Vaher}, {Weaver}, {Tocknell}, {Jamieson}, {van Kerkwijk},
  {Robitaille}, {Merry}, {Bachetti}, {G{"u}nther}, {Aldcroft},
  {Alvarado-Montes}, {Archibald}, {B{'o}di}, {Bapat}, {Barentsen}, {Baz{'a}n},
  {Biswas}, {Boquien}, {Burke}, {Cara}, {Cara}, {Conroy}, {Conseil}, {Craig},
  {Cross}, {Cruz}, {D'Eugenio}, {Dencheva}, {Devillepoix}, {Dietrich},
  {Eigenbrot}, {Erben}, {Ferreira}, {Foreman-Mackey}, {Fox}, {Freij}, {Garg},
  {Geda}, {Glattly}, {Gondhalekar}, {Gordon}, {Grant}, {Greenfield}, {Groener},
  {Guest}, {Gurovich}, {Handberg}, {Hart}, {Hatfield-Dodds}, {Homeier},
  {Hosseinzadeh}, {Jenness}, {Jones}, {Joseph}, {Kalmbach}, {Karamehmetoglu},
  {Ka{l}uszy{'n}ski}, {Kelley}, {Kern}, {Kerzendorf}, {Koch}, {Kulumani},
  {Lee}, {Ly}, {Ma}, {MacBride}, {Maljaars}, {Muna}, {Murphy}, {Norman},
  {O'Steen}, {Oman}, {Pacifici}, {Pascual}, {Pascual-Granado}, {Patil},
  {Perren}, {Pickering}, {Rastogi}, {Roulston}, {Ryan}, {Rykoff}, {Sabater},
  {Sakurikar}, {Salgado}, {Sanghi}, {Saunders}, {Savchenko}, {Schwardt},
  {Seifert-Eckert}, {Shih}, {Jain}, {Shukla}, {Sick}, {Simpson},
  {Singanamalla}, {Singer}, {Singhal}, {Sinha}, {Sip{H{o}}cz}, {Spitler},
  {Stansby}, {Streicher}, {{{S}}umak}, {Swinbank}, {Taranu}, {Tewary},
  {Tremblay}, {Val-Borro}, {Van Kooten}, {Vasovi{'c}}, {Verma}, {de Miranda
  Cardoso}, {Williams}, {Wilson}, {Winkel}, {Wood-Vasey}, {Xue}, {Yoachim},
  {Zhang}, {Zonca}, \& {Astropy Project Contributors}}]{astropy:2022}
{Astropy Collaboration}, {Price-Whelan}, A.~M., {Lim}, P.~L., {et~al.} 2022,
  \apj, 935, 167, \dodoi{10.3847/1538-4357/ac7c74}

\bibitem[{{Barbary}(2016)}]{Barbary2016}
{Barbary}, K. 2016, The Journal of Open Source Software, 1, 58,
  \dodoi{10.21105/joss.00058}

\bibitem[{{Bertin} \& {Arnouts}(1996)}]{Bertin1996}
{Bertin}, E., \& {Arnouts}, S. 1996, \aaps, 117, 393,
  \dodoi{10.1051/aas:1996164}

\bibitem[{{Bielby} {et~al.}(2016){Bielby}, {Tummuangpak}, {Shanks}, {Francke},
  {Crighton}, {Ba{\~n}ados}, {Gonz{\'a}lez-L{\'o}pez}, {Infante}, \&
  {Orsi}}]{Bielby2016}
{Bielby}, R.~M., {Tummuangpak}, P., {Shanks}, T., {et~al.} 2016, \mnras, 456,
  4061, \dodoi{10.1093/mnras/stv2914}

\bibitem[{{Bleem} {et~al.}(2015){Bleem}, {Stalder}, {de Haan}, {Aird}, {Allen},
  {Applegate}, {Ashby}, {Bautz}, {Bayliss}, {Benson}, {Bocquet}, {Brodwin},
  {Carlstrom}, {Chang}, {Chiu}, {Cho}, {Clocchiatti}, {Crawford}, {Crites},
  {Desai}, {Dietrich}, {Dobbs}, {Foley}, {Forman}, {George}, {Gladders},
  {Gonzalez}, {Halverson}, {Hennig}, {Hoekstra}, {Holder}, {Holzapfel},
  {Hrubes}, {Jones}, {Keisler}, {Knox}, {Lee}, {Leitch}, {Liu}, {Lueker},
  {Luong-Van}, {Mantz}, {Marrone}, {McDonald}, {McMahon}, {Meyer}, {Mocanu},
  {Mohr}, {Murray}, {Padin}, {Pryke}, {Reichardt}, {Rest}, {Ruel}, {Ruhl},
  {Saliwanchik}, {Saro}, {Sayre}, {Schaffer}, {Schrabback}, {Shirokoff},
  {Song}, {Spieler}, {Stanford}, {Staniszewski}, {Stark}, {Story}, {Stubbs},
  {Vanderlinde}, {Vieira}, {Vikhlinin}, {Williamson}, {Zahn}, \&
  {Zenteno}}]{Bleem2015}
{Bleem}, L.~E., {Stalder}, B., {de Haan}, T., {et~al.} 2015, \apjs, 216, 27,
  \dodoi{10.1088/0067-0049/216/2/27}

\bibitem[{{Bond} {et~al.}(1996){Bond}, {Kofman}, \& {Pogosyan}}]{bond96}
{Bond}, J.~R., {Kofman}, L., \& {Pogosyan}, D. 1996, \nat, 380, 603,
  \dodoi{10.1038/380603a0}

\bibitem[{{Boylan-Kolchin} {et~al.}(2009){Boylan-Kolchin}, {Springel}, {White},
  {Jenkins}, \& {Lemson}}]{Bolyan_Kolchin2009}
{Boylan-Kolchin}, M., {Springel}, V., {White}, S. D.~M., {Jenkins}, A., \&
  {Lemson}, G. 2009, \mnras, 398, 1150,
  \dodoi{10.1111/j.1365-2966.2009.15191.x}

\bibitem[{{B{\u{a}}descu} {et~al.}(2017){B{\u{a}}descu}, {Yang}, {Bertoldi},
  {Zabludoff}, {Karim}, \& {Magnelli}}]{Badescu2017}
{B{\u{a}}descu}, T., {Yang}, Y., {Bertoldi}, F., {et~al.} 2017, \apj, 845, 172,
  \dodoi{10.3847/1538-4357/aa8220}

\bibitem[{{Byrohl} \& {Nelson}(2023)}]{Byrohl2023}
{Byrohl}, C., \& {Nelson}, D. 2023, \mnras, 523, 5248,
  \dodoi{10.1093/mnras/stad1779}

\bibitem[{Campello {et~al.}(2013)Campello, Moulavi, \& Sander}]{Campello2013}
Campello, R. J. G.~B., Moulavi, D., \& Sander, J. 2013, in Advances in
  Knowledge Discovery and Data Mining, ed. J.~Pei, V.~S. Tseng, L.~Cao,
  H.~Motoda, \& G.~Xu (Berlin, Heidelberg: Springer Berlin Heidelberg),
  160--172

\bibitem[{{Casey} {et~al.}(2015){Casey}, {Cooray}, {Capak}, {Fu}, {Kovac},
  {Lilly}, {Sanders}, {Scoville}, \& {Treister}}]{Casey2015}
{Casey}, C.~M., {Cooray}, A., {Capak}, P., {et~al.} 2015, \apjl, 808, L33,
  \dodoi{10.1088/2041-8205/808/2/L33}

\bibitem[{{Cautun} \& {van de Weygaert}(2011)}]{Cautun2011}
{Cautun}, M.~C., \& {van de Weygaert}, R. 2011, arXiv e-prints,
  arXiv:1105.0370, \dodoi{10.48550/arXiv.1105.0370}

\bibitem[{{Chiang} {et~al.}(2013){Chiang}, {Overzier}, \&
  {Gebhardt}}]{Chiang2013}
{Chiang}, Y.-K., {Overzier}, R., \& {Gebhardt}, K. 2013, \apj, 779, 127,
  \dodoi{10.1088/0004-637X/779/2/127}

\bibitem[{{Chiang} {et~al.}(2014){Chiang}, {Overzier}, \&
  {Gebhardt}}]{Chiang2014}
---. 2014, \apjl, 782, L3, \dodoi{10.1088/2041-8205/782/1/L3}

\bibitem[{{Ciardullo} {et~al.}(2002){Ciardullo}, {Feldmeier}, {Krelove},
  {Jacoby}, \& {Gronwall}}]{Ciardullo2002}
{Ciardullo}, R., {Feldmeier}, J.~J., {Krelove}, K., {Jacoby}, G.~H., \&
  {Gronwall}, C. 2002, \apj, 566, 784, \dodoi{10.1086/338230}

\bibitem[{{Ciardullo} {et~al.}(2012){Ciardullo}, {Gronwall}, {Wolf},
  {McCathran}, {Bond}, {Gawiser}, {Guaita}, {Feldmeier}, {Treister}, {Padilla},
  {Francke}, {Matkovi{\'c}}, {Altmann}, \& {Herrera}}]{Ciardullo2012}
{Ciardullo}, R., {Gronwall}, C., {Wolf}, C., {et~al.} 2012, \apj, 744, 110,
  \dodoi{10.1088/0004-637X/744/2/110}

\bibitem[{{Cucciati} {et~al.}(2018){Cucciati}, {Lemaux}, {Zamorani}, {Le
  F{\`e}vre}, {Tasca}, {Hathi}, {Lee}, {Bardelli}, {Cassata}, {Garilli}, {Le
  Brun}, {Maccagni}, {Pentericci}, {Thomas}, {Vanzella}, {Zucca}, {Lubin},
  {Amorin}, {Cassar{\`a}}, {Cimatti}, {Talia}, {Vergani}, {Koekemoer}, {Pforr},
  \& {Salvato}}]{Cucciati2018}
{Cucciati}, O., {Lemaux}, B.~C., {Zamorani}, G., {et~al.} 2018, \aap, 619, A49,
  \dodoi{10.1051/0004-6361/201833655}

\bibitem[{{Daddi} {et~al.}(2021){Daddi}, {Valentino}, {Rich}, {Neill},
  {Gronke}, {O'Sullivan}, {Elbaz}, {Bournaud}, {Finoguenov}, {Marchal},
  {Delvecchio}, {Jin}, {Liu}, {Strazzullo}, {Calabro}, {Coogan}, {D'Eugenio},
  {Gobat}, {Kalita}, {Laursen}, {Martin}, {Puglisi}, {Schinnerer}, \&
  {Wang}}]{Daddi2021}
{Daddi}, E., {Valentino}, F., {Rich}, R.~M., {et~al.} 2021, \aap, 649, A78,
  \dodoi{10.1051/0004-6361/202038700}

\bibitem[{{Darvish} {et~al.}(2015){Darvish}, {Mobasher}, {Sobral}, {Scoville},
  \& {Aragon-Calvo}}]{Darvish2015}
{Darvish}, B., {Mobasher}, B., {Sobral}, D., {Scoville}, N., \& {Aragon-Calvo},
  M. 2015, \apj, 805, 121, \dodoi{10.1088/0004-637X/805/2/121}

\bibitem[{{Dekel} {et~al.}(2009){Dekel}, {Birnboim}, {Engel}, {Freundlich},
  {Goerdt}, {Mumcuoglu}, {Neistein}, {Pichon}, {Teyssier}, \&
  {Zinger}}]{Dekel2009}
{Dekel}, A., {Birnboim}, Y., {Engel}, G., {et~al.} 2009, \nat, 457, 451,
  \dodoi{10.1038/nature07648}

\bibitem[{{Dey} {et~al.}(2016){Dey}, {Lee}, {Reddy}, {Cooper}, {Inami}, {Hong},
  {Gonzalez}, \& {Jannuzi}}]{Dey2016}
{Dey}, A., {Lee}, K.-S., {Reddy}, N., {et~al.} 2016, \apj, 823, 11,
  \dodoi{10.3847/0004-637X/823/1/11}

\bibitem[{{Dijkstra} \& {Westra}(2010)}]{Dijkstra2010}
{Dijkstra}, M., \& {Westra}, E. 2010, \mnras, 401, 2343,
  \dodoi{10.1111/j.1365-2966.2009.15859.x}

\bibitem[{{Eisenhardt} {et~al.}(2008){Eisenhardt}, {Brodwin}, {Gonzalez},
  {Stanford}, {Stern}, {Barmby}, {Brown}, {Dawson}, {Dey}, {Doi}, {Galametz},
  {Jannuzi}, {Kochanek}, {Meyers}, {Morokuma}, \& {Moustakas}}]{Eisenhardt2008}
{Eisenhardt}, P. R.~M., {Brodwin}, M., {Gonzalez}, A.~H., {et~al.} 2008, \apj,
  684, 905, \dodoi{10.1086/590105}

\bibitem[{{Firestone} {et~al.}(2023){Firestone}, {Gawiser}, {Ramakrishnan},
  {Lee}, {Valdes}, {Park}, {Yang}, {Ciardullo}, {Artale}, {Benda}, {Broussard},
  {Eid}, {Farooq}, {Gronwall}, {Guaita}, {Gwyn}, {Hwang}, {Hyeok Im}, {Jeong},
  {Karthikeyan}, {Lang}, {Moon}, {Padilla}, {Sawicki}, {Seo}, {Singh}, {Song},
  \& {Troncoso Iribarren}}]{Firestone2023}
{Firestone}, N.~M., {Gawiser}, E., {Ramakrishnan}, V., {et~al.} 2023, arXiv
  e-prints, arXiv:2312.16075, \dodoi{10.48550/arXiv.2312.16075}

\bibitem[{{Gawiser} {et~al.}(2007){Gawiser}, {Francke}, {Lai}, {Schawinski},
  {Gronwall}, {Ciardullo}, {Quadri}, {Orsi}, {Barrientos}, {Blanc}, {Fazio},
  {Feldmeier}, {Huang}, {Infante}, {Lira}, {Padilla}, {Taylor}, {Treister},
  {Urry}, {van Dokkum}, \& {Virani}}]{Gawiser2007}
{Gawiser}, E., {Francke}, H., {Lai}, K., {et~al.} 2007, \apj, 671, 278,
  \dodoi{10.1086/522955}

\bibitem[{{Gebhardt} {et~al.}(2021){Gebhardt}, {Mentuch Cooper}, {Ciardullo},
  {Acquaviva}, {Bender}, {Bowman}, {Castanheira}, {Dalton}, {Davis}, {de Jong},
  {DePoy}, {Devarakonda}, {Dongsheng}, {Drory}, {Fabricius}, {Farrow},
  {Feldmeier}, {Finkelstein}, {Froning}, {Gawiser}, {Gronwall}, {Herold},
  {Hill}, {Hopp}, {House}, {Janowiecki}, {Jarvis}, {Jeong}, {Jogee}, {Kakuma},
  {Kelz}, {Kollatschny}, {Komatsu}, {Krumpe}, {Landriau}, {Liu}, {Niemeyer},
  {MacQueen}, {Marshall}, {Mawatari}, {McLinden}, {Mukae}, {Nagaraj}, {Ono},
  {Ouchi}, {Papovich}, {Sakai}, {Saito}, {Schneider}, {Schulze},
  {Shanmugasundararaj}, {Shetrone}, {Sneden}, {Snigula}, {Steinmetz}, {Thomas},
  {Thomas}, {Tuttle}, {Urrutia}, {Wisotzki}, {Wold}, {Zeimann}, \&
  {Zhang}}]{gebhardt21}
{Gebhardt}, K., {Mentuch Cooper}, E., {Ciardullo}, R., {et~al.} 2021, \apj,
  923, 217, \dodoi{10.3847/1538-4357/ac2e03}

\bibitem[{{Gonzalez} {et~al.}(2019){Gonzalez}, {Gettings}, {Brodwin},
  {Eisenhardt}, {Stanford}, {Wylezalek}, {Decker}, {Marrone}, {Moravec},
  {O'Donnell}, {Stalder}, {Stern}, {Abdulla}, {Brown}, {Carlstrom}, {Chambers},
  {Hayden}, {Lin}, {Magnier}, {Masci}, {Mantz}, {McDonald}, {Mo}, {Perlmutter},
  {Wright}, \& {Zeimann}}]{Gonzalez2019}
{Gonzalez}, A.~H., {Gettings}, D.~P., {Brodwin}, M., {et~al.} 2019, \apjs, 240,
  33, \dodoi{10.3847/1538-4365/aafad2}

\bibitem[{{Gronwall} {et~al.}(2007){Gronwall}, {Ciardullo}, {Hickey},
  {Gawiser}, {Feldmeier}, {van Dokkum}, {Urry}, {Herrera}, {Lehmer}, {Infante},
  {Orsi}, {Marchesini}, {Blanc}, {Francke}, {Lira}, \&
  {Treister}}]{Gronwall2007}
{Gronwall}, C., {Ciardullo}, R., {Hickey}, T., {et~al.} 2007, \apj, 667, 79,
  \dodoi{10.1086/520324}

\bibitem[{{Guaita} {et~al.}(2010){Guaita}, {Gawiser}, {Padilla}, {Francke},
  {Bond}, {Gronwall}, {Ciardullo}, {Feldmeier}, {Sinawa}, {Blanc}, \&
  {Virani}}]{Guaita2010}
{Guaita}, L., {Gawiser}, E., {Padilla}, N., {et~al.} 2010, \apj, 714, 255,
  \dodoi{10.1088/0004-637X/714/1/255}

\bibitem[{{Hagen} {et~al.}(2014){Hagen}, {Ciardullo}, {Gronwall}, {Acquaviva},
  {Bridge}, {Zeimann}, {Blanc}, {Bond}, {Finkelstein}, {Song}, {Gawiser},
  {Fox}, {Gebhardt}, {Malz}, {Schneider}, {Drory}, {Gebhardt}, \&
  {Hill}}]{Hagen2014}
{Hagen}, A., {Ciardullo}, R., {Gronwall}, C., {et~al.} 2014, \apj, 786, 59,
  \dodoi{10.1088/0004-637X/786/1/59}

\bibitem[{{Harikane} {et~al.}(2019){Harikane}, {Ouchi}, {Ono}, {Fujimoto},
  {Donevski}, {Shibuya}, {Faisst}, {Goto}, {Hatsukade}, {Kashikawa}, {Kohno},
  {Hashimoto}, {Higuchi}, {Inoue}, {Lin}, {Martin}, {Overzier}, {Smail},
  {Toshikawa}, {Umehata}, {Ao}, {Chapman}, {Clements}, {Im}, {Jing},
  {Kawaguchi}, {Lee}, {Lee}, {Lin}, {Matsuoka}, {Marinello}, {Nagao},
  {Onodera}, {Toft}, \& {Wang}}]{Harikane2019}
{Harikane}, Y., {Ouchi}, M., {Ono}, Y., {et~al.} 2019, \apj, 883, 142,
  \dodoi{10.3847/1538-4357/ab2cd5}

\bibitem[{{Hayashi} {et~al.}(2018){Hayashi}, {Tanaka}, {Shimakawa}, {Furusawa},
  {Momose}, {Koyama}, {Silverman}, {Kodama}, {Komiyama}, {Leauthaud}, {Lin},
  {Miyazaki}, {Nagao}, {Nishizawa}, {Ouchi}, {Shibuya}, {Tadaki}, \&
  {Yabe}}]{Hayashi2018}
{Hayashi}, M., {Tanaka}, M., {Shimakawa}, R., {et~al.} 2018, \pasj, 70, S17,
  \dodoi{10.1093/pasj/psx088}

\bibitem[{{Hayashino} {et~al.}(2004){Hayashino}, {Matsuda}, {Tamura},
  {Yamauchi}, {Yamada}, {Ajiki}, {Fujita}, {Murayama}, {Nagao}, {Ohta},
  {Okamura}, {Ouchi}, {Shimasaku}, {Shioya}, \& {Taniguchi}}]{Hayashino2004}
{Hayashino}, T., {Matsuda}, Y., {Tamura}, H., {et~al.} 2004, \aj, 128, 2073,
  \dodoi{10.1086/424935}

\bibitem[{{Hopkins} {et~al.}(2006){Hopkins}, {Hernquist}, {Cox}, {Di Matteo},
  {Robertson}, \& {Springel}}]{hopkins06}
{Hopkins}, P.~F., {Hernquist}, L., {Cox}, T.~J., {et~al.} 2006, \apjs, 163, 1,
  \dodoi{10.1086/499298}

\bibitem[{{Huang} {et~al.}(2022){Huang}, {Lee}, {Cucciati}, {Lemaux},
  {Sawicki}, {Malavasi}, {Ramakrishnan}, {Xue}, {Cassara}, {Chiang}, {Dey},
  {Gwyn}, {Hathi}, {Pentericci}, {Prescott}, \& {Zamorani}}]{huang22}
{Huang}, Y., {Lee}, K.-S., {Cucciati}, O., {et~al.} 2022, \apj, 941, 134,
  \dodoi{10.3847/1538-4357/ac9ea4}

\bibitem[{{Hung} {et~al.}(2020){Hung}, {Lemaux}, {Gal}, {Tomczak}, {Lubin},
  {Cucciati}, {Pelliccia}, {Shen}, {Le F{\`e}vre}, {Wu}, {Kocevski}, {Mei}, \&
  {Squires}}]{Hung2020}
{Hung}, D., {Lemaux}, B.~C., {Gal}, R.~R., {et~al.} 2020, \mnras, 491, 5524,
  \dodoi{10.1093/mnras/stz3164}

\bibitem[{{Im} {et~al.}(2024){Im}, {Hwang}, {Park}, {Lee}, {Song}, {Appleby},
  {Dubois}, {Few}, {Gibson}, {Kim}, {Kim}, {Park}, {Pichon}, {Shin}, {Snaith},
  {Artale}, {Gawiser}, {Guaita}, {Jeong}, {Lee}, {Padilla}, {Ramakrishnan},
  {Troncoso}, \& {Yang}}]{Im2024}
{Im}, S.~H., {Hwang}, H.~S., {Park}, J., {et~al.} 2024, arXiv e-prints,
  arXiv:2407.18602, \dodoi{10.48550/arXiv.2407.18602}

\bibitem[{{Ito} {et~al.}(2023){Ito}, {Tanaka}, {Valentino}, {Toft}, {Brammer},
  {Gould}, {Ilbert}, {Kashikawa}, {Kubo}, {Liang}, {McCracken}, \&
  {Weaver}}]{Ito2023}
{Ito}, K., {Tanaka}, M., {Valentino}, F., {et~al.} 2023, \apjl, 945, L9,
  \dodoi{10.3847/2041-8213/acb49b}

\bibitem[{{Kikuta} {et~al.}(2019){Kikuta}, {Matsuda}, {Cen}, {Steidel}, {Yagi},
  {Hayashino}, {Imanishi}, {Komiyama}, {Momose}, \& {Saito}}]{Kikuta2019}
{Kikuta}, S., {Matsuda}, Y., {Cen}, R., {et~al.} 2019, \pasj, 71, L2,
  \dodoi{10.1093/pasj/psz055}

\bibitem[{{Koyama} {et~al.}(2014){Koyama}, {Kodama}, {Tadaki}, {Hayashi},
  {Tanaka}, \& {Shimakawa}}]{Koyama2014}
{Koyama}, Y., {Kodama}, T., {Tadaki}, K.-i., {et~al.} 2014, \apj, 789, 18,
  \dodoi{10.1088/0004-637X/789/1/18}

\bibitem[{Kraljic {et~al.}(2017)Kraljic, Arnouts, Pichon, Laigle, de~la Torre,
  Vibert, Cadiou, Dubois, Treyer, Schimd, Codis, de~Lapparent, Devriendt,
  Hwang, Borgne, Malavasi, Milliard, Musso, Pogosyan, Alpaslan, Bland-Hawthorn,
  \& Wright}]{Kraljic2017}
Kraljic, K., Arnouts, S., Pichon, C., {et~al.} 2017, Monthly Notices of the
  Royal Astronomical Society, 474, 547, \dodoi{10.1093/mnras/stx2638}

\bibitem[{Kriegel {et~al.}(2011)Kriegel, Kr{\"o}ger, Sander, \&
  Zimek}]{Kriegel2011}
Kriegel, H.-P., Kr{\"o}ger, P., Sander, J., \& Zimek, A. 2011, Wiley
  interdisciplinary reviews: data mining and knowledge discovery, 1, 231

\bibitem[{{Kuchner} {et~al.}(2022){Kuchner}, {Haggar}, {Arag{\'o}n-Salamanca},
  {Pearce}, {Gray}, {Rost}, {Cui}, {Knebe}, \& {Yepes}}]{Kuchner2022}
{Kuchner}, U., {Haggar}, R., {Arag{\'o}n-Salamanca}, A., {et~al.} 2022, \mnras,
  510, 581, \dodoi{10.1093/mnras/stab3419}

\bibitem[{{Kusakabe} {et~al.}(2018){Kusakabe}, {Shimasaku}, {Ouchi},
  {Nakajima}, {Goto}, {Hashimoto}, {Konno}, {Harikane}, {Silverman}, \&
  {Capak}}]{Kusakabe2018}
{Kusakabe}, H., {Shimasaku}, K., {Ouchi}, M., {et~al.} 2018, \pasj, 70, 4,
  \dodoi{10.1093/pasj/psx148}

\bibitem[{{Laigle} {et~al.}(2018){Laigle}, {Pichon}, {Arnouts}, {McCracken},
  {Dubois}, {Devriendt}, {Slyz}, {Le Borgne}, {Benoit-L{\'e}vy}, {Hwang},
  {Ilbert}, {Kraljic}, {Malavasi}, {Park}, \& {Vibert}}]{Laigle2018}
{Laigle}, C., {Pichon}, C., {Arnouts}, S., {et~al.} 2018, \mnras, 474, 5437,
  \dodoi{10.1093/mnras/stx3055}

\bibitem[{{Lee} {et~al.}(2014){Lee}, {Dey}, {Hong}, {Reddy}, {Wilson},
  {Jannuzi}, {Inami}, \& {Gonzalez}}]{Lee2014}
{Lee}, K.-S., {Dey}, A., {Hong}, S., {et~al.} 2014, \apj, 796, 126,
  \dodoi{10.1088/0004-637X/796/2/126}

\bibitem[{{Lee} {et~al.}(2024){Lee}, {Gawiser}, {Park}, {Yang}, {Valdes},
  {Lang}, {Ramakrishnan}, {Moon}, {Firestone}, {Appleby}, {Artale}, {Andrews},
  {Bauer}, {Benda}, {Broussard}, {Chiang}, {Ciardullo}, {Dey}, {Farooq},
  {Gronwall}, {Guaita}, {Huang}, {Hwang}, {Im}, {Jeong}, {Karthikeyan}, {Kim},
  {Kim}, {Kumar}, {Nagaraj}, {Nantais}, {Padilla}, {Park}, {Pope}, {Popescu},
  {Schlegel}, {Seo}, {Singh}, {Song}, {Troncoso}, {Vivas}, {Zabludoff}, \&
  {Zenteno}}]{Lee2024}
{Lee}, K.-S., {Gawiser}, E., {Park}, C., {et~al.} 2024, \apj, 962, 36,
  \dodoi{10.3847/1538-4357/ad165e}

\bibitem[{{Lemaux} {et~al.}(2018){Lemaux}, {Le F{\`e}vre}, {Cucciati},
  {Ribeiro}, {Tasca}, {Zamorani}, {Ilbert}, {Thomas}, {Bardelli}, {Cassata},
  {Hathi}, {Pforr}, {Smol{\v{c}}i{\'c}}, {Delvecchio}, {Novak}, {Berta},
  {McCracken}, {Koekemoer}, {Amor{\'\i}n}, {Garilli}, {Maccagni}, {Schaerer},
  \& {Zucca}}]{Lemaux2018}
{Lemaux}, B.~C., {Le F{\`e}vre}, O., {Cucciati}, O., {et~al.} 2018, \aap, 615,
  A77, \dodoi{10.1051/0004-6361/201730870}

\bibitem[{{Lemaux} {et~al.}(2022){Lemaux}, {Cucciati}, {Le F{\`e}vre},
  {Zamorani}, {Lubin}, {Hathi}, {Ilbert}, {Pelliccia}, {Amor{\'\i}n},
  {Bardelli}, {Cassata}, {Gal}, {Garilli}, {Guaita}, {Giavalisco}, {Hung},
  {Koekemoer}, {Maccagni}, {Pentericci}, {Ribeiro}, {Schaerer}, {Shah}, {Shen},
  {Staab}, {Talia}, {Thomas}, {Tomczak}, {Tresse}, {Vanzella}, {Vergani}, \&
  {Zucca}}]{Lemaux2022}
{Lemaux}, B.~C., {Cucciati}, O., {Le F{\`e}vre}, O., {et~al.} 2022, \aap, 662,
  A33, \dodoi{10.1051/0004-6361/202039346}

\bibitem[{{Madau} \& {Dickinson}(2014)}]{Madau2014}
{Madau}, P., \& {Dickinson}, M. 2014, \araa, 52, 415,
  \dodoi{10.1146/annurev-astro-081811-125615}

\bibitem[{{Malavasi} {et~al.}(2020){Malavasi}, {Aghanim}, {Douspis},
  {Tanimura}, \& {Bonjean}}]{Malavasi2020}
{Malavasi}, N., {Aghanim}, N., {Douspis}, M., {Tanimura}, H., \& {Bonjean}, V.
  2020, \aap, 642, A19, \dodoi{10.1051/0004-6361/202037647}

\bibitem[{{Malavasi} {et~al.}(2021){Malavasi}, {Lee}, {Dey}, {Xue}, {Huang}, \&
  {Shi}}]{Malavasi2021}
{Malavasi}, N., {Lee}, K.-S., {Dey}, A., {et~al.} 2021, \apj, 921, 103,
  \dodoi{10.3847/1538-4357/ac1c6e}

\bibitem[{{Malavasi} {et~al.}(2017){Malavasi}, {Arnouts}, {Vibert}, {de la
  Torre}, {Moutard}, {Pichon}, {Davidzon}, {Kraljic}, {Bolzonella}, {Guzzo},
  {Garilli}, {Scodeggio}, {Granett}, {Abbas}, {Adami}, {Bottini}, {Cappi},
  {Cucciati}, {Franzetti}, {Fritz}, {Iovino}, {Krywult}, {Le Brun}, {Le
  F{\`e}vre}, {Maccagni}, {Ma{\l}ek}, {Marulli}, {Polletta}, {Pollo}, {Tasca},
  {Tojeiro}, {Vergani}, {Zanichelli}, {Bel}, {Branchini}, {Coupon}, {De Lucia},
  {Dubois}, {Hawken}, {Ilbert}, {Laigle}, {Moscardini}, {Sousbie}, {Treyer}, \&
  {Zamorani}}]{Malavasi2017}
{Malavasi}, N., {Arnouts}, S., {Vibert}, D., {et~al.} 2017, \mnras, 465, 3817,
  \dodoi{10.1093/mnras/stw2864}

\bibitem[{{Matsuda} {et~al.}(2004){Matsuda}, {Yamada}, {Hayashino}, {Tamura},
  {Yamauchi}, {Ajiki}, {Fujita}, {Murayama}, {Nagao}, {Ohta}, {Okamura},
  {Ouchi}, {Shimasaku}, {Shioya}, \& {Taniguchi}}]{Matsuda2004}
{Matsuda}, Y., {Yamada}, T., {Hayashino}, T., {et~al.} 2004, \aj, 128, 569,
  \dodoi{10.1086/422020}

\bibitem[{{Matsuda} {et~al.}(2009){Matsuda}, {Nakamura}, {Morimoto}, {Smail},
  {De Breuck}, {Ohta}, {Kodama}, {Inoue}, {Hayashino}, {Kousai}, {Nakamura},
  {Horie}, {Yamada}, {Kitamura}, {Saito}, {Taniguchi}, {Tanaka}, \&
  {Hibon}}]{Matsuda2009}
{Matsuda}, Y., {Nakamura}, Y., {Morimoto}, N., {et~al.} 2009, \mnras, 400, L66,
  \dodoi{10.1111/j.1745-3933.2009.00764.x}

\bibitem[{{Matthee} {et~al.}(2017){Matthee}, {Sobral}, {Best}, {Smail}, {Bian},
  {Darvish}, {R{\"o}ttgering}, \& {Fan}}]{Matthee2017}
{Matthee}, J., {Sobral}, D., {Best}, P., {et~al.} 2017, \mnras, 471, 629,
  \dodoi{10.1093/mnras/stx1569}

\bibitem[{McInnes {et~al.}(2017)McInnes, Healy, \& Astels}]{McInnes2017}
McInnes, L., Healy, J., \& Astels, S. 2017, The Journal of Open Source
  Software, 2, \dodoi{10.21105/joss.00205}

\bibitem[{{Momose} {et~al.}(2021){Momose}, {Shimasaku}, {Kashikawa},
  {Nagamine}, {Shimizu}, {Nakajima}, {Terao}, {Kusakabe}, {Ando}, {Motohara},
  \& {Spitler}}]{Momose2021}
{Momose}, R., {Shimasaku}, K., {Kashikawa}, N., {et~al.} 2021, \apj, 909, 117,
  \dodoi{10.3847/1538-4357/abd2af}

\bibitem[{{Nelson} {et~al.}(2019){Nelson}, {Springel}, {Pillepich},
  {Rodriguez-Gomez}, {Torrey}, {Genel}, {Vogelsberger}, {Pakmor}, {Marinacci},
  {Weinberger}, {Kelley}, {Lovell}, {Diemer}, \& {Hernquist}}]{Nelson2019}
{Nelson}, D., {Springel}, V., {Pillepich}, A., {et~al.} 2019, Computational
  Astrophysics and Cosmology, 6, 2, \dodoi{10.1186/s40668-019-0028-x}

\bibitem[{{Oteo} {et~al.}(2018){Oteo}, {Ivison}, {Dunne}, {Manilla-Robles},
  {Maddox}, {Lewis}, {de Zotti}, {Bremer}, {Clements}, {Cooray}, {Dannerbauer},
  {Eales}, {Greenslade}, {Omont}, {Perez{\textendash}Fourn{\'o}n}, {Riechers},
  {Scott}, {van der Werf}, {Weiss}, \& {Zhang}}]{Oteo2018}
{Oteo}, I., {Ivison}, R.~J., {Dunne}, L., {et~al.} 2018, \apj, 856, 72,
  \dodoi{10.3847/1538-4357/aaa1f1}

\bibitem[{{Ouchi} {et~al.}(2008){Ouchi}, {Shimasaku}, {Akiyama}, {Simpson},
  {Saito}, {Ueda}, {Furusawa}, {Sekiguchi}, {Yamada}, {Kodama}, {Kashikawa},
  {Okamura}, {Iye}, {Takata}, {Yoshida}, \& {Yoshida}}]{Ouchi2008}
{Ouchi}, M., {Shimasaku}, K., {Akiyama}, M., {et~al.} 2008, \apjs, 176, 301,
  \dodoi{10.1086/527673}

\bibitem[{{Overzier}(2016)}]{Overzier2016}
{Overzier}, R.~A. 2016, \aapr, 24, 14, \dodoi{10.1007/s00159-016-0100-3}

\bibitem[{{Peng} {et~al.}(2010){Peng}, {Lilly}, {Kova{\v{c}}}, {Bolzonella},
  {Pozzetti}, {Renzini}, {Zamorani}, {Ilbert}, {Knobel}, {Iovino}, {Maier},
  {Cucciati}, {Tasca}, {Carollo}, {Silverman}, {Kampczyk}, {de Ravel},
  {Sanders}, {Scoville}, {Contini}, {Mainieri}, {Scodeggio}, {Kneib}, {Le
  F{\`e}vre}, {Bardelli}, {Bongiorno}, {Caputi}, {Coppa}, {de la Torre},
  {Franzetti}, {Garilli}, {Lamareille}, {Le Borgne}, {Le Brun}, {Mignoli},
  {Perez Montero}, {Pello}, {Ricciardelli}, {Tanaka}, {Tresse}, {Vergani},
  {Welikala}, {Zucca}, {Oesch}, {Abbas}, {Barnes}, {Bordoloi}, {Bottini},
  {Cappi}, {Cassata}, {Cimatti}, {Fumana}, {Hasinger}, {Koekemoer},
  {Leauthaud}, {Maccagni}, {Marinoni}, {McCracken}, {Memeo}, {Meneux}, {Nair},
  {Porciani}, {Presotto}, \& {Scaramella}}]{Peng2010}
{Peng}, Y.-j., {Lilly}, S.~J., {Kova{\v{c}}}, K., {et~al.} 2010, \apj, 721,
  193, \dodoi{10.1088/0004-637X/721/1/193}

\bibitem[{{Pillepich} {et~al.}(2018{\natexlab{a}}){Pillepich}, {Springel},
  {Nelson}, {Genel}, {Naiman}, {Pakmor}, {Hernquist}, {Torrey}, {Vogelsberger},
  {Weinberger}, \& {Marinacci}}]{Pillepich2018a}
{Pillepich}, A., {Springel}, V., {Nelson}, D., {et~al.} 2018{\natexlab{a}},
  \mnras, 473, 4077, \dodoi{10.1093/mnras/stx2656}

\bibitem[{{Pillepich} {et~al.}(2018{\natexlab{b}}){Pillepich}, {Nelson},
  {Hernquist}, {Springel}, {Pakmor}, {Torrey}, {Weinberger}, {Genel}, {Naiman},
  {Marinacci}, \& {Vogelsberger}}]{Pillepich2018b}
{Pillepich}, A., {Nelson}, D., {Hernquist}, L., {et~al.} 2018{\natexlab{b}},
  \mnras, 475, 648, \dodoi{10.1093/mnras/stx3112}

\bibitem[{{Planck Collaboration} {et~al.}(2016){Planck Collaboration}, {Ade},
  {Aghanim}, {Arnaud}, {Ashdown}, {Aumont}, {Baccigalupi}, {Banday},
  {Barreiro}, {Bartlett}, {Bartolo}, {Battaner}, {Battye}, {Benabed},
  {Beno{\^\i}t}, {Benoit-L{\'e}vy}, {Bernard}, {Bersanelli}, {Bielewicz},
  {Bock}, {Bonaldi}, {Bonavera}, {Bond}, {Borrill}, {Bouchet}, {Boulanger},
  {Bucher}, {Burigana}, {Butler}, {Calabrese}, {Cardoso}, {Catalano},
  {Challinor}, {Chamballu}, {Chary}, {Chiang}, {Chluba}, {Christensen},
  {Church}, {Clements}, {Colombi}, {Colombo}, {Combet}, {Coulais}, {Crill},
  {Curto}, {Cuttaia}, {Danese}, {Davies}, {Davis}, {de Bernardis}, {de Rosa},
  {de Zotti}, {Delabrouille}, {D{\'e}sert}, {Di Valentino}, {Dickinson},
  {Diego}, {Dolag}, {Dole}, {Donzelli}, {Dor{\'e}}, {Douspis}, {Ducout},
  {Dunkley}, {Dupac}, {Efstathiou}, {Elsner}, {En{\ss}lin}, {Eriksen},
  {Farhang}, {Fergusson}, {Finelli}, {Forni}, {Frailis}, {Fraisse},
  {Franceschi}, {Frejsel}, {Galeotta}, {Galli}, {Ganga}, {Gauthier}, {Gerbino},
  {Ghosh}, {Giard}, {Giraud-H{\'e}raud}, {Giusarma}, {Gjerl{\o}w},
  {Gonz{\'a}lez-Nuevo}, {G{\'o}rski}, {Gratton}, {Gregorio}, {Gruppuso},
  {Gudmundsson}, {Hamann}, {Hansen}, {Hanson}, {Harrison}, {Helou},
  {Henrot-Versill{\'e}}, {Hern{\'a}ndez-Monteagudo}, {Herranz}, {Hildebrandt},
  {Hivon}, {Hobson}, {Holmes}, {Hornstrup}, {Hovest}, {Huang}, {Huffenberger},
  {Hurier}, {Jaffe}, {Jaffe}, {Jones}, {Juvela}, {Keih{\"a}nen}, {Keskitalo},
  {Kisner}, {Kneissl}, {Knoche}, {Knox}, {Kunz}, {Kurki-Suonio}, {Lagache},
  {L{\"a}hteenm{\"a}ki}, {Lamarre}, {Lasenby}, {Lattanzi}, {Lawrence}, {Leahy},
  {Leonardi}, {Lesgourgues}, {Levrier}, {Lewis}, {Liguori}, {Lilje},
  {Linden-V{\o}rnle}, {L{\'o}pez-Caniego}, {Lubin}, {Mac{\'\i}as-P{\'e}rez},
  {Maggio}, {Maino}, {Mandolesi}, {Mangilli}, {Marchini}, {Maris}, {Martin},
  {Martinelli}, {Mart{\'\i}nez-Gonz{\'a}lez}, {Masi}, {Matarrese}, {McGehee},
  {Meinhold}, {Melchiorri}, {Melin}, {Mendes}, {Mennella}, {Migliaccio},
  {Millea}, {Mitra}, {Miville-Desch{\^e}nes}, {Moneti}, {Montier}, {Morgante},
  {Mortlock}, {Moss}, {Munshi}, {Murphy}, {Naselsky}, {Nati}, {Natoli},
  {Netterfield}, {N{\o}rgaard-Nielsen}, {Noviello}, {Novikov}, {Novikov},
  {Oxborrow}, {Paci}, {Pagano}, {Pajot}, {Paladini}, {Paoletti}, {Partridge},
  {Pasian}, {Patanchon}, {Pearson}, {Perdereau}, {Perotto}, {Perrotta},
  {Pettorino}, {Piacentini}, {Piat}, {Pierpaoli}, {Pietrobon}, {Plaszczynski},
  {Pointecouteau}, {Polenta}, {Popa}, {Pratt}, {Pr{\'e}zeau}, {Prunet},
  {Puget}, {Rachen}, {Reach}, {Rebolo}, {Reinecke}, {Remazeilles}, {Renault},
  {Renzi}, {Ristorcelli}, {Rocha}, {Rosset}, {Rossetti}, {Roudier},
  {Rouill{\'e} d'Orfeuil}, {Rowan-Robinson}, {Rubi{\~n}o-Mart{\'\i}n},
  {Rusholme}, {Said}, {Salvatelli}, {Salvati}, {Sandri}, {Santos},
  {Savelainen}, {Savini}, {Scott}, {Seiffert}, {Serra}, {Shellard}, {Spencer},
  {Spinelli}, {Stolyarov}, {Stompor}, {Sudiwala}, {Sunyaev}, {Sutton},
  {Suur-Uski}, {Sygnet}, {Tauber}, {Terenzi}, {Toffolatti}, {Tomasi},
  {Tristram}, {Trombetti}, {Tucci}, {Tuovinen}, {T{\"u}rler}, {Umana},
  {Valenziano}, {Valiviita}, {Van Tent}, {Vielva}, {Villa}, {Wade}, {Wandelt},
  {Wehus}, {White}, {White}, {Wilkinson}, {Yvon}, {Zacchei}, \&
  {Zonca}}]{Planck2016_cosmology}
{Planck Collaboration}, {Ade}, P.~A.~R., {Aghanim}, N., {et~al.} 2016, \aap,
  594, A13, \dodoi{10.1051/0004-6361/201525830}

\bibitem[{{Quadri} {et~al.}(2012){Quadri}, {Williams}, {Franx}, \&
  {Hildebrandt}}]{Quadri2012}
{Quadri}, R.~F., {Williams}, R.~J., {Franx}, M., \& {Hildebrandt}, H. 2012,
  \apj, 744, 88, \dodoi{10.1088/0004-637X/744/2/88}

\bibitem[{Ramakrishnan {et~al.}(2023)Ramakrishnan, Moon, Im, Farooq, Lee,
  Gawiser, Yang, Park, Hwang, Valdes, Artale, Ciardullo, Dey, Gronwall, Guaita,
  Jeong, Padilla, Singh, \& Zabludoff}]{Ramakrishnan2023}
Ramakrishnan, V., Moon, B., Im, S.~H., {et~al.} 2023, The Astrophysical
  Journal, 951, 119, \dodoi{10.3847/1538-4357/acd341}

\bibitem[{{Ravi} {et~al.}(2024){Ravi}, {Hadzhiyska}, {White}, {Hernquist}, \&
  {Bose}}]{Ravi2024}
{Ravi}, J., {Hadzhiyska}, B., {White}, M., {Hernquist}, L., \& {Bose}, S. 2024,
  arXiv e-prints, arXiv:2403.02414, \dodoi{10.48550/arXiv.2403.02414}

\bibitem[{{Reddy} \& {Steidel}(2009)}]{reddy09}
{Reddy}, N.~A., \& {Steidel}, C.~C. 2009, \apj, 692, 778,
  \dodoi{10.1088/0004-637X/692/1/778}

\bibitem[{{Rykoff} {et~al.}(2014){Rykoff}, {Rozo}, {Busha}, {Cunha},
  {Finoguenov}, {Evrard}, {Hao}, {Koester}, {Leauthaud}, {Nord}, {Pierre},
  {Reddick}, {Sadibekova}, {Sheldon}, \& {Wechsler}}]{Rykoff2014}
{Rykoff}, E.~S., {Rozo}, E., {Busha}, M.~T., {et~al.} 2014, \apj, 785, 104,
  \dodoi{10.1088/0004-637X/785/2/104}

\bibitem[{{Rykoff} {et~al.}(2016){Rykoff}, {Rozo}, {Hollowood},
  {Bermeo-Hernandez}, {Jeltema}, {Mayers}, {Romer}, {Rooney}, {Saro}, {Vergara
  Cervantes}, {Wechsler}, {Wilcox}, {Abbott}, {Abdalla}, {Allam}, {Annis},
  {Benoit-L{\'e}vy}, {Bernstein}, {Bertin}, {Brooks}, {Burke}, {Capozzi},
  {Carnero Rosell}, {Carrasco Kind}, {Castander}, {Childress}, {Collins},
  {Cunha}, {D'Andrea}, {da Costa}, {Davis}, {Desai}, {Diehl}, {Dietrich},
  {Doel}, {Evrard}, {Finley}, {Flaugher}, {Fosalba}, {Frieman}, {Glazebrook},
  {Goldstein}, {Gruen}, {Gruendl}, {Gutierrez}, {Hilton}, {Honscheid}, {Hoyle},
  {James}, {Kay}, {Kuehn}, {Kuropatkin}, {Lahav}, {Lewis}, {Lidman}, {Lima},
  {Maia}, {Mann}, {Marshall}, {Martini}, {Melchior}, {Miller}, {Miquel},
  {Mohr}, {Nichol}, {Nord}, {Ogando}, {Plazas}, {Reil}, {Sahl{\'e}n},
  {Sanchez}, {Santiago}, {Scarpine}, {Schubnell}, {Sevilla-Noarbe}, {Smith},
  {Soares-Santos}, {Sobreira}, {Stott}, {Suchyta}, {Swanson}, {Tarle},
  {Thomas}, {Tucker}, {Uddin}, {Viana}, {Vikram}, {Walker}, {Zhang}, \& {DES
  Collaboration}}]{Rykoff2016}
{Rykoff}, E.~S., {Rozo}, E., {Hollowood}, D., {et~al.} 2016, \apjs, 224, 1,
  \dodoi{10.3847/0067-0049/224/1/1}

\bibitem[{{Salerno} {et~al.}(2019){Salerno}, {Mart{\'\i}nez}, \&
  {Muriel}}]{Salerno2019}
{Salerno}, J.~M., {Mart{\'\i}nez}, H.~J., \& {Muriel}, H. 2019, \mnras, 484, 2,
  \dodoi{10.1093/mnras/sty3456}

\bibitem[{{Sarron} {et~al.}(2019){Sarron}, {Adami}, {Durret}, \&
  {Laigle}}]{Sarron2019}
{Sarron}, F., {Adami}, C., {Durret}, F., \& {Laigle}, C. 2019, \aap, 632, A49,
  \dodoi{10.1051/0004-6361/201935394}

\bibitem[{{Sarron} \& {Conselice}(2021)}]{Sarron2021}
{Sarron}, F., \& {Conselice}, C.~J. 2021, \mnras, 506, 2136,
  \dodoi{10.1093/mnras/stab1844}

\bibitem[{{Sawicki} {et~al.}(2019){Sawicki}, {Arnouts}, {Huang}, {Coupon},
  {Golob}, {Gwyn}, {Foucaud}, {Moutard}, {Iwata}, {Liu}, {Chen}, {Desprez},
  {Harikane}, {Ono}, {Strauss}, {Tanaka}, {Thibert}, {Balogh}, {Bundy},
  {Chapman}, {Gunn}, {Hsieh}, {Ilbert}, {Jing}, {LeF{\`e}vre}, {Li}, {Matsuda},
  {Miyazaki}, {Nagao}, {Nishizawa}, {Ouchi}, {Shimasaku}, {Silverman}, {de la
  Torre}, {Tresse}, {Wang}, {Willott}, {Yamada}, {Yang}, \&
  {Yee}}]{Sawicki2019}
{Sawicki}, M., {Arnouts}, S., {Huang}, J., {et~al.} 2019, \mnras, 489, 5202,
  \dodoi{10.1093/mnras/stz2522}

\bibitem[{{Schaap} \& {van de Weygaert}(2000)}]{Schaap2000}
{Schaap}, W.~E., \& {van de Weygaert}, R. 2000, \aap, 363, L29,
  \dodoi{10.48550/arXiv.astro-ph/0011007}

\bibitem[{{Shi} {et~al.}(2021){Shi}, {Toshikawa}, {Lee}, {Wang}, {Cai}, \&
  {Fang}}]{Shi2021}
{Shi}, K., {Toshikawa}, J., {Lee}, K.-S., {et~al.} 2021, \apj, 911, 46,
  \dodoi{10.3847/1538-4357/abe62e}

\bibitem[{{Shi} {et~al.}(2019){Shi}, {Huang}, {Lee}, {Toshikawa}, {Bowen},
  {Malavasi}, {Lemaux}, {Cucciati}, {Le Fevre}, \& {Dey}}]{Shi2019}
{Shi}, K., {Huang}, Y., {Lee}, K.-S., {et~al.} 2019, \apj, 879, 9,
  \dodoi{10.3847/1538-4357/ab2118}

\bibitem[{{Shimakawa} {et~al.}(2017){Shimakawa}, {Kodama}, {Hayashi}, {Tanaka},
  {Matsuda}, {Kashikawa}, {Shibuya}, {Tadaki}, {Koyama}, {Suzuki}, \&
  {Yamamoto}}]{Shimakawa2017}
{Shimakawa}, R., {Kodama}, T., {Hayashi}, M., {et~al.} 2017, \mnras, 468, L21,
  \dodoi{10.1093/mnrasl/slx019}

\bibitem[{Sousbie(2011)}]{Sousbie2011a}
Sousbie, T. 2011, Monthly Notices of the Royal Astronomical Society, 414, 350,
  \dodoi{10.1111/j.1365-2966.2011.18394.x}

\bibitem[{Sousbie {et~al.}(2011)Sousbie, Pichon, \& Kawahara}]{Sousbie2011b}
Sousbie, T., Pichon, C., \& Kawahara, H. 2011, Monthly Notices of the Royal
  Astronomical Society, 414, 384, \dodoi{10.1111/j.1365-2966.2011.18395.x}

\bibitem[{{Springel} {et~al.}(2005){Springel}, {White}, {Jenkins}, {Frenk},
  {Yoshida}, {Gao}, {Navarro}, {Thacker}, {Croton}, {Helly}, {Peacock}, {Cole},
  {Thomas}, {Couchman}, {Evrard}, {Colberg}, \& {Pearce}}]{Springel2005}
{Springel}, V., {White}, S. D.~M., {Jenkins}, A., {et~al.} 2005, \nat, 435,
  629, \dodoi{10.1038/nature03597}

\bibitem[{{Stark} {et~al.}(2010){Stark}, {Ellis}, {Chiu}, {Ouchi}, \&
  {Bunker}}]{stark10}
{Stark}, D.~P., {Ellis}, R.~S., {Chiu}, K., {Ouchi}, M., \& {Bunker}, A. 2010,
  \mnras, 408, 1628, \dodoi{10.1111/j.1365-2966.2010.17227.x}

\bibitem[{{Steidel} {et~al.}(1998){Steidel}, {Adelberger}, {Dickinson},
  {Giavalisco}, {Pettini}, \& {Kellogg}}]{Steidel1998}
{Steidel}, C.~C., {Adelberger}, K.~L., {Dickinson}, M., {et~al.} 1998, \apj,
  492, 428, \dodoi{10.1086/305073}

\bibitem[{{Steidel} {et~al.}(2000){Steidel}, {Adelberger}, {Shapley},
  {Pettini}, {Dickinson}, \& {Giavalisco}}]{Steidel2000}
{Steidel}, C.~C., {Adelberger}, K.~L., {Shapley}, A.~E., {et~al.} 2000, \apj,
  532, 170, \dodoi{10.1086/308568}

\bibitem[{{Tempel} {et~al.}(2014){Tempel}, {Stoica}, {Mart{\'\i}nez},
  {Liivam{\"a}gi}, {Castellan}, \& {Saar}}]{Tempel2014}
{Tempel}, E., {Stoica}, R.~S., {Mart{\'\i}nez}, V.~J., {et~al.} 2014, \mnras,
  438, 3465, \dodoi{10.1093/mnras/stt2454}

\bibitem[{{Toshikawa} {et~al.}(2016){Toshikawa}, {Kashikawa}, {Overzier},
  {Malkan}, {Furusawa}, {Ishikawa}, {Onoue}, {Ota}, {Tanaka}, {Niino}, \&
  {Uchiyama}}]{Toshikawa2016}
{Toshikawa}, J., {Kashikawa}, N., {Overzier}, R., {et~al.} 2016, \apj, 826,
  114, \dodoi{10.3847/0004-637X/826/2/114}

\bibitem[{{Toshikawa} {et~al.}(2018){Toshikawa}, {Uchiyama}, {Kashikawa},
  {Ouchi}, {Overzier}, {Ono}, {Harikane}, {Ishikawa}, {Kodama}, {Matsuda},
  {Lin}, {Onoue}, {Tanaka}, {Nagao}, {Akiyama}, {Komiyama}, {Goto}, \&
  {Lee}}]{Toshikawa2018}
{Toshikawa}, J., {Uchiyama}, H., {Kashikawa}, N., {et~al.} 2018, \pasj, 70,
  S12, \dodoi{10.1093/pasj/psx102}

\bibitem[{{Umehata} {et~al.}(2015){Umehata}, {Tamura}, {Kohno}, {Ivison},
  {Alexander}, {Geach}, {Hatsukade}, {Hughes}, {Ikarashi}, {Kato}, {Izumi},
  {Kawabe}, {Kubo}, {Lee}, {Lehmer}, {Makiya}, {Matsuda}, {Nakanishi}, {Saito},
  {Smail}, {Yamada}, {Yamaguchi}, \& {Yun}}]{Umehata2015}
{Umehata}, H., {Tamura}, Y., {Kohno}, K., {et~al.} 2015, \apjl, 815, L8,
  \dodoi{10.1088/2041-8205/815/1/L8}

\bibitem[{{van de Weygaert} \& {Schaap}(2009)}]{van_de_Weygaert2009}
{van de Weygaert}, R., \& {Schaap}, W. 2009, in Data Analysis in Cosmology, ed.
  V.~J. {Mart{\'\i}nez}, E.~{Saar}, E.~{Mart{\'\i}nez-Gonz{\'a}lez}, \& M.~J.
  {Pons-Border{\'\i}a}, Vol. 665, 291--413,
  \dodoi{10.1007/978-3-540-44767-2_11}

\bibitem[{{van der Burg} {et~al.}(2013){van der Burg}, {Muzzin}, {Hoekstra},
  {Lidman}, {Rettura}, {Wilson}, {Yee}, {Hildebrandt}, {Marchesini},
  {Stefanon}, {Demarco}, \& {Kuijken}}]{van_der_Burg2013}
{van der Burg}, R.~F.~J., {Muzzin}, A., {Hoekstra}, H., {et~al.} 2013, \aap,
  557, A15, \dodoi{10.1051/0004-6361/201321237}

\bibitem[{{Vargas} {et~al.}(2014){Vargas}, {Bish}, {Acquaviva}, {Gawiser},
  {Finkelstein}, {Ciardullo}, {Ashby}, {Feldmeier}, {Ferguson}, {Gronwall},
  {Guaita}, {Hagen}, {Koekemoer}, {Kurczynski}, {Newman}, \&
  {Padilla}}]{Vargas2014}
{Vargas}, C.~J., {Bish}, H., {Acquaviva}, V., {et~al.} 2014, \apj, 783, 26,
  \dodoi{10.1088/0004-637X/783/1/26}

\bibitem[{{Wang} {et~al.}(2016){Wang}, {Elbaz}, {Daddi}, {Finoguenov}, {Liu},
  {Schreiber}, {Mart{\'\i}n}, {Strazzullo}, {Valentino}, {van der Burg},
  {Zanella}, {Ciesla}, {Gobat}, {Le Brun}, {Pannella}, {Sargent}, {Shu}, {Tan},
  {Cappelluti}, \& {Li}}]{wang16}
{Wang}, T., {Elbaz}, D., {Daddi}, E., {et~al.} 2016, \apj, 828, 56,
  \dodoi{10.3847/0004-637X/828/1/56}

\bibitem[{{Weiss} {et~al.}(2021){Weiss}, {Bowman}, {Ciardullo}, {Zeimann},
  {Gronwall}, {Mentuch Cooper}, {Gebhardt}, {Hill}, {Blanc}, {Farrow},
  {Finkelstein}, {Gawiser}, {Janowiecki}, {Jogee}, {Schneider}, \&
  {Wisotzki}}]{weiss21}
{Weiss}, L.~H., {Bowman}, W.~P., {Ciardullo}, R., {et~al.} 2021, \apj, 912,
  100, \dodoi{10.3847/1538-4357/abedb9}

\bibitem[{{White} {et~al.}(2024){White}, {Raichoor}, {Dey}, {Garrison},
  {Gawiser}, {Lang}, {Lee}, {Myers}, {Schlegel}, {Valdes}, {Aguilar}, {Ahlen},
  {Brooks}, {Chaussidon}, {Claybaugh}, {Dawson}, {de la Macorra}, {Dey},
  {Doel}, {Fanning}, {Font-Ribera}, {Forero-Romero}, {Gontcho}, {Gutierrez},
  {Guy}, {Honscheid}, {Kirkby}, {Kremin}, {Landriau}, {Le Guillou}, {Levi},
  {Magneville}, {Manera}, {Martini}, {Meisner}, {Miquel}, {Moon}, {Newman},
  {Niz}, {Palanque-Delabrouille}, {Park}, {Percival}, {Prada}, {Rossi},
  {Ruhlmann-Kleider}, {Sanchez}, {Schlafly}, {Schubnell}, {Seo}, {Sprayberry},
  {Tarl{\'e}}, {Weaver}, {Yang}, {Y{\`e}che}, \& {Zou}}]{White2024}
{White}, M., {Raichoor}, A., {Dey}, A., {et~al.} 2024, arXiv e-prints,
  arXiv:2406.01803, \dodoi{10.48550/arXiv.2406.01803}

\bibitem[{{Yamada} {et~al.}(2012){Yamada}, {Nakamura}, {Matsuda}, {Hayashino},
  {Yamauchi}, {Morimoto}, {Kousai}, \& {Umemura}}]{Yamada2012}
{Yamada}, T., {Nakamura}, Y., {Matsuda}, Y., {et~al.} 2012, \aj, 143, 79,
  \dodoi{10.1088/0004-6256/143/4/79}

\bibitem[{{Zhu} {et~al.}(2021){Zhu}, {Zhang}, \& {Feng}}]{Zhu2021}
{Zhu}, W., {Zhang}, F., \& {Feng}, L.-L. 2021, \apj, 920, 2,
  \dodoi{10.3847/1538-4357/ac15f1}

\end{thebibliography}
\bibliographystyle{aasjournal}

\end{document}